\documentclass[useAMS,usenatbib]{mnras}
\usepackage{graphicx,natbib,amssymb,amsmath,stmaryrd,makecell}
\usepackage{txfonts}
\usepackage{xcolor}
\usepackage{hyperref}

\newcommand{\expf}[1]{{{\rm e}^{#1}}}

\newcommand{\Tz}{{T_{z}}}

\newcommand{\TCMB}{T_{\rm CMB}}

\newcommand{\nbb}{{n^{\rm pl}}}

\newcommand{\muc}{\mu_{\rm c}}

\newcommand{\xe}{x_{\rm e}}

\newcommand{\id}{{\,\rm d}}

\newcommand{\beq}{\begin{equation}}   %

\newcommand{\eeq}{\end{equation}}   %

\newcommand{\beqa}{\begin{eqnarray}}   %

\newcommand{\eeqa}{\end{eqnarray}}   %

\newcommand{\beal}{\begin{align}}

\newcommand{\enal}{\end{align}}

\newcommand{\bspl}{\begin{split}}

\newcommand{\espl}{\end{split}}

\newcommand{\bsub}{\begin{subequations}}

\newcommand{\esub}{\end{subequations}}

\newcommand{\bmulti}{\begin{multline}}   %

\newcommand{\beqm}{\begin{mathletters}}   %

\newcommand{\eeqm}{\end{mathletters}}   %

\newcommand{\me}{m_{\rm e}}

\newcommand{\Ne}{N_{\rm e}}

\newcommand{\Te}{T_{\rm e}}

\newcommand{\The}{\theta_{\rm e}}

\newcommand{\sigT}{\sigma_{\rm T}}

\newcommand{\vek} [1]{\mbox{\boldmath${#1}$\unboldmath}}

\newcommand{\pot}[2]{#1 \times 10^{#2}}

\newcommand{\Thz}{\theta_{z}}

\title[Repeated Compton scattering]{Comparison of numerical methods for computing the repeated Compton scattering of photons in isotropic media}

\begin{document}

\author[Acharya et al.]{Sandeep Kumar Acharya$^1$\thanks{E-mail:sandeep.acharya@manchester.ac.uk}
Jens Chluba$^1$\thanks{E-mail:jens.chluba@manchester.ac.uk} and
Abir Sarkar$^1$\thanks{E-mail:abirsarkar.150490@gmail.com}
\\
$^1$Jodrell Bank Centre for Astrophysics, School of Physics and Astronomy, The University of Manchester, Manchester M13 9PL, U.K.
}

\date{\vspace{-0mm}{Accepted 2021 --. Received 2021 --}}

\maketitle

\begin{abstract}
%--------------------------------------------------------------
Repeated Compton scattering of photons with thermal electrons is one of the fundamental processes at work in many astrophysical plasma. Solving the exact evolution equations is hard and one common simplification is based on Fokker-Planck (FP) approximations of the Compton collision term. Here we carry out a detailed numerical comparison of several FP approaches with the exact scattering kernel solution for a range of test problems assuming isotropic media and thermal electrons at various temperatures. The Kompaneets equation, being one of the most widely used FP approximations, fails to account for Klein-Nishina corrections and enhanced Doppler boosts and recoil at high energies. These can be accounted for with an alternative FP approach based on the exact first and second moments of the scattering kernel. As demonstrated here, the latter approach works very well in dilute media, but inherently fails to reproduce the correct equilibrium solution in the limit of many scattering. Conditions for the applicability of the FP approximations are clarified, overall showing that the Kompaneets equation provides the most robust approximation to the full problem, even if inaccurate in many cases. 
We close our numerical analysis by briefly illustrating the solutions for the spectral distortions of the cosmic microwave background (CMB) after photon injection at redshift $z\lesssim 10^5$, when double Compton and Bremsstrahlung emission can be omitted. We demonstrate that the exact treatment using the scattering kernel computed with {\tt CSpack} is often needed. This work should provide an important step towards accurate computations of the CMB spectral distortions from high-energy particle cascades.
%--------------------------------------------------------------
\end{abstract}

\begin{keywords}
Cosmology - Cosmic Microwave Background; Cosmology - Theory 
\end{keywords}

\section{Introduction}
%--------------------------------------------------------------
Compton scattering is one of the fundamental processes in many astrophysical plasma \citep[e.g.,][]{Blumenthal1970, Rybicki1979}. It has important applications in active galactic nuclei, supernovae and $\gamma$-ray bursts \citep[e.g.,][]{Giannios2006, Mimica2009, Giannios2010}. It furthermore strongly affects the dynamics of accretion-flows \citep[e.g.,][]{Shakura1973, Abramowicz1988, Narayan2003, McKinney2017} and determines the evolution of electromagnetic particle cascades present during multiple phases in the evolution of the Universe \citep{Zdziarski1988, Shull1985, Slatyer2009, Valdes2010, Slatyer2015, Liu2019}. The Sunyaev-Zeldovich (SZ) effect \citep{Zeldovich1969, Carlstrom2002, SZreview2019} and thermalization of spectral distortions of the cosmic microwave background (CMB) caused by heating \citep{Sunyaev1970mu, Burigana1991, Hu1993, Chluba2011therm, Khatri2012mix, Chluba2020large}, photon injection \citep{Chluba2015GreensII, Bolliet2020PI} and interactions with non-thermal electron populations \citep{Ensslin2000, Colafrancesco2003, Acharya2019a} provide additional examples in which Comptonization plays an important role.
It is therefore crucial to understand this process for a wide range of energies and physical conditions. 

Solving the exact Compton scattering evolution equation is computationally extremely challenging in most cases. In many situations (especially in the optically-thick regime), locally one encounters a quasi-isotropic medium with thermal electrons scattering photons of various energies. These conditions greatly simplify the calculations, making the process mainly an energy redistribution/exchange problem. 

One common approach to simplifying the Compton collision term is using a Fokker-Planck approximation. In this scheme, the main assumption is that the photon distribution is sufficiently smooth across the region over which they are redistributed in one scattering event. This means that the mean energy and width of the scattering kernel have to be small in comparison to the frequency derivatives of the photon field. Taking into account only leading order terms in the electron temperature, this leads to the famous Kompaneets equation \citep{Kompa56, Weymann1965}, which describes the problem using an energy diffusion equation. It takes into account leading order Doppler boosting and broadening, recoil and stimulated recoil, which together ensure that the equilibrium spectrum is given by a Bose-Einstein distribution with constant chemical potential \citep[e.g., see][]{Sunyaev1970mu}.

Given the assumptions for the derivation of the Kompaneets equation it is clear that for problems in which the photon distribution is very narrow in comparison to the scattering kernel the Kompaneets equation is not expected to give reliable results. It is also clear that at high electron temperatures relativistic corrections become important \citep{Buchler1976, Wright1979, Sazonov1998}. Furthermore, for photons with energies $h\nu\gg k\Te$ recoil broadening and Klein-Nishina corrections start to be significant \citep[e.g..][]{Sazonov2000}. All these effects necessitate refined computational schemes. 

One possibility is to use a scattering kernel formulation of the problem \citep[e.g.,][]{Pozdniakov1979, Sazonov2000}. Here, the challenge is that simple analytic expressions for the exact thermally-averaged scattering kernel do not exist. Various approximations can be obtained in limiting cases \citep[see][for comprehensive overview]{Sazonov2000}, but even these have limitations with respect to conservation of photons and energy for many scatterings and numerical stability \citep{Chluba2020large}, such that there is no simple one-fits-it-all analytic description. This makes evaluations of the scattering integrals numerically expensive such that FP approximations, also with improvements over the Kompaneets equation \citep[e.g.,][]{Sazonov1998, Itoh98, Belmont2008}, are often used instead. 

However, quasi-exact and highly-efficient representations of the scattering kernel are now possible using {\tt CSpack}\footnote{{\tt CSpack} is available at \url{www.chluba.de/CSpack}} \citep{CSpack2019}. {\tt CSpack} is based on exact expressions for the scattering kernel given by \citet{1968PhRv..167.1159J} written in a numerically stable way and combined with efficient quadrature schemes to obtain a highly accurate representation. This scheme was already applied to the thermalization problem of CMB spectral distortions at high redshifts \citep{Chluba2020large}, for the first time directly showing how relativistic temperature and Klein-Nishina corrections affect energy release distortions. Here we now explicitly solve several photon scattering problems to demonstrate in which regimes other approximations fail. One important comparison is with solutions to the Kompaneets equation, for which we demonstrate the shortcomings mentioned above. We also show that extended FP approximations have severe limitations in particular for many scatterings, as the correct equilibrium solution is not automatically ensured (see Fig.~\ref{fig:equilibrium_soln}). The numerical scheme based on {\tt CSpack} and presented here together with modern computing and parallelization allows us to overcome these limitations, providing and important step towards an exact modeling of general Comptonization problems.

The paper is structured as follows: After introducing the photon evolution equations and discussing some of its properties in Sect.~\ref{sec:kinetic_equation}, we consider various FP approaches in Sect.~\ref{sec:FP_expansion}. Having the modeling of CMB spectral distortions in mind, we also extend the description of the diffusion coefficients to include stimulated scattering terms, which have been neglected previously.
A detailed comparison of the exact solution with various approximation schemes is presented in Sect.~\ref{sec:num_sols} for test problems with single photon injection at various energies and electron temperatures. This allows us to clearly highlight the shortcomings of FP schemes. As an example, we apply our exact solver to the thermalization problem of CMB spectral distortions at $z\lesssim 10^5$, here with a focus on illustrating the shortcomings of the Kompaneets solutions. These solutions were previously discussed in \citet{Chluba2015GreensII}, but our analysis clearly motivates the necessity for a more comprehensive follow-up study in this context. Overall, our study shows that the Kompaneets equation is highly robust but approximate in detail, and that the kernel approach is needed for obtaining accurate solutions.

%\vspace{-3mm}
%--------------------------------------------------------------
\section{Photon evolution equation}
\label{sec:kinetic_equation}
%--------------------------------------------------------------
The repeated scattering of photons by thermal electrons inside an isotropic medium can be described using the kinetic equation for the photon occupation number, $n_0=n(\omega_0)$, at energy\footnote{We shall use the common definitions of physical constants ($c, h,\me$, etc).} $\omega_0=h\nu_0/\me c^2$ \citep[e.g., for comparison see][albeit with slightly different conventions]{Sazonov2000}
%-------------------------------------------------------------------
\begin{align}
\frac{\text{d}n_0}{\text{d}\tau} 
&= \int P^{\rm th}(\omega_0 \rightarrow \omega) \,\Big[\expf{\frac{\omega-\omega_0}{\The}} \,n(1+n_0) - n_0(1+n) \Big]\,\id\omega.
\label{eq:kin_eq2_starting_point}
\end{align}
%-------------------------------------------------------------------
Here, we introduced the Thomson optical depth, $\tau = \int c \Ne \sigT \id t$ and also used the Compton scattering kernel, $P^{\rm th}(\omega_0 \rightarrow \omega)$ for thermal electrons of temperature $\The=k\Te/\me c^2$, with normalization $\int P^{\rm th}(\omega_0 \rightarrow \omega)\id\omega=\sigma(\omega_0, \The)/\sigT$, where $\sigma(\omega_0, \The)$ is the total Compton scattering cross section. The kernel describes the redistribution of photons from the initial energy $\omega_0$ to $\omega$, while the reverse process is given by the detailed balance relation
%-------------------------------------------------------------------
\begin{align}
P^{\rm th}(\omega \rightarrow \omega_0)&=\frac{\omega_0^2}{\omega^2}\,\expf{\frac{\omega-\omega_0}{\The}}\,P^{\rm th}(\omega_0 \rightarrow \omega).
\end{align}
%-------------------------------------------------------------------
These kernels are obtained from the single-momentum scattering kernel, $P(\omega_0 \rightarrow \omega, p_0)$ after integrating over the electron momenta and can be computed efficiently using {\tt CSpack}.

Solving the kinetic equation, Eq.~\eqref{eq:kin_eq2_starting_point}, in general is difficult. It has the equilibrium solution
%-------------------------------------------------------------------
\begin{align}
n^{\rm eq}&=\frac{1}{\expf{x+\muc}-1},
\end{align}
%-------------------------------------------------------------------
where $x=\omega/\The$ and $\muc$ is a constant photon chemical potential that allows fixing the total number of photons in the spectrum, since Compton scattering is a photon number conserving process. 
In full thermodynamic equilibrium, one can has $\muc=0$, such that the Bose-Einstein spectrum $n^{\rm eq}$ becomes a blackbody at temperature $\The$.
Compton scattering alone cannot establish $\muc=0$, but 
photon number non-conserving processes such as Bremsstrahlung and double Compton emission\footnote{See \citet{BRpack} and \citet{Ravenni2020DC} for most recent discussion of these classical processes.} are required to achieve this .
For $\muc>0$, a lack of photons with respect to the equilibrium blackbody is present, while for $\muc<0$, it is an excess. The solution $\muc={\rm const}<0$ is singular at $x=-\muc$; however, this case is not reached in physical problems due to processes like Bremsstrahlung and double Compton scattering.

If the total number of photons is small, we have $\muc\gg 1$ and thus $n^{\rm eq}\ll 1$, such that the stimulated terms $(1+n)$ and $(1+n_0)$ in Eq.~\eqref{eq:kin_eq2_starting_point} can be omitted. In this case, we obtain a Wien spectrum
%-------------------------------------------------------------------
\begin{align}
n^{\rm eq}&\approx \expf{-x-\muc}=n_{\rm c}\,\expf{-x}
\end{align}
%-------------------------------------------------------------------
in equilibrium. Here, $n_{\rm c}$ is a constant that is directly related to the total number of photons of the distribution. This solution is frequently encountered in diluted astrophysical plasmas \citep{Blumenthal1970, Rybicki1979}.

%--------------------------------------------------------------
\subsection{Linearized evolution equation with stimulated terms}
\label{sec:lin_stim_evol_eq}
%--------------------------------------------------------------
For problems in astrophysical plasmas, it is often the case that either stimulated scattering terms can be entirely neglected or that one is close to equilibrium with $n=n^{\rm eq}+\Delta n$ and small distortion $\Delta n\ll n^{\rm eq}$. In both cases the general evolution equation can be linearized to solve for the distortion $\Delta n$. The kinetic equation, Eq.~\eqref{eq:kin_eq2_starting_point} can then be cast into the form (see Appendix~\ref{app:intermediate_Eq5})
%-------------------------------------------------------------------
\bsub
\label{eq:col_CS_Dn_lin}
\begin{align}
\frac{\text{d}n_0}{\text{d}\tau} 
&=\int P(\omega_0 \rightarrow \omega, \The) \,\Big[\expf{x-x_0} \,\Delta n \, f_{\omega,\omega_0} - \Delta n_0\, f_{\omega_0,\omega}\Big]\,\id\omega
\\[1mm]
f_{\omega_0,\omega}&=\frac{1+n^{\rm eq}}{1+n^{\rm eq}_0}=\frac{1-\expf{-x_0-\muc}}{1-\expf{-x-\muc}}.
\end{align}
\esub
%-------------------------------------------------------------------
The factor $f_{\omega,\omega'}$ is caused by stimulated scattering effects and can be set to $1$, should these be neglected. This latter case is equivalent to a Wien equilibrium with $n^{\rm eq}\ll 1$ and $n\approx \Delta n\ll 1$. 

In kinetic equilibrium, the solution of Eq.~\eqref{eq:col_CS_Dn_lin} is determined by
%-------------------------------------------------------------------
\begin{align}
\label{eq:Equil_gen}
\Delta n^{\rm eq}\,\expf{x-x_0} f_{\omega,\omega_0}=\Delta n^{\rm eq}_0\,f_{\omega_0,\omega}\longleftrightarrow \frac{\Delta n^{\rm eq}\expf{x}}{(1+n^{\rm eq})^2}=\frac{\Delta n^{\rm eq}_0\expf{x_0}}{(1+n^{\rm eq}_0)^2},
\end{align}
%-------------------------------------------------------------------
which implies the equilibrium solution
%-------------------------------------------------------------------
\begin{align}
\label{eq:Dneq_stim_equ}
\Delta n^{\rm eq}(x) &= \Delta n_{\rm c}\,\frac{\expf{x}}{(\expf{x}-1)^2},
\end{align}
%-------------------------------------------------------------------
defining a Bose-Einstein spectrum with small chemical potential. When simulated terms are neglected, a Wien distribution is again found in equilibrium, $\Delta n^{\rm eq}(x) \approx \Delta n_{\rm c}\,\expf{-x}$.

%--------------------------------------------------------------
\subsection{Moments of the evolution equation}
\label{sec:kinetic_equation_moments}
%--------------------------------------------------------------
Using the detailed balance relation it is straightforward to show that the moments of the evolution equation are given by (see Appendix~\ref{app:intermediate_Eq8} for details)
%-------------------------------------------------------------------
\bsub
\label{eq:target_moments_kin}
\begin{align}
\int \omega_0^{k+2} \frac{\text{d}n_0}{\text{d}\tau} \id \omega_0
&=\int \omega_0^{k+2} \,\Lambda_k(\omega_0, \The, n)\,n_0 \id \omega_0
\\
\Lambda_k(\omega_0, \The, n)&=\left<\left[\frac{\omega^k}{\omega_0^k}  - 1\right] (1+n) \right>
\\
\left<X(\omega)\right>&=\int X(\omega)\,P^{\rm th}(\omega_0 \rightarrow \omega) \id\omega.
\end{align}
\esub
%-------------------------------------------------------------------
These expressions are exact given a solution for the photon field.
For $k=0$, one naturally has $\int \omega_0^{k+2} \frac{\text{d}n_0}{\text{d}\tau} \id \omega_0=0$, which reflects photon number conservation. For $k=1$, one obtains the net energy transfer between the electrons and the photons under Compton scattering, while $k=2$ describes the change of the dispersion of the photon field.

Assuming that stimulated scattering terms are negligible [i.e., $(1+n)\approx 1$ in the Wien limit $n\ll 1$], with the identity
%-------------------------------------------------------------------
\begin{align}
\label{eq:moments_id}
\frac{\omega^k}{\omega_0^k}  - 1
&=\sum_{m=1}^{k}\,\binom{k}{m}\,\left(\frac{\omega-\omega_0}{\omega_0}\right)^m
\end{align}
%-------------------------------------------------------------------
we then may write $\Lambda_k(\omega_0, \The, n)\approx \Lambda^{\rm w}_k(\omega_0, \The, n)$ with
%-------------------------------------------------------------------
\bsub
\label{eq:moments}
\beal
\Lambda^{\rm w}_k(\omega_0, \The, n)&
=\sum_{m=1}^{k}\,\binom{k}{m}\,\,\Sigma_m(\omega_0, \The)
\\
\Sigma_m(\omega_0, \The) &= \left<\left(\frac{\omega-\omega_0}{\omega_0}\right)^k\right>
\end{align}
\esub
%-------------------------------------------------------------------
Here, $\Sigma_m(\omega_0, \The)$ are the moments of the scattering kernel, which can be computed efficiently with {\tt CSpack}.

If instead we are close to kinetic equilibrium, i.e., we have a distribution $n=n^{\rm eq}+\Delta n$ with $\Delta n\ll 1$, such that stimulated effects related to $n^{\rm eq}$ are relevant but terms $\mathcal{O}(\Delta n)^2$ can be omitted, from Eq.~\eqref{eq:target_moments_kin} we have (Appendix~\ref{app:intermediate_Eq11} for details)
%-------------------------------------------------------------------
\begin{align}
\label{eq:target_moments_kin_lin_sim}
\int \omega_0^{k+2} \frac{\text{d}n_0}{\text{d}\tau} \id \omega_0
&\approx 
\int \omega_0^{k+2}\left<\left[\frac{\omega^{k}}{\omega_0^{k}}-1\right] f_{\omega_0,\omega}\right> 
\,\Delta n_0 \,\text{d}\omega_0.
\end{align}
%-------------------------------------------------------------------
This means that 
$\Lambda_k(\omega_0, \The)\approx \Lambda^*_k(\omega_0, \The)$ with $\Lambda^*_k(\omega_0, \The)$ given by Eq.~\eqref{eq:moments} after replacing $\Sigma_m(\omega_0, \The)$ by
%-------------------------------------------------------------------
\beal
\label{eq:moments_stim}
\Sigma^*_m(\omega_0, \The) 
&= \left<\Bigg(\frac{\omega-\omega_0}{\omega_0}\Bigg)^m \,f_{\omega_0,\omega}\right>.
\end{align}
%-------------------------------------------------------------------
The aforementioned moments will become important for writing the Fokker-Planck expansions of the kinetic equation. We will also compute the evolution of the moments for various numerical solutions below to compare the precision of different approaches. An important role falls to the first moment of the kernel, as it defines the net energy exchange between electrons and photons and thus determines the Compton equilibrium temperature.

%\vspace{-3mm}
\section{Fokker-Planck approaches}
\label{sec:FP_expansion}
%--------------------------------------------------------------
Solving the general evolution equation, Eq.~\eqref{eq:kin_eq2_starting_point}, can be quite difficult. Below we will discuss a method that directly discretizes the integro-differential equation, however, the resultant matrix equation, Eq.~\eqref{eq:gen_matrix_eq}, can become very dense. For numerical applications it is thus beneficial to perform a Fokker-Planck (FP) approximation of the problem. 
The standard approach is to replace $n(\omega)$ by a Taylor-series around $\omega_0$ and then compute all the moments of the kernel \citep[e.g.,][]{Sazonov1998, Itoh98}. Starting from Eq.~\eqref{eq:kin_eq2_starting_point}, in compact form the exact kinetic equation can also be written as \citep{Chluba2001Diploma}
%-------------------------------------------------------------------
\begin{align}
\frac{\text{d}n_0}{\text{d}\tau} 
%&= \int P^{\rm th}(\omega_0 \rightarrow \omega) \Big[\expf{\frac{\omega-\omega_0}{\The}} \,n(1+n_0) - n_0(1+n) \Big]\,\text{d}\omega
%\nonumber
%\\
&=\sum_{k=1}^\infty  
\frac{x_0^k\,\Sigma_k}{k!} 
\left\{
(1+n_0)\,(1+\partial_{x_0})^k -n_0 \,\partial^k_{x_0} 
\right\} n_0.
\label{eq:kin_eq2_FP1}
\end{align}
%-------------------------------------------------------------------
with the moments $\Sigma_k=\left<(\Delta \nu/\nu_0)^k\right>$. This approach yields a partial differential equation, which describes the evolution as a diffusion process in frequency space. 
Including only terms up to first order in $\The\ll 1$, we have \citep[see][]{Sazonov2000}
%-------------------------------------------------------------------
\begin{align}
\label{eq:kompaneets_mom}
\Sigma_0\approx 1,
\quad
\Sigma_1\approx (4-x_0)\The,
\quad
\Sigma_2\approx 2 \The
\end{align}
%-------------------------------------------------------------------
and $\Sigma_k\approx 0$ for $k>2$. Inserting this into Eq.~\eqref{eq:kin_eq2_FP1} then yields the well-known Kompaneets equation \citep{Kompa56}
%-------------------------------------------------------------------
\begin{align}
\frac{\text{d}n_0}{\text{d}\tau} 
&\approx \frac{\The}{x_0^2} \frac{\partial}{\partial x_0} x_0^4 \,\left[\frac{\partial}{\partial x_0} n_0+ n_0(1+n_0)\right].
\label{eq:Kompaneets}
\end{align}
%-------------------------------------------------------------------
Including higher order corrections in $\The$ implies presence of higher order derivatives $\partial^k_{x_0} n_0$. Some explicit expressions are given in the literature \citep{Sazonov1998, Challinor1998, Itoh98} and can also be directly obtained using Eq.~\eqref{eq:kin_eq2_FP1}. However, for practical applications, this approach is not easy to implement and solving the full kernel equation is preferred.

The Kompaneets equation has been applied to many physical problems. It naturally conserves photon number and a blackbody spectrum at the temperature of the electrons. However, one shortcoming of the Kompaneets equation is that it does not correctly include the reduction of the scattering probability at high energies. In addition, the evolution of the first and second moment of the photon distribution function can only be captured to leading order in temperature. The question now is whether one can improve some of these short-comings and which inevitable trade-offs remain.

%\vspace{-3mm}
\subsection{First improved Fokker-Planck equation}
\label{sec:New_attempt}
%--------------------------------------------------------------
The first attempt to improve over the Kompaneets equation is to try and fix both the first and second moment of the collision term in addition to ensuring photon number conservation. We shall start with the linearized problem, Eq.~\eqref{eq:col_CS_Dn_lin}, neglecting stimulated terms ($f_{\omega, \omega_0}=1$). The first few moments of this equation then are
%-------------------------------------------------------------------
\bsub
\label{eq:target_moments}
\begin{align}
\int \omega_0^2 \frac{\text{d}n_0}{\text{d}\tau} \id \omega_0
&= 0
\\
\int \omega_0^3 \frac{\text{d}n_0}{\text{d}\tau} \id \omega_0
&=\int \Sigma_1(\omega_0,\The)\,\omega_0^3 \,\Delta n_0 \id \omega_0
\\
\int \omega_0^4 \frac{\text{d}n_0}{\text{d}\tau} \id \omega_0
&=\int \left[\Sigma_2(\omega_0,\The)+2\Sigma_1(\omega_0,\The)\right]\,\omega_0^4 \,\Delta n_0 \id \omega_0.
\end{align}
\esub
%-------------------------------------------------------------------
To obtain a diffusion equation (second order only) that conserves the moments Eq.~\eqref{eq:target_moments} of the kinetic equation, we can use the Ansatz
%-------------------------------------------------------------------
\begin{align}
\label{eq:Ansatz_lin}
\frac{\text{d}n_0}{\text{d}\tau} 
&= \frac{1}{\omega_0^2}\partial_{\omega_0} \left[  \mathcal{D}(\omega_0) \,\partial_{\omega_0} \Delta n_0 +\mathcal{A}(\omega_0) \Delta n_0 \right].
\end{align}
%-------------------------------------------------------------------
In this form, photon number conservation is directly built in and the two free functions provide the freedom to conserve the first two moments of the evolution equation. However, the equilibrium solution is not automatically driven towards a Wien spectrum, $\Delta n(\omega)\propto \expf{-\omega/\The}$, as we will discuss below.

Taking the moments of Eq.~\eqref{eq:Ansatz_lin} and integrating a few times by parts one then has
%-------------------------------------------------------------------
\bsub
\label{eq:target_moments_Ansatz}
\begin{align}
\int \omega_0^3 \frac{\text{d}n_0}{\text{d}\tau} \id \omega_0
&=\int  \left[ \frac{\mathcal{D}'}{\omega_0^3} - \frac{\mathcal{A}}{\omega_0^3} \right]
\omega_0^3 \,\Delta n_0 \id \omega_0
\\
\int \omega_0^4 \frac{\text{d}n_0}{\text{d}\tau} \id \omega_0
&=\int  2\left[ \frac{\mathcal{D}}{\omega_0^4}+\frac{\mathcal{D}'}{\omega_0^3} - \frac{\mathcal{A}}{\omega_0^3} \right]
\omega_0^4 \,\Delta n_0 \id \omega_0,
\end{align}
\esub
%-------------------------------------------------------------------
where the prime indicates derivatives with respect to $\omega_0$. Comparing this with Eq.~\eqref{eq:target_moments}, we then have
%-------------------------------------------------------------------
\begin{align}
\label{eq:Coefficients}
\mathcal{A}&=\omega_0^4 \left\{\frac{2\Sigma_2-\Sigma_1}{\omega_0}+\frac{\Sigma'_2}{2}\right\},
\qquad
\mathcal{D}=\omega_0^4 \,\frac{\Sigma_2}{2},
\end{align}
%-------------------------------------------------------------------
such that we can finally write the first improved FP equation
%-------------------------------------------------------------------
\begin{align}
\label{eq:FP_approach1}
\frac{\text{d}n_0}{\text{d}\tau} 
&= \frac{1}{\omega_0^2}\partial_{\omega_0} \omega_0^4 \left[\frac{\Sigma_2}{2} \partial_{\omega_0} \Delta n_0 
+ \left\{\frac{2\Sigma_2-\Sigma_1}{\omega_0}+\frac{\Sigma'_2}{2}\right\}  \Delta n_0 \right].
\end{align}
%-------------------------------------------------------------------
This expression is equivalent to the second order FP approach presented in \citet{Belmont2008} and correctly captures the evolution of the first and second moments of the photon field at intermediate temperatures and number of scatterings, as we demonstrate below. However, it omits stimulated scattering effects and in general does not yield the correct equilibrium solution in the limit of many scatterings. In addition, higher order moment terms become relevant at high temperatures, as we demonstrate below. 

%---------------
\begin{figure}
\centering 
\includegraphics[width=\columnwidth]{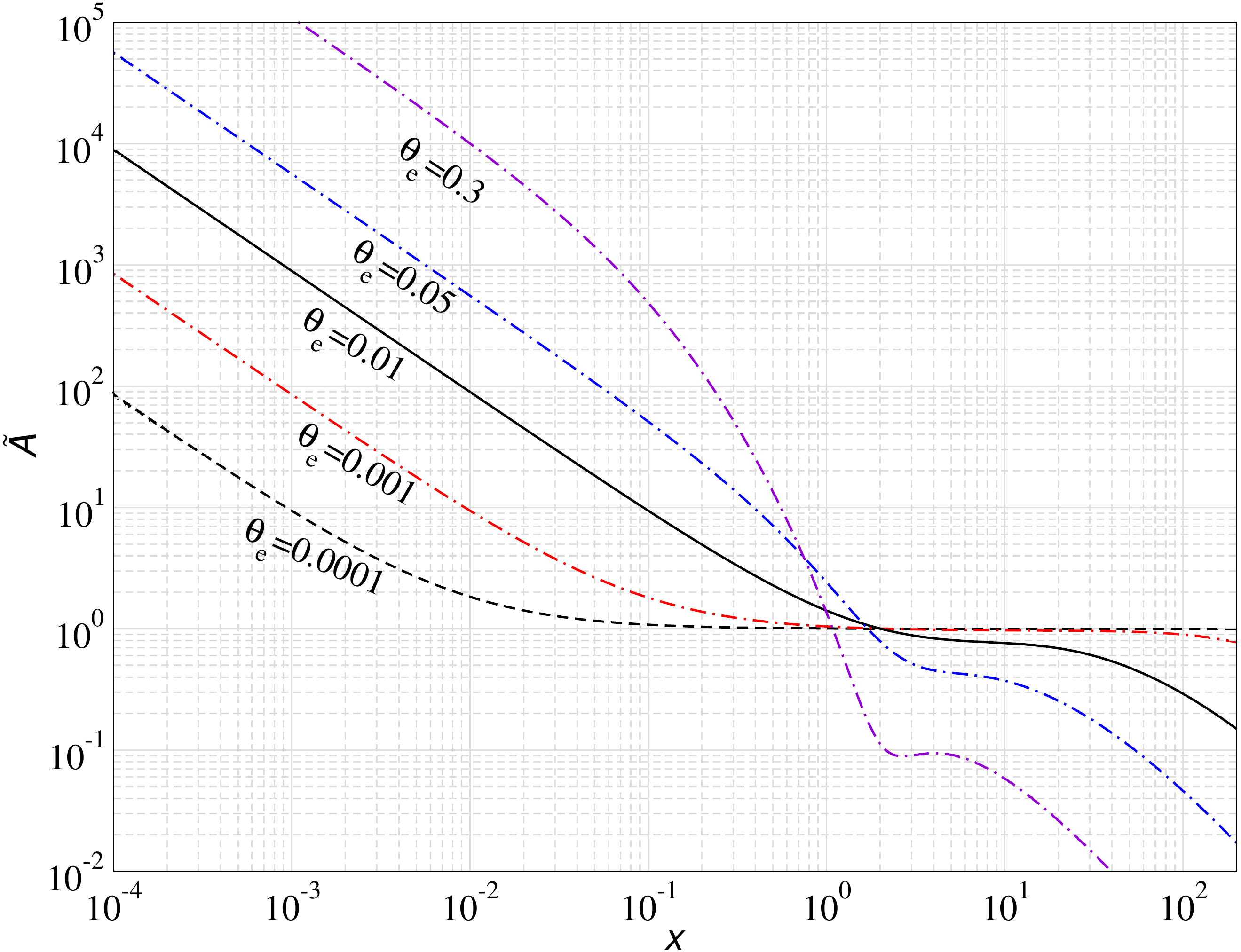}
\\[3mm]
\includegraphics[width=\columnwidth]{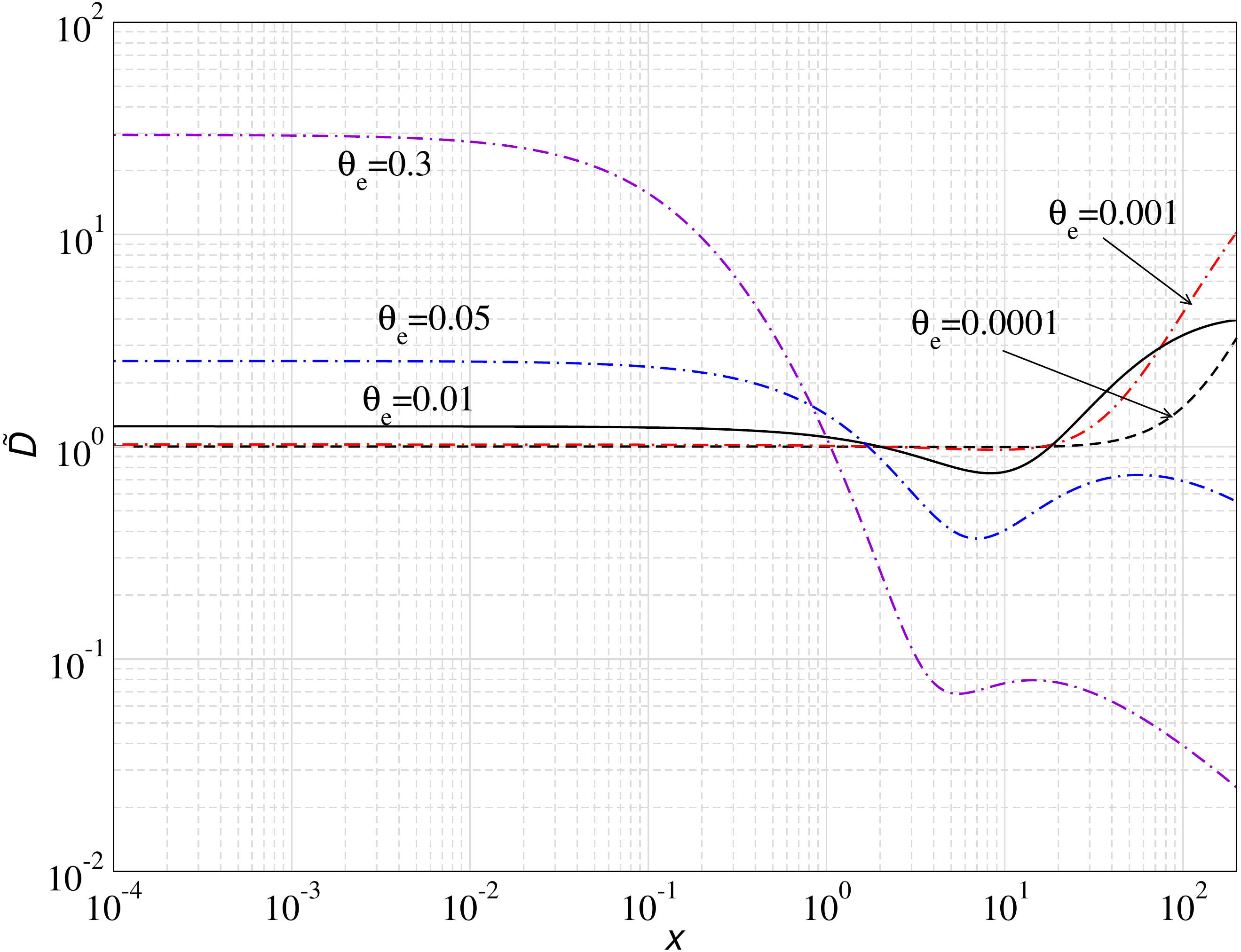}
\\
\caption{Diffusion coefficients ($\tilde{A}=\mathcal{A}/[\omega_0^4]$ and $\tilde{D}=\mathcal{D}/[\theta_e\omega_0^4]$ in the upper and lower panel, respectively) for the improved Fokker-Planck approximation without stimulated terms [Eq.~\eqref{eq:FP_approach1}] for different temperatures. The limit $\tilde{A}=1$ and $\tilde{D}=1$ corresponds to Kompaneets approximation.}
\label{fig:Diffusion_coeff_nostim}
\end{figure}
%---------------
In Fig.~\ref{fig:Diffusion_coeff_nostim}, we illustrate the improved FP coefficients as a function of photon energy for several electron temperatures. At low temperature, one can see that $\mathcal{A}\rightarrow \omega_0^4$ and $\mathcal{D}\rightarrow \omega_0^4\,\The$ (i.e., $\tilde{A}=1$ and $\tilde{D}=1$ in the figure). This can be readily confirmed using the low temperature expansion for the moments, which imply
$\Sigma_1\approx 4\The - \omega_0$ and $\Sigma_2\approx2\The$, $2\Sigma_2-\Sigma_1\approx\omega_0$, $\Sigma'_2\approx0$ and thus $\mathcal{A}\approx \omega_0^4$ and $\mathcal{D}\approx \omega_0^4 \The$.
Inserting this into Eq.~\eqref{eq:FP_approach1}, yields
%-------------------------------------------------------------------
\begin{align}
\nonumber
\frac{\text{d}n_0}{\text{d}\tau} 
&= \frac{1}{\omega_0^2}\partial_{\omega_0} \omega_0^4 \left[\The\partial_{\omega_0} \Delta n_0 
+ \Delta n_0 \right]
= \frac{\The}{x_0^2}\partial_{x_0} x_0^4 \left[\partial_{x_0} \Delta n_0 + \Delta n_0\right],
\end{align}
%-------------------------------------------------------------------
which is the well-known Kompaneets equation in the Wien-limit, i.e., assuming $\Delta n\ll1$. In this limit, we find 
$\mathcal{D}/\mathcal{A} \approx \The$, such that the equilibrium solution is correctly given by $\Delta n=\Delta n_{\rm c} \expf{-\omega/\The}$. 

However, for none of the cases illustrated in Fig.~\ref{fig:Diffusion_coeff_nostim} is the Kompaneets limit exactly reached. Even for the lowest temperature (i.e., $\The=10^{-4}$), one can see visible departures from $\mathcal{A}_{\rm K} = \omega_0^4$ at $x\lesssim 0.01$. This is the regime where Doppler terms strongly exceed recoil terms. By carefully computing $\mathcal{A}$ using the Taylor series expressions for the first and second kernel moments \citep[see][]{CSpack2019}, we find the extended approximation
%-------------------------------------------------------------------
\begin{align}
\nonumber
\mathcal{A}&\approx
\omega_0^4\left[1-\frac{79}{2}\The+\frac{85}{\omega_0}\The^2\right]
\end{align}
%-------------------------------------------------------------------
at very low frequencies. This shows that at $\omega_0< 85\The^2$ or $x < 85\The$ one expect $\mathcal{A}>\mathcal{A}_{\rm K}$. For $\The=10^{-4}$ this yields $x\simeq 10^{-2}$ for the transition, which is in good agreement with Fig.~\ref{fig:Diffusion_coeff_nostim}.

In a similar manner, we observe in Fig.~\ref{fig:Diffusion_coeff_nostim} that even at low temperature $\mathcal{D}>\mathcal{D}_{\rm K}=\omega_0^4\The$ at high frequencies. The origin of the departures in this case is the transition to recoil-dominated scattering, which increases line-broadening. Including the next terms in $\omega_0$ and $\The$ for the second kernel moment we find
%-------------------------------------------------------------------
\begin{align}
\nonumber
\mathcal{D}&\approx
\omega_0^4 \The\left[1+\frac{47}{2}\The+\frac{7}{10}\frac{\omega_0^2}{\The}\right],
\end{align}
%-------------------------------------------------------------------
implying that at $\omega_0\gtrsim 1.2\sqrt{\The}$ or $x\gtrsim 1.2/\sqrt{\The}$ one has $\mathcal{D}>\mathcal{D}_{\rm K}$. For $\The=10^{-4}$, this implies a transition at $x\simeq 10^2$, again in good agreement with the numerical result (see Fig.~\ref{fig:Diffusion_coeff_nostim}). At very low frequencies, we can also observe the increase of $\mathcal{D}$ over $\mathcal{D}_{\rm K}$, caused by higher order temperature corrections.

%---------------
\begin{figure}
\centering 
\includegraphics[width=\columnwidth]{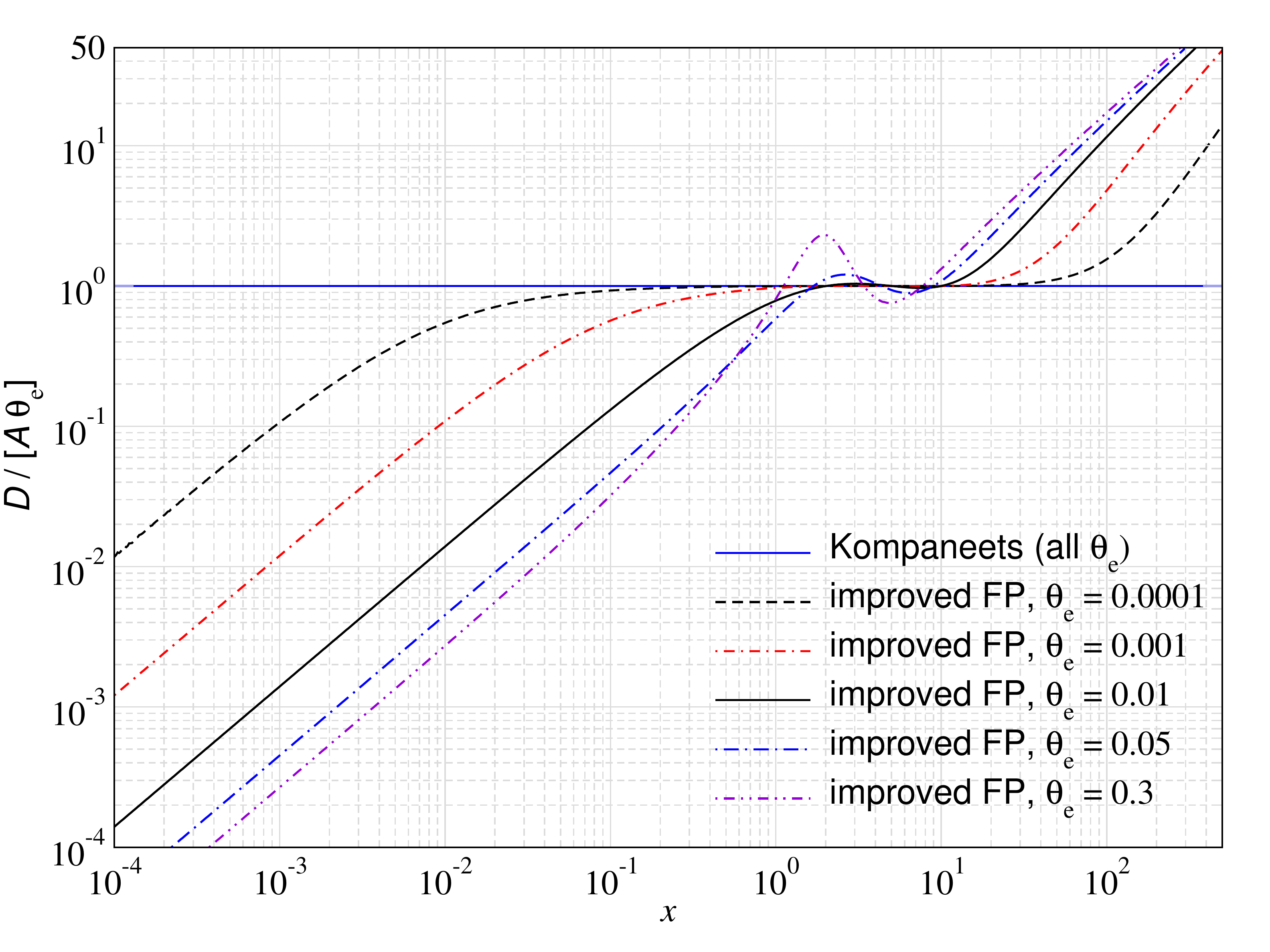}
\\
\caption{Ratio of diffusion coefficients $\mathcal{D}/[\mathcal{A}\The]$. To ensure detailed balance this ratio has to equal unity as in the Kompaneets limit. At higher temperature, departures become important at very low and high frequencies. The curves were computed using {\tt CSpack}.}
\label{fig:D_A_ratio}
\end{figure}
%-------------------------------------

%---------------
\begin{figure}
\centering 
\includegraphics[angle=0,width=\columnwidth]{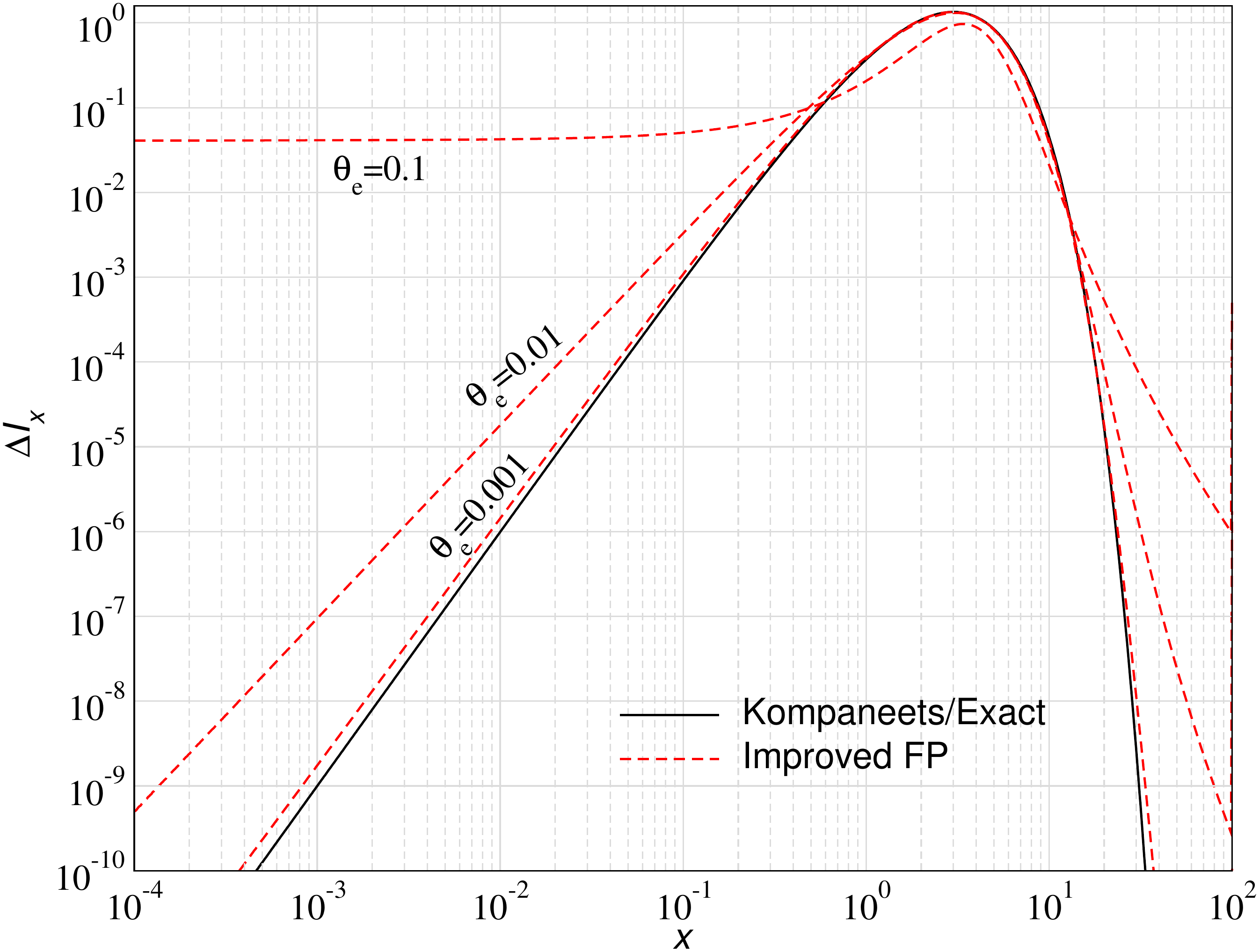}
\caption{Solution of Eq.~\eqref{eq:FP_approach1_eq} for different temperatures in comparison to the Kompaneets/Exact solution, $n(x)=\expf{-x}$. In each case, the total number of photons is fixed to the same value.}
\label{fig:equilibrium_soln}
\end{figure}
%-------------------------------------

One important consequence of the departures for $\mathcal{A}$ and $\mathcal{D}$ from the Kompaneets values is that the equilibrium solution generally departs from a Wien spectrum. This can be easily shown by computing the equilibrium solution of Eq.~\eqref{eq:FP_approach1}, which is determined by the condition
%-------------------------------------------------------------------
\begin{align}
\label{eq:FP_approach1_eq}
\frac{\partial \Delta n_0}{\partial {\omega_0}}
&= -\frac{\mathcal{A}(\omega_0)}{\mathcal{D}(\omega_0)}\, \Delta n_0
\nonumber
\\
\rightarrow&\quad
\Delta n^{\rm eq}(\omega)=\Delta n_{\rm c}\exp\left(-\int_0^\omega\frac{\mathcal{A}(\omega_0)}{\mathcal{D}(\omega_0)}\id \omega_0\right).
\end{align}
%-------------------------------------------------------------------
For $\mathcal{D}/\mathcal{A}= \The$, one readily finds $\Delta n^{\rm eq}(\omega)=n_{\rm c}\,\expf{-\omega/\The}$. 
However, by plotting the ratio $\mathcal{D}/[\mathcal{A}\The]$ using the exact FP coefficients (see Fig.~\ref{fig:D_A_ratio}), we can see that this condition is not generally fulfilled in the improved FP approach.

To further illustrate this point, in Fig.~\ref{fig:equilibrium_soln}, we plot the equilibrium solution obtained from Eq.~\eqref{eq:FP_approach1_eq} for several temperatures. For comparison, we also show the Wien spectrum, $n(x)=\expf{-\omega/\theta_e}=\expf{-x}$, which is obtained in the Kompaneets limit and also when using the exact scattering kernel approach.
For $\The=10^{-3}$, the improved FP solution is fairly close to the exact equilibrium solution and corrections remain small in the considered computational domain. 
However, there are visible changes due to $\mathcal{D}/\mathcal{A}\neq\The$ for the other cases. To leading order, the coefficient $\mathcal{A}$ captures the recoil of photons while $\mathcal{D}$ captures Doppler broadening and boosting. At lower frequencies, $\mathcal{A}>\mathcal{D}$ (Fig. \ref{fig:D_A_ratio}), which stretches the low-frequency tail of the equilibrium solution. This effect is amplified for higher temperatures as the ratio of $\mathcal{D}/\mathcal{A}$ departs further from the Kompaneets limit. 
At higher frequencies, $\mathcal{A}<\mathcal{D}$. This results in slower decay of the equilibrium solution as compared to $\expf{-\omega/\theta_e}$. We will see that these solutions are approached in the limit of many scatterings.

%\vspace{-3mm}
\subsection{Second improved Fokker-Planck equation}
\label{sec:New_attempt_II}
%--------------------------------------------------------------
As shown above, simultaneously conserving first and second moment for the linearized collision term does not lead to the correct equilibrium solution. A modified Ansatz that ensures the correct equilibrium case can thus be
%-------------------------------------------------------------------
\begin{align}
\label{eq:Ansatz}
\frac{\text{d}n_0}{\text{d}\tau} 
&= \frac{1}{\omega_0^2}\partial_{\omega_0} \mathcal{D}(\omega_0) \left[  \partial_{\omega_0} \Delta n_0 + \frac{\Delta n_0}{\The} \right].
\end{align}
%-------------------------------------------------------------------
This expression again directly conserves photon number and also automatically ensures that a Wien spectrum is reached in full equilibrium. We now in principle have the freedom to demand that the first {\it or} the second moment of the collision term is conserved. However, using the conditions for the first moment leads to unphysical cases, where injection at certain critical frequencies do not cause any evolution as the diffusion coefficient vanishes. Thus, using similar steps as above, the only alternative solution is
%-------------------------------------------------------------------
\begin{align}
\label{eq:Coefficients_D2}
\mathcal{D}(\omega_0)&=\omega_0^4 \,\frac{\Sigma_2}{2},
\end{align}
%-------------------------------------------------------------------
which reproduces the correct evolution of the second moment. At low temperature this reduces to $\mathcal{D}\approx \omega_0^4 \The$, again giving the Kompaneets equation. As we will show below, this second improved FP approach does not yield a good general scheme, even if it does approach the correct equilibrium solution.

We also mention that another possibility that comes to mind for fixing the problem with the equilibrium solution is to set %-------------------------------------------------------------------
\begin{align}
\label{eq:Coefficients_D2_b}
\mathcal{D}(\omega_0)&=\omega_0^4 \left\{\frac{2\Sigma_2-\Sigma_1}{\omega_0}+\frac{\Sigma'_2}{2}\right\}.
\end{align}
%-------------------------------------------------------------------
Indeed, we found that in some regimes this does improve matters over the choice in Eq.~\eqref{eq:Coefficients_D2}, but it does not deliver a general robust solution. We therefore omit this option in the discussion below.

%\vspace{-3mm}
\subsection{Adding stimulated scattering terms}
\label{sec:New_attempt_stim}
%--------------------------------------------------------------
As we have seen for Eq.~\eqref{eq:target_moments_kin_lin_sim}, simulated terms can be added to the linearized evolution equation by simply replacing $\Sigma_k\rightarrow \Sigma_k^*$ given by Eq.~\eqref{eq:moments_stim}. This then results in
%-------------------------------------------------------------------
\begin{align}
\label{eq:FP_new_stim}
\frac{\text{d}n_0}{\text{d}\tau} 
&= \frac{1}{\omega_0^2}\partial_{\omega_0} \omega_0^4 \left[\frac{\Sigma^*_2}{2} \partial_{\omega_0} \Delta n_0 
+ \left\{\frac{2\Sigma^*_2-\Sigma^*_1}{\omega_0}+\frac{(\Sigma^*_2)'}{2}\right\}  \Delta n_0 \right].
\end{align}
%-------------------------------------------------------------------
This equation provides a second order diffusion approximation for the problem of repeated scattering of photons assuming that higher order derivatives of the departure from equilibrium, $\Delta n$, are negligible. It allows including stimulated scattering terms due to the presence of equilibrium radiation to all orders, and conserves the first and second moments of the transfer equation.

%-------------------------------------
\begin{figure}
\centering 
\includegraphics[width=0.95\columnwidth]{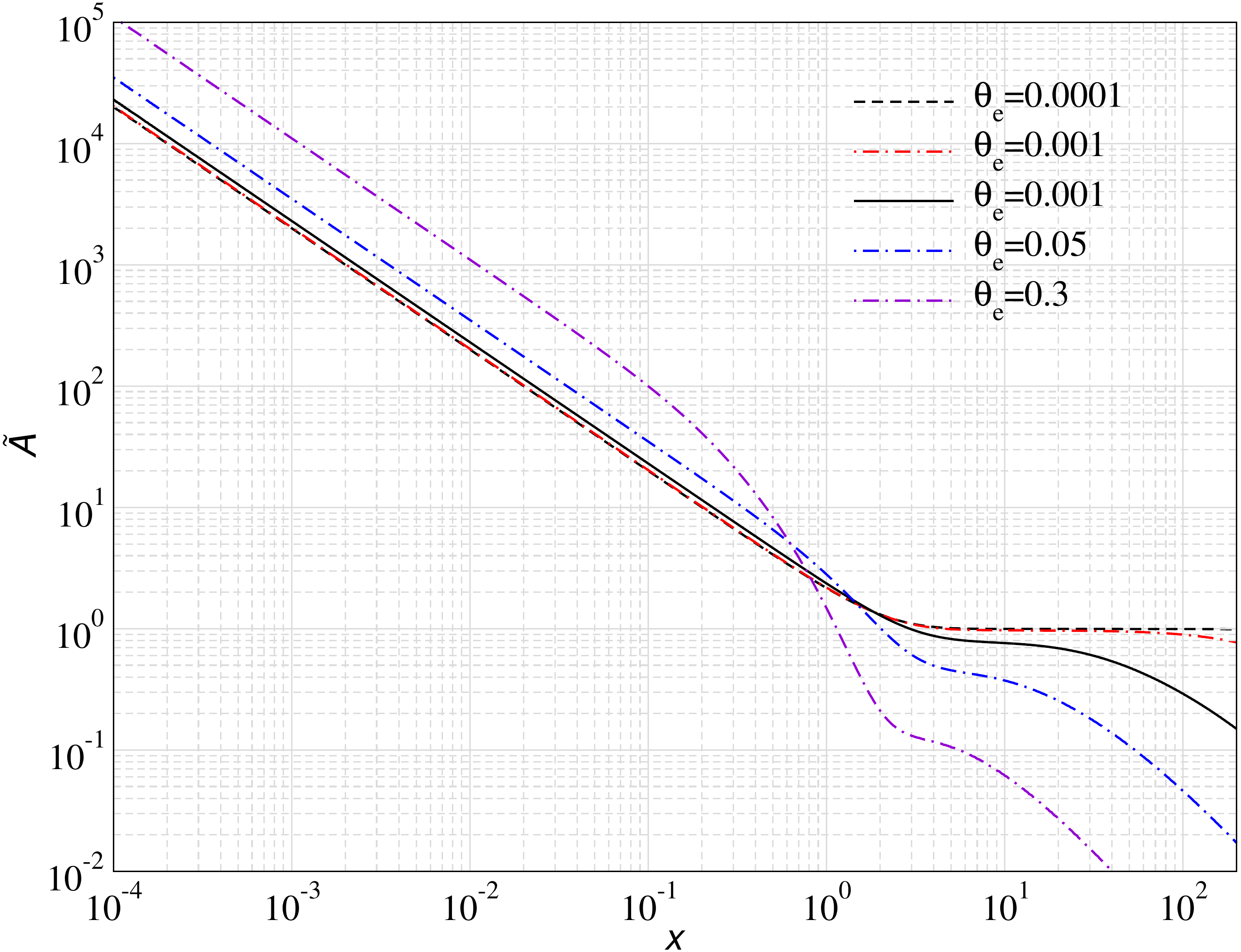}
\\[2mm]
\includegraphics[width=0.95\columnwidth]{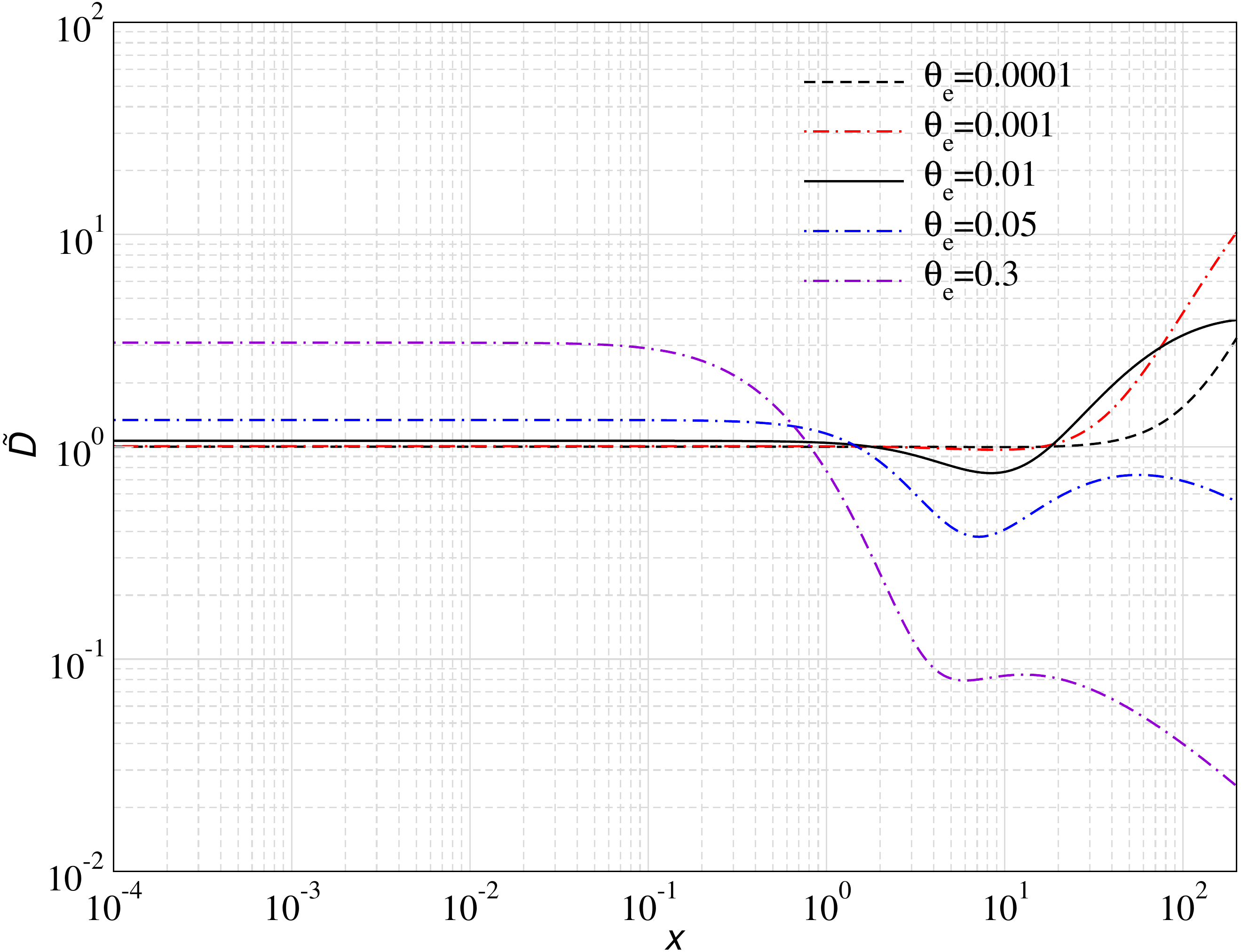}
\\
\caption{Same as Fig.~\ref{fig:Diffusion_coeff_nostim} but with stimulated scattering effects. The main effect of stimulated terms is visible at $x\lesssim 1$.}
\label{fig:Diffusion_coeff_stim}
\end{figure}
%-------------------------------------

\begin{figure}
\centering 
\includegraphics[width=\columnwidth]{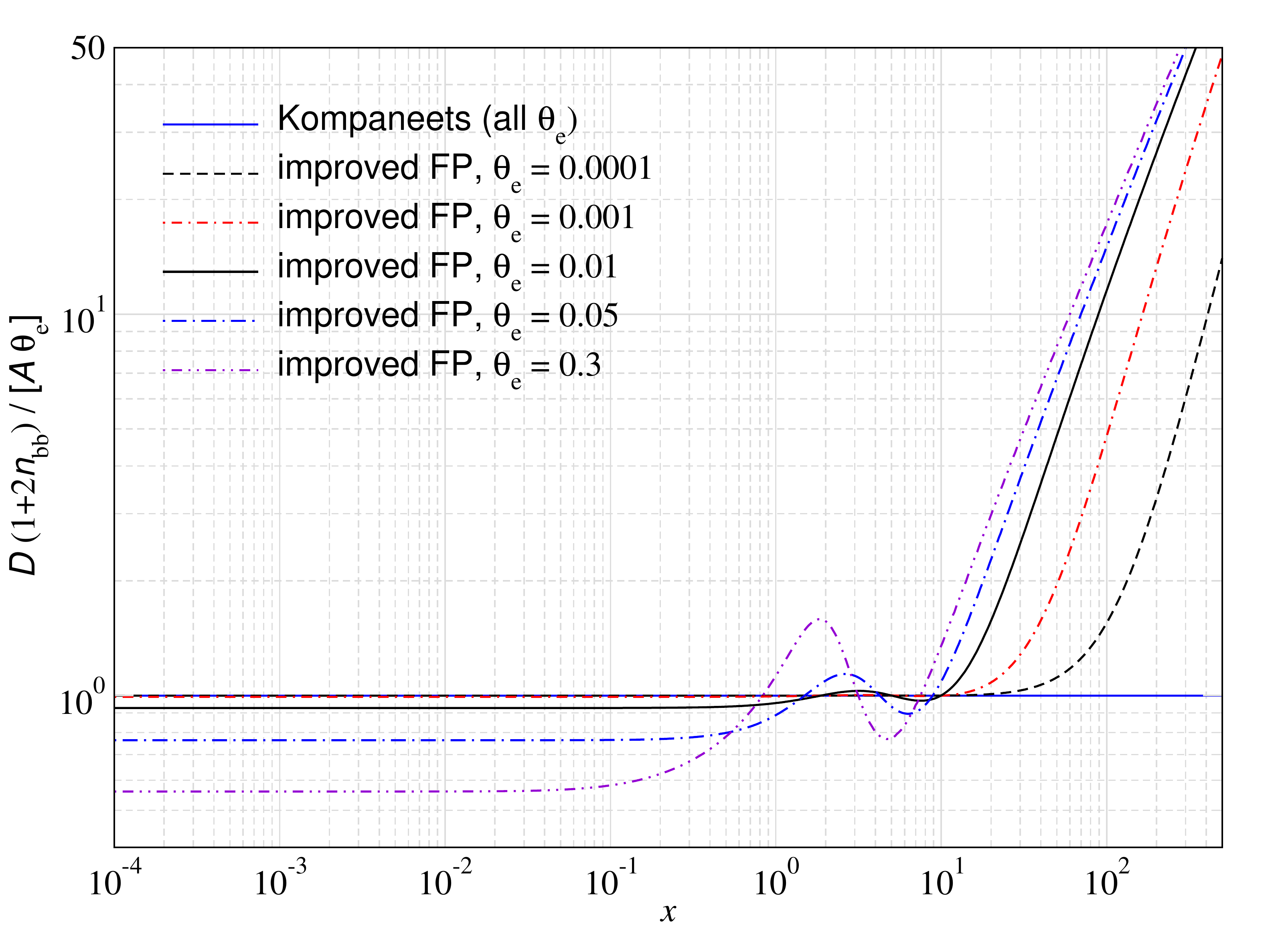}
\\
\caption{Ratio of diffusion coefficients $\mathcal{D}[1+2\,\nbb(\xe)]/[\mathcal{A}\The]$ in the presence of stimulated effects. To ensure detailed balance this ratio has to equal unity as in the linearize Kompaneets equation. At higher temperature, departures become important at very low and high frequencies. The curves were computed using {\tt CSpack}.}
\label{fig:D_A_ratio_stim}
\end{figure}

\begin{figure}
\centering 
\includegraphics[angle=0,width=0.98\columnwidth]{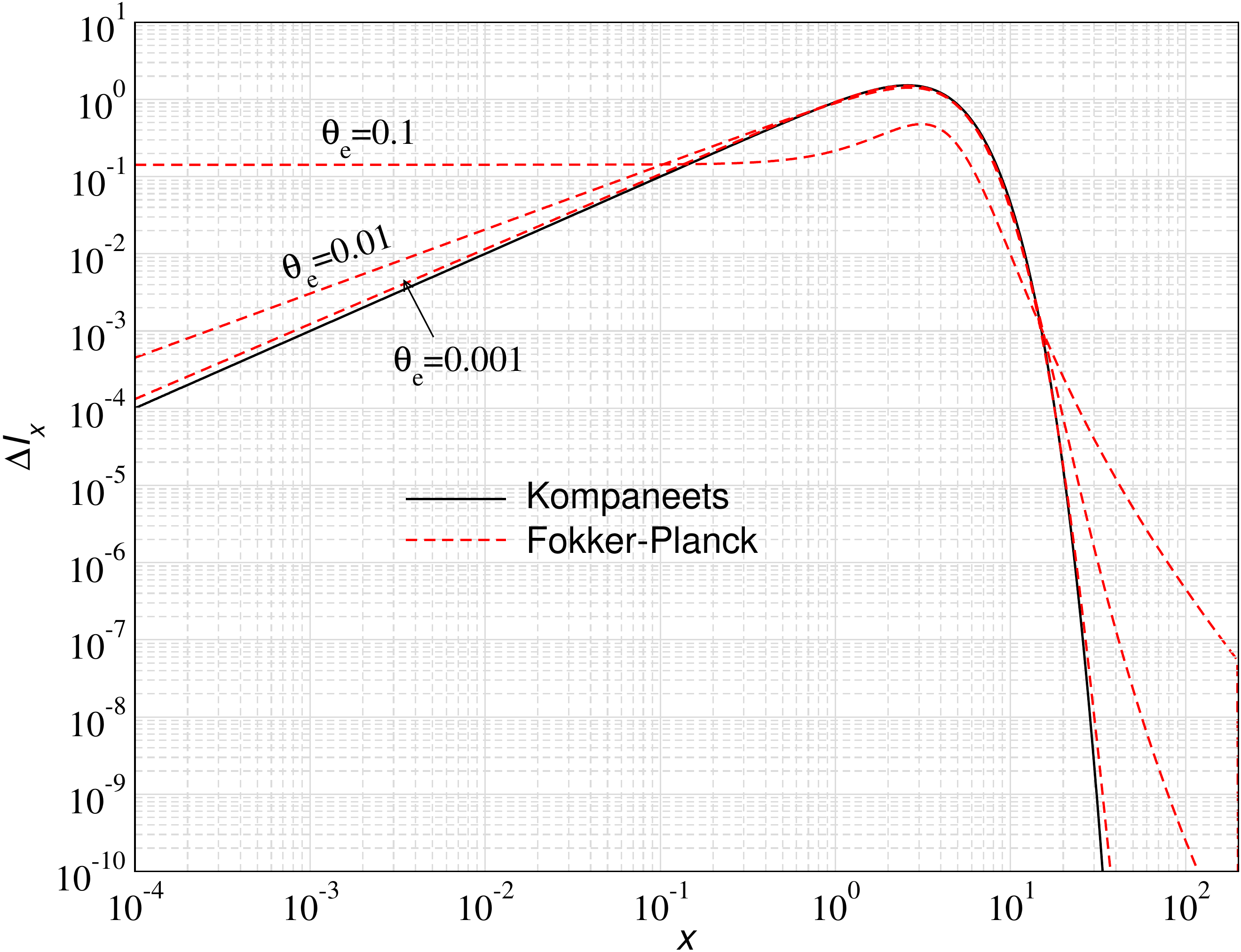}
\\
\caption{Solution of Eq.~\eqref{eq:FP_approach1_eq} for different temperatures and with stimulated scattering terms included. For comparison, we show the Kompaneets/Exact solution, $n(x)=\expf{x}/(\expf{x}-1)^2$. In each case, the total number of photons is fixed to the same value.}
\label{fig:nx_stim}
\end{figure}

To appreciate the effect of stimulated scatterings, let us first consider the modifications to the diffusion coefficients (see Fig.~\ref{fig:Diffusion_coeff_stim}). The main effect of stimulated terms should appear at low frequencies ($x\lesssim 1$), which in comparison to the case without stimulated terms brings the curves closer together.
To understand the general behavior, we compute the moments $\Sigma_k^*$, defined by Eq.~\eqref{eq:moments_stim} at lowest order in the electron temperature. We first compute the Taylor series for $f_{\omega_0\omega}=n^{\rm eq}
(1-\expf{(\omega-\omega_0)/\The})$ around $\omega_0$:
%-------------------------------------------------------------------
\begin{align}
n^{\rm eq}
\left(1-\expf{\frac{\omega-\omega_0}{\The}}\right)&
\approx  - n^{\rm eq}_0 (x-x_0) + n^{\rm eq}_0 (1+2 n^{\rm eq}_0) \frac{(x-x_0)^2}{2},
\end{align}
%-------------------------------------------------------------------
which we can truncate at second order since the usual moments are $\Sigma_k=\mathcal{O}(\The^2)$ for $k>2$. Inserting this into Eq.~\eqref{eq:moments_stim} we then have
%-------------------------------------------------------------------
\begin{align}
\Sigma_1^*
&\approx \Sigma_1- x_0 n^{\rm eq}_0 \Sigma_2 + \frac{x_0^2}{2} n^{\rm eq}_0(1+2 n^{\rm eq}_0)\Sigma_3
\approx (4-x_0)\The - 2 x_0\,n^{\rm eq}_0 \The
\nonumber
\\
\nonumber
\Sigma_2^*
&\approx \Sigma_2-x_0 n^{\rm eq}_0 \Sigma_3 + \frac{x_0^2}{2} n^{\rm eq}_0(1+2 n^{\rm eq}_0)\Sigma_4
\approx 2 \The
\\
&\!\!\!\!\!\!\!\!
\frac{2\Sigma^*_2-\Sigma^*_1}{\omega_0}+\frac{(\Sigma^*_2)'}{2}\approx \frac{4\The-(4-x_0)\The + 2 x_0\,n^{\rm eq}_0 \The}{\omega_0}=1+2\,n^{\rm eq}_0.
\label{eq:stimulated-moments_approx}
\end{align}
%-------------------------------------------------------------------
This shows that $\mathcal{A}^*\approx \omega_0^4 (1+2n^{\rm eq}_0)$, while $\mathcal{D}^*\approx\mathcal{D}\approx \omega_0^4\The$. In the Kompaneets limit, stimulated terms leave $\mathcal{D}$ unaltered while they enhance $\mathcal{A}$ by a factor of $(1+2n^{\rm eq}_0)\approx 2/x$ at low frequencies.
Inserting these diffusion coefficients back into Eq.~\eqref{eq:FP_new_stim} then gives
%-------------------------------------------------------------------
\begin{align}
\frac{\text{d}n_0}{\text{d}\tau} 
&= \frac{1}{\omega_0^2}\partial_{\omega_0} \omega_0^4 \left[ \The \partial_{\omega_0} \Delta n_0 
+ \Delta n_0 (1+2\,n^{\rm eq}_0) \right]
\nonumber\\
&= \frac{\The}{x^2}\partial_{x} x^4 \left[  \partial_{x} \Delta n_0 
+ \Delta n_0 (1+2\,n^{\rm eq}_0) \right],
\end{align}
%-------------------------------------------------------------------
which is the correct linearized Kompaneets equation with induced scattering terms included. The equilibrium spectrum is given by 
%-------------------------------------------------------------------
\begin{align}
\label{eq:Dneq_stim}
\Delta n^{\rm eq}(x)
&= \Delta n_{\rm c}\,\frac{\expf{x}}{(\expf{x}-1)^2},
\end{align}
%-------------------------------------------------------------------
which defines a chemical potential distortion for small $\Delta n_{\rm c}$. To obtain this solution with general diffusion coefficients then implies the condition $\mathcal{D}^*(\omega_0)=\The \mathcal{A}^*(\omega_0)/[1+2\,\nbb(x)]$, which in general again is not fulfilled (see Fig.~\ref{fig:D_A_ratio_stim}). This means that also the improved Fokker-Planck equation, Eq.~\eqref{eq:FP_new_stim}, does not approach the correct equilibrium solution, Eq.~\eqref{eq:Dneq_stim}, in the limit of many scatterings, as illustrated in Fig.~\ref{fig:nx_stim}. 
%-------------------------------------
%-------------------------------------
However, since the departures from $\mathcal{D}^*(\omega_0)=\The \mathcal{A}^*(\omega_0)/[1+2\,\nbb(x)]$ remain more moderate at low frequencies (see Fig.~\ref{fig:D_A_ratio_stim}), the equilibrium solution is relatively closer to $\Delta n^{\rm eq}(x)
\propto \expf{x}/(\expf{x}-1)^2$ at low frequencies than when stimulated terms are neglected. Nevertheless, one expects the improved FP approximation to perform badly in the limit of many scatterings. An improvement to the asymptotic behavior that can again be achieved with the method described in Sect.~\ref{sec:New_attempt_II} by including stimulated effects in the diffusion coefficient Eq.~\eqref{eq:Coefficients_D2}.

%---------------
%-------------------------------------

\section{Numerical solutions for test problems}
\label{sec:num_sols}
%--------------------------------------------------------------
We are now ready to start comparing the solutions using the various FP approaches presented in Sect.~\ref{sec:FP_expansion}. The main discussion starts with illustrations for the evolution of the first and second moments of the solution (Sect.~\ref{sec:evol_moments}), which shows that the first improved FP approach outperforms the Kompaneets equation in several regimes.
However, by studying the solutions in detail, we can show that none of these approximate schemes work well while the solution is very narrow.
Before starting the numerical discussion, we briefly explain how to solve the diffusion equations and then give our general numerical scheme to the integro-differential equation.

%--------------------------------------------------------------
\subsection{Solving the Fokker-Planck equations}
\label{sec:FP_solvers}
%--------------------------------------------------------------
For all problems considered in this section, we shall assume that $\The={\rm const}$ through the computation. Physically, this has limited applicability and assumes that the electrons are in contact with a heat bath that can either provide or absorb energy that is exchanged with the photons. However, it allows us to focus on the main features relating to the scattering physics.

To follow the evolution for the photon field, the best time variable to monitor is the scattering $y$-parameter
%-------------------------------------------------------------------
\begin{align}
\label{eq:y-parameter}
y&=\int \frac{k\Te}{\me c^2} \Ne \sigT c \id t =\int \The \id \tau.
\end{align}
%-------------------------------------------------------------------
This naturally provides a time-coordinate as the broadening of lines depends on $\Delta \nu / \nu \simeq \sqrt{y}$ \citep{Zeldovich1969}. Time-steps can then be chosen such that $\Delta y\ll 1$. We shall furthermore use $x=h\nu/k\Te=\omega/\The$ as our energy variable. In this case, the FP equations all take the form
%-------------------------------------------------------------------
\begin{align}
\label{eq:FP_solver}
\frac{\text{d}\Delta n_0}{\text{d}y} 
&= \frac{1}{x_0^2}\partial_{x_0} x_0^4\,\left[\mathcal{D}_{x_0}\,\partial_{x_0} \Delta n_0 +\mathcal{A}_{x_0}\Delta n_0 \right].
\end{align}
%-------------------------------------------------------------------
In the Kompaneets limit one has
%-------------------------------------------------------------------
\bsub
\begin{align}
\mathcal{D}_{x_0}&= 1,\qquad\mathcal{A}_{x_0}=1
\\
\mathcal{D}^*_{x_0}&= 1,\qquad \mathcal{A}^*_{x_0}=1+2n^{\rm eq}_0,
\end{align}
\esub
%-------------------------------------------------------------------
where the asterisk indicates that stimulated scattering is included. 
In the first improved FP approach outlined in Sect.~\ref{sec:New_attempt}, we have
%-------------------------------------------------------------------
\begin{align}
\mathcal{D}_{x_0}&= \frac{\Sigma_2}{2\The},\qquad\mathcal{A}_{x_0}=\frac{2\Sigma_2-\Sigma_1}{x_0\The}+\frac{1}{2\The}\frac{\partial\Sigma_2}{\partial x_0}.
\label{eq:imFPI}
\end{align}
%-------------------------------------------------------------------
The FP coefficients with stimulated effects are simply obtained by replacing the first and second moments with their stimulated equivalents, $\Sigma_k\rightarrow\Sigma_k^*$. Finally, in the second FP approach described in Sect.~\ref{sec:New_attempt_II}, we have $\mathcal{D}_{x_0}=\mathcal{A}_{x_0} =\frac{\Sigma_2}{2\The}$ and the stimulated equivalent by replacing $\Sigma_2\rightarrow\Sigma_2^*$.

To discretize Eq.~\eqref{eq:FP_solver}, we first convert it into the form
%-------------------------------------------------------------------
\begin{align}
\label{eq:FP_solver_prep}
\frac{\text{d}\Delta n_0}{\text{d}y} 
&= A(x_0, \The)\,\partial^2_{x_0} \Delta n_0 +B(x_0, \The)\,\partial_{x_0} \Delta n_0+
C(x_0, \The) \,\Delta n_0.
\end{align}
%-------------------------------------------------------------------
The required coefficients $A$, $B$ and $C$ are summarized in Appendix~\ref{app:diff_solver_coefficients} for each of the considered cases. The first and second derivatives of the solution $\Delta n_0$ are computed using a 5-point stencil. This leads to a banded matrix equation that can be solved explicitly within a semi-implicit Crank-Nicolson scheme, as also done for the cosmological recombination problem \citep{Chluba2010b}. Here, we set the solution to zero at the boundaries, which we choose to be far from the main frequency domain of interest. For extremely long runs (i.e., large $y$) this has a small effect close to the boundaries, but we do not rely heavily on these results.

%--------------------------------------------------------------
\subsection{Discretization of the kernel equation}
\label{sec:formulation}
%--------------------------------------------------------------
To numerically solve the general evolution equation we will assume that stimulated terms $\mathcal{O}(\Delta n)^2$ can be omitted and thus start from Eq.~\eqref{eq:col_CS_Dn_lin}.
Using appropriate weight factors\footnote{In computation we use a fifth order Lagrange interpolation coefficient and their integrals to discretize the solution \citep{Chluba2010}.} to write the integral of any function $g(x)$ as $\int g(\omega') \id \omega'=\sum_{i} g(x_i) \,w_i$, we can then cast this equation into the matrix form
%-------------------------------------------------------------------
\begin{align}
\label{eq:evol_Dn}
\frac{\text{d}\Delta n_i}{\text{d}\tau} 
&\approx \sum_j  \mathcal{S}_{ij} \Delta n_j,
\nonumber
\\[1mm]
\mathcal{S}_{ij}&=w_j\,P(\omega_i \rightarrow \omega_j, \The) \,\expf{x_j-x_i} \, f_{\omega_j,\omega_i} - \delta_{ij}\,\sigma_i^*,
\nonumber
\\[1mm]
\sigma_i^*&=\sum_j w_j\,P(\omega_i \rightarrow \omega_j, \The)\, f_{\omega_i,\omega_j},
\end{align}
%-------------------------------------------------------------------
where $\mathcal{S}_{ij}$ defines the redistribution of photons due to the electron scattering process, and $\sigma_i^*$ can be thought of as the total scattering cross section at frequency $x_i$ with the inclusion of stimulated scattering effects when present. Using a simple implicit Euler step in time, we then have matrix equation
%-------------------------------------------------------------------
\begin{align}
\label{eq:gen_matrix_eq}
\sum_j (\delta_{ij}-\Delta \tau \mathcal{S}_{ij}) \, \Delta n^{(t+1)}_j 
&\approx \Delta n^{(t)}_i,
\end{align}
%-------------------------------------------------------------------
where $\Delta n^{(t)}_i$ is the solution at time $t$ and $\Delta \tau =\tau^{(t+1)}-\tau^{(t)}$. This system can be solved for $n^{(t+1)}_i$ using iterative methods \citep{Chluba2010}. Since the electron temperature is assumed to be fixed, the scattering matrix only has to be computed once, which greatly simplifies the problem. We typically use $\Delta \ln y\simeq 10^{-3}$ to achieve high precision. The $y$-parameter is monitored using a simple update $y^{(t+1)}=y^{(t)}+\Delta y$ at each step. We checked the precision by changing the number of intermediates steps. For the frequency grid, we used log-spacing with a typical log-density of 500 points per decade.

One of the big benefits of the scattering matrix approach is that it perfectly conserves photon number within the computational domain. In addition, {\it no} extra boundary conditions have to be given, avoiding problematic choices required in FP schemes.

%-------------------------------------
\begin{figure*}
\centering 
\includegraphics[width=\columnwidth]{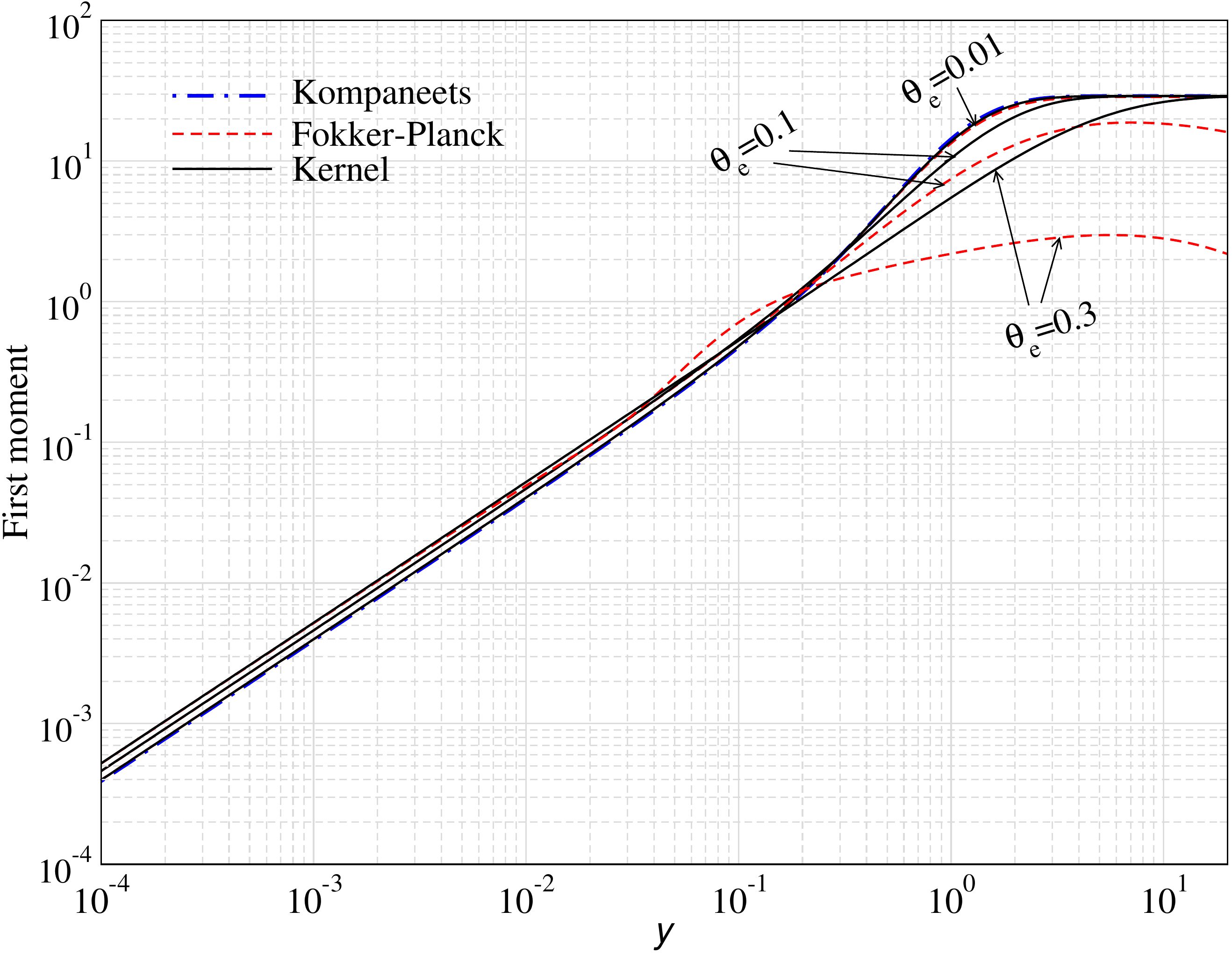}
\hspace{4mm}
\includegraphics[width=\columnwidth]{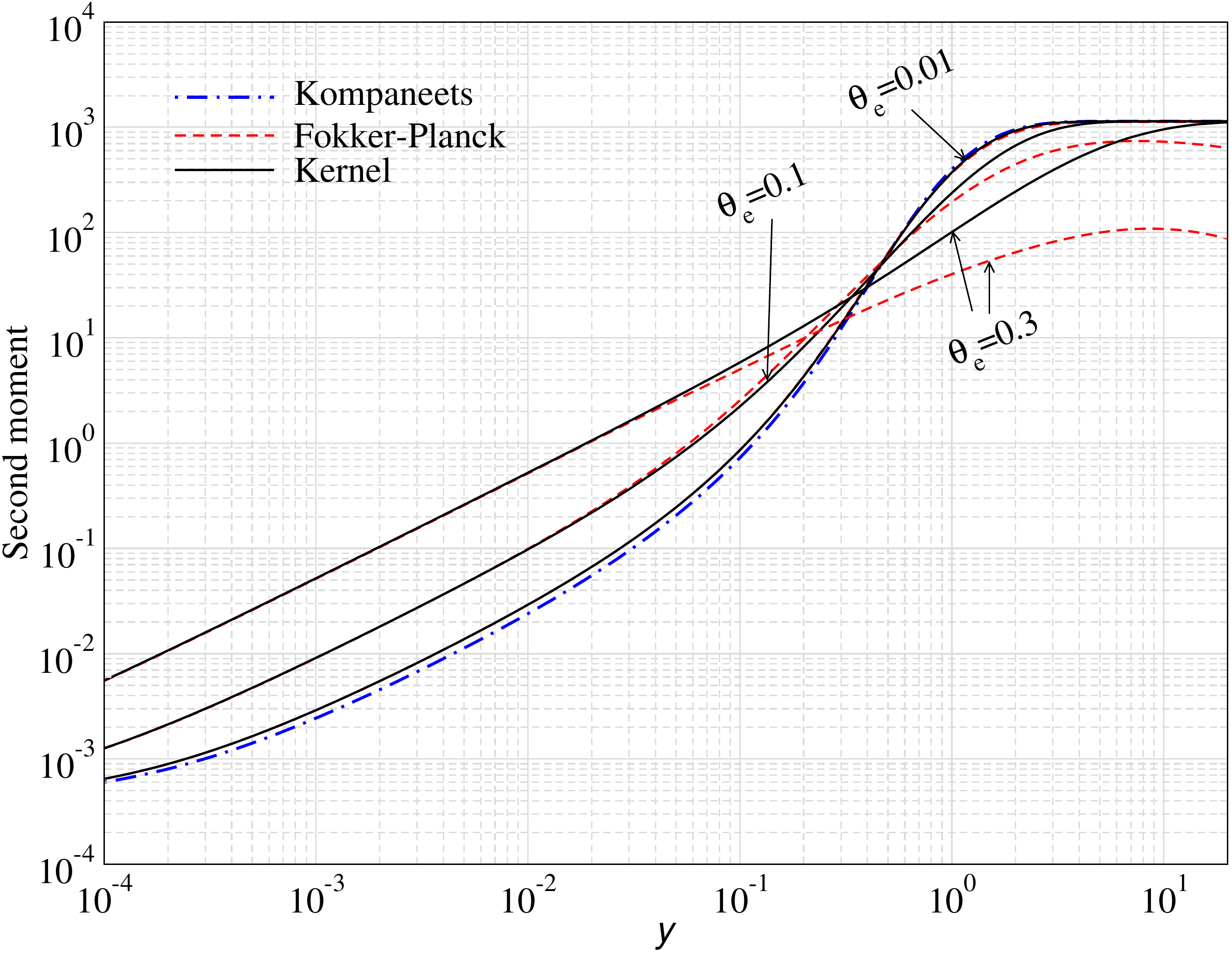}
\\[3mm]
\includegraphics[width=\columnwidth]{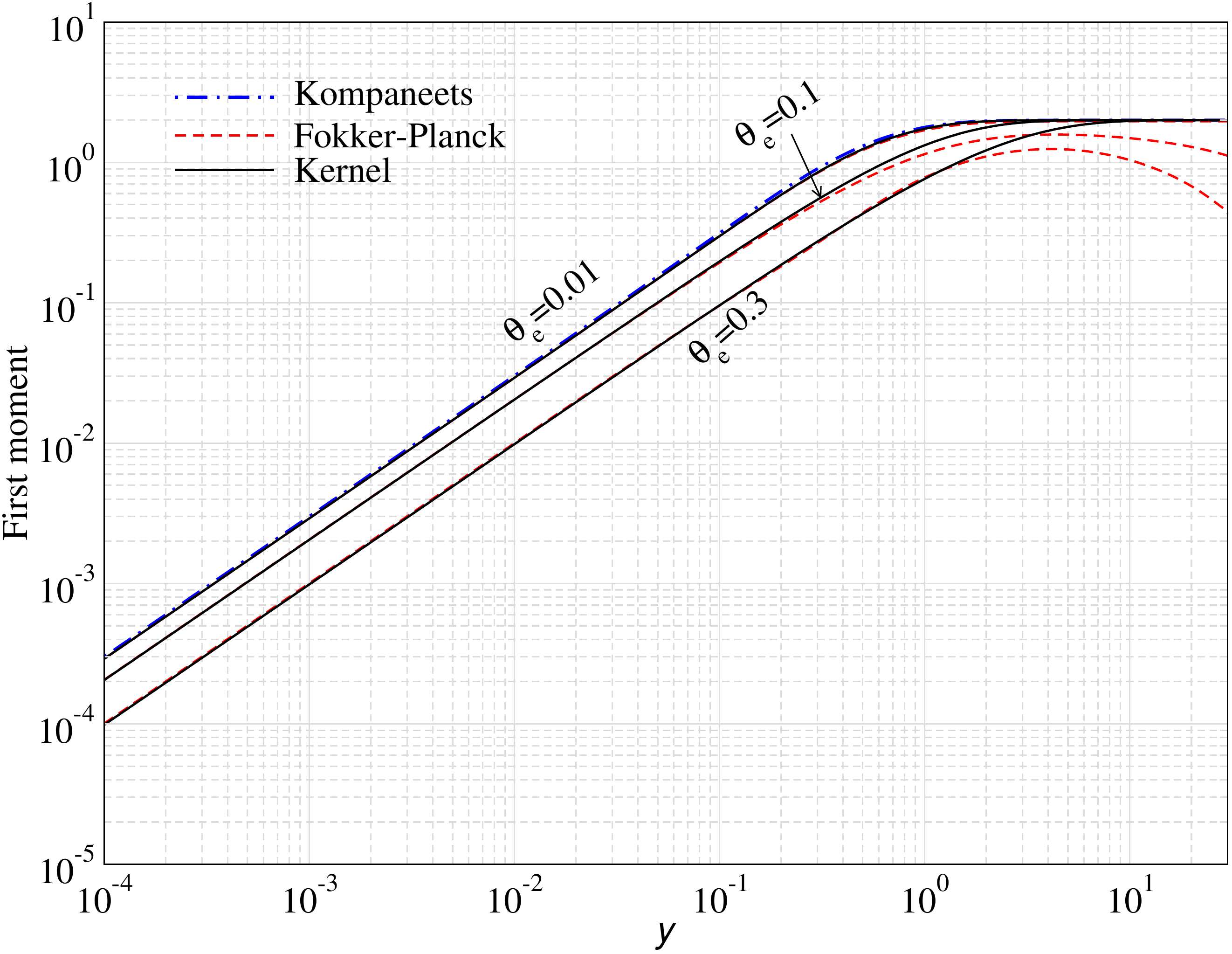}
\hspace{4mm}
\includegraphics[width=\columnwidth]{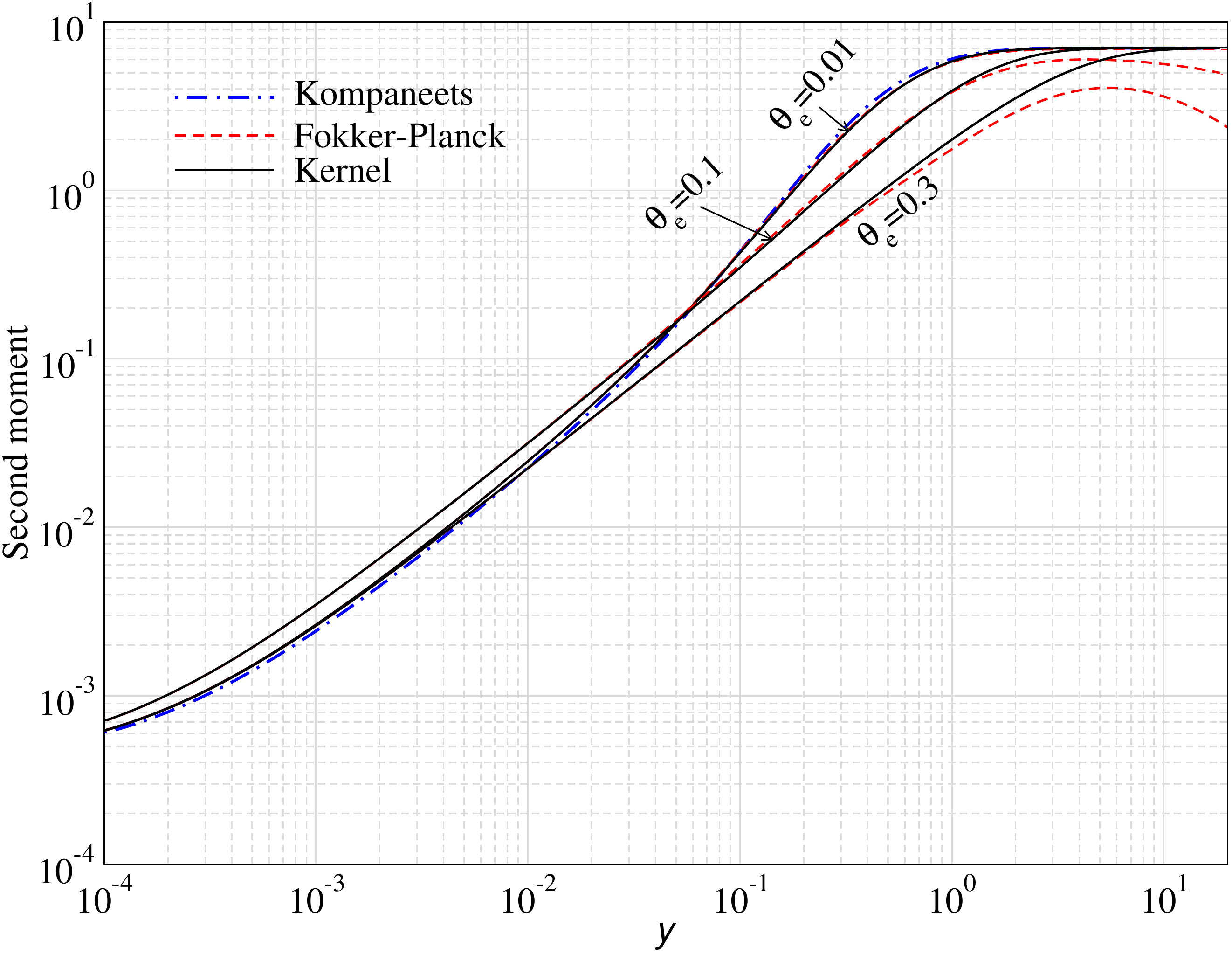}
\\[3mm]
\includegraphics[width=\columnwidth]{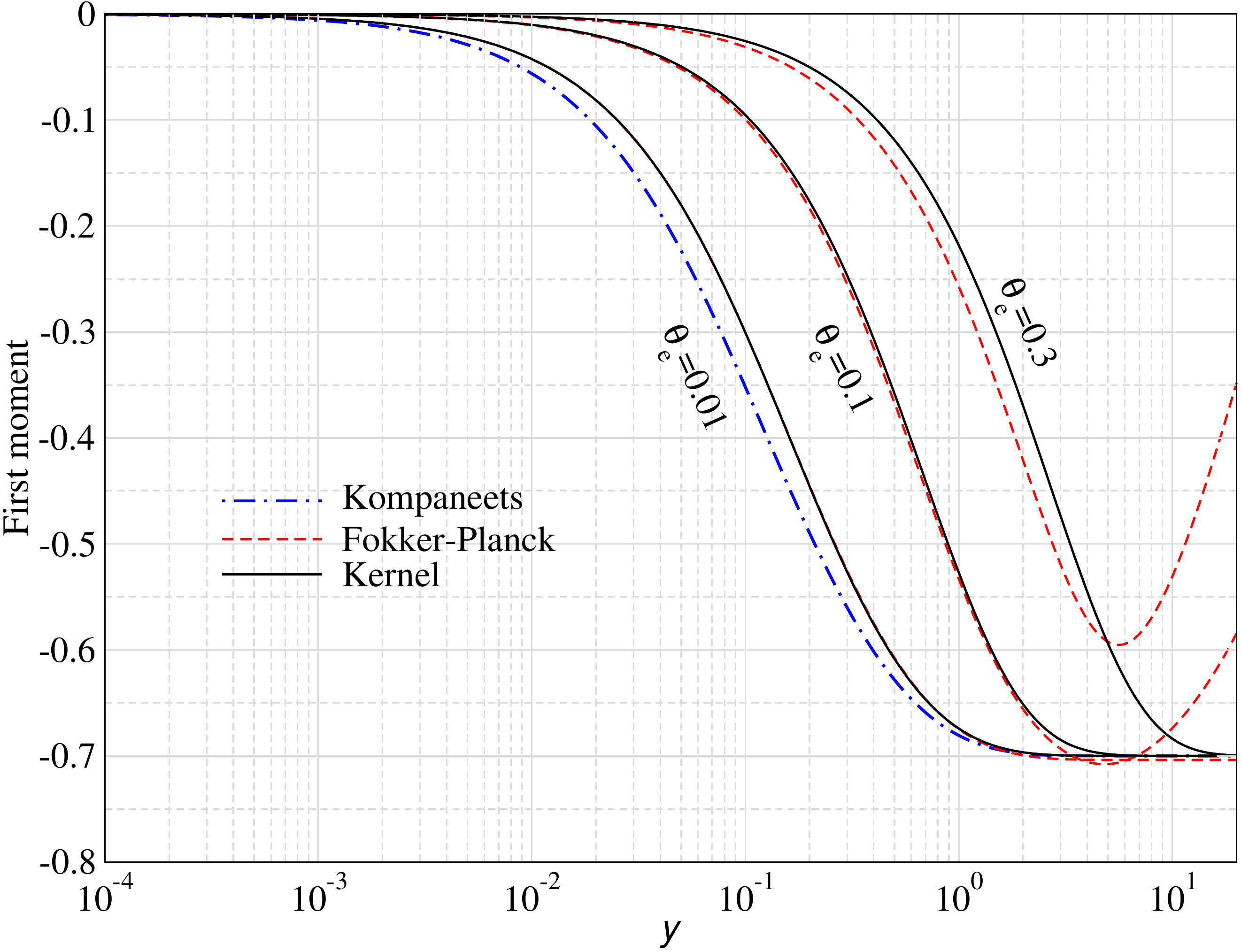}
\hspace{4mm}
\includegraphics[width=\columnwidth]{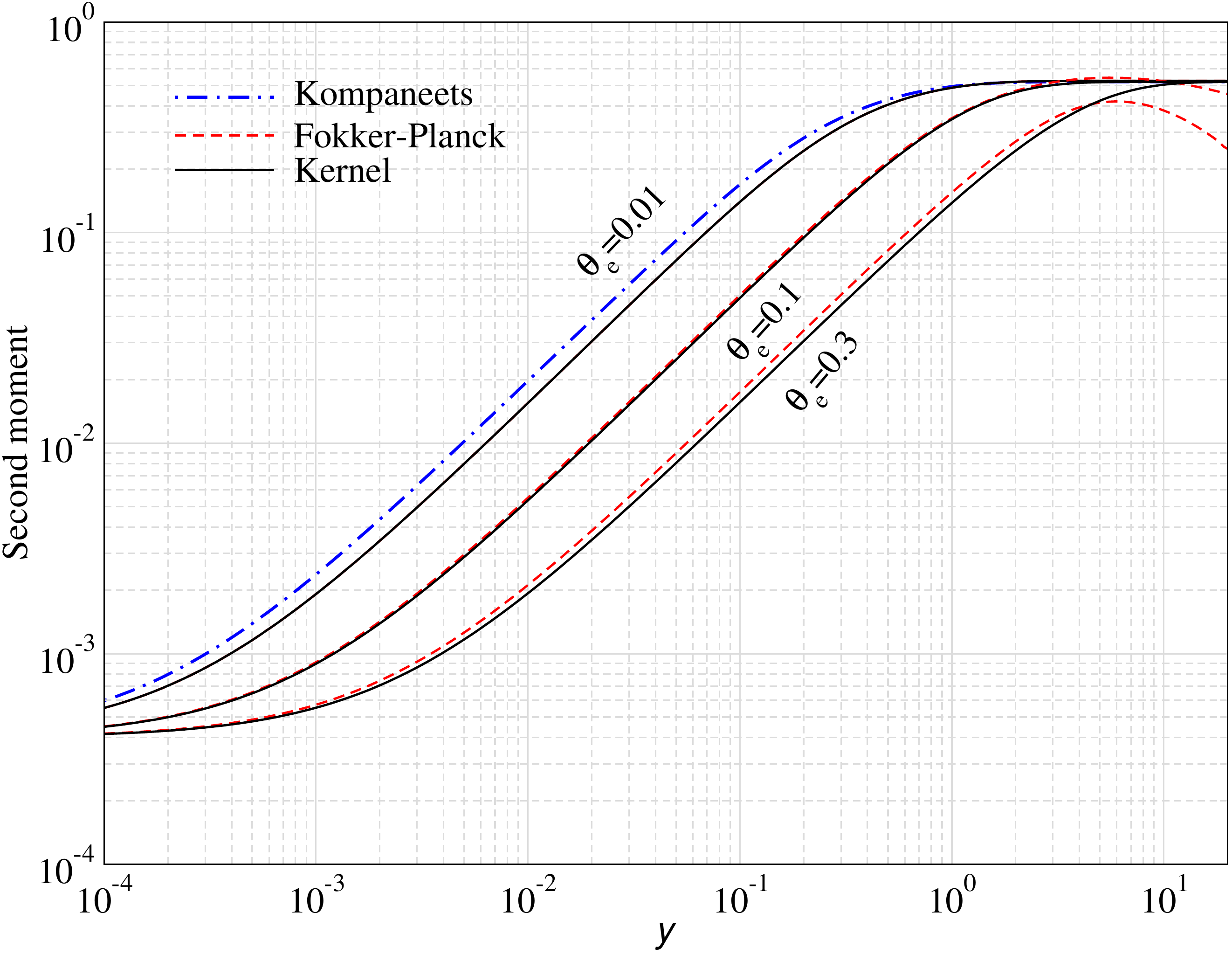}
\\
\caption{Evolution of first (left panels) and second (right panels) moments of photon field for injection at $x_{\rm inj}=0.1$ (upper panels), $x_{\rm inj}=1$ (central panels) and $x_{\rm inj}=10$ (lower panels). Stimulated terms were neglected. In each panel, the electrons are kept at a constant temperature $\The$ with values as annotated. We compare the solutions of the Kompaneets equation and the improved FP approach [diffusion coefficients as in Eq.~\eqref{eq:imFPI}] with the solution for the exact scattering kernel. The Kompaneets solutions are independent of the temperature. The temperature dependence is captured well by the improved FP approach at low values of the $y$ parameter. However, this approach fails as expected once equilibrium is reached.}
\label{fig:x_0.1_10_moment_evol}
\end{figure*}
%-------------------------------------

%%-------------------------------------
%\begin{figure}
%\centering 
%\includegraphics[width=\columnwidth]{./eps/x_0.1_1st_moment.pdf}
%\\
%\includegraphics[width=\columnwidth]{./eps/x_0.1_2nd_moment.pdf}
%\\
%\caption{Evolution of first and second moment of photon field for injection at $x_{\rm inj}=0.1$ (no stimulated terms). The electrons are kept at a constant temperature $\The$ with values as annotated. We compared the solutions of the Kompaneets equation and the improved FP approach [diffusion coefficients as in Eq.~\eqref{eq:imFPI}] with the solution for the exact scattering kernel.}
%\label{fig:x_0.1_moment_evol}
%\end{figure}
%%-------------------------------------

%%-------------------------------------
%\begin{figure}
%\centering 
%\includegraphics[width=\columnwidth]{./eps/x_1.0_1st_moment.pdf}
%\\
%\includegraphics[width=\columnwidth]{./eps/x_1.0_2nd_moment.pdf}
%\\
%\caption{As in Fig.~\ref{fig:x_0.1_moment_evol} but for $x_{\rm inj}=1$}
%\label{fig:x_1_moment_evol}
%\end{figure}
%%-------------------------------------
%
%%-------------------------------------
%\begin{figure}
%\centering 
%\includegraphics[width=\columnwidth]{./eps/x_10.0_1st_moment.pdf}
%\\
%\includegraphics[width=\columnwidth]{./eps/x_10.0_2nd_moment.pdf}
%\\
%\caption{As in Fig.~\ref{fig:x_0.1_moment_evol} but for $x_{\rm inj}=10$}
%\label{fig:x_10_moment_evol}
%\end{figure}
%%-------------------------------------

%--------------------------------------------------------------
\subsection{Evolution of the first and second moments}
\label{sec:evol_moments}
%--------------------------------------------------------------
To understand the performance of the various approaches we start by comparing the evolution of the first and second moments of the photon field computed numerically. These are all functions of the $y$ parameter as the mean energy and dispersion of the photon field evolve with the number of scatterings. We start by injecting a narrow photon line at various values of $x_{\rm inj}$. For injection around $x_{\rm inj}\simeq 4/(1+76\The)^{0.1}$, we expect the first moment to remain constant for a relatively long time. This frequency corresponds to the null of the first moment as a function of the temperature \citep{CSpack2019}. When including stimulated scattering terms, we expect this number to reduce to $x_{\rm inj}\simeq 3.6$ for $\The\ll 1$ \citep[see][for discussion]{Chluba2015GreensII}.
Below and above this energy, the photon field gains or looses energy, respectively.
The second moment of the photon field is initially dominated by the width of the photon line. However, by construction we always expect the second moment of the photon field to grow as a function of $y$.

In Figure~\ref{fig:x_0.1_10_moment_evol} we compare the evolution of first and second moment of the photon field as a function of the $y$ parameter for $x_{\rm inj}=\{0.1, 1, 10\}$. 
For $x_{\rm inj}=0.1$ and $1$, the first moment grows as photons on average gain energy from the electrons, while for $x_{\rm inj}=10$, energy is extracted from photons by the less energetic electrons. The second moment is positive and grows for all the cases. We also confirmed that for $x_{\rm inj}\simeq 3-4$, the first moment evolves slowly for the full computation. All these findings are in agreement with the expectations. 

For non-relativistic approximation (Kompaneets limit), the scattering cross-section is just given by Thomson cross section, which does not depend upon the energy of photons and electrons. Equation~\eqref{eq:kompaneets_mom} suggests that the first and second moments of the solution are proportional to electron temperature. However, here we are working with dimensionless frequency $x=\omega/\The$ and with the $y$ parameter, which scales out the factor of $\The$. Therefore, the evolution of the first two moments is independent of electron temperature in the Kompaneets limit and only a function of $x_{\rm inj}$ and $y$. For $\The\lesssim 0.01$, the Kompaneets equation reproduces the evolution of the moments for the shown cases fairly well, with the differences increasing with injection energy. This is mainly because Klein-Nishina corrections to the total cross section and temperature corrections to the Doppler broadening and recoil are not accounted for by the Kompaneets equation.

Moving to the improved Fokker-Planck approximation and the exact kernel calculations, in which relativistic kinematics are taken into account, the explicit temperature dependence of the moments becomes apparent.  
To a large part, this is due to the fact that the total scattering cross section is a function of the energy of the photons and electrons. This difference is more pronounced for higher photon energy (at fixed electron temperature) as we go deeper into relativistic regime. The improved FP treatment  matches the exact kernel calculation very well in all shown cases while $y\lesssim 0.1$. However, it dramatically fails to reproduce the evolution of the first two moments of the solution at larger $y$ values, when a stationary state is reached. In this regime, the Kompaneets equation approaches the exact solution, correctly describing the asymptotic behavior. 
 
One of the main differences between the solutions for varying temperatures is the time-scale on which the photon distribution evolves. For higher temperature, the stationary state is reached on a longer timescale. In all considered cases, the stationary solution requires $y\gtrsim 1-2$. During the early phases of the evolution, the result depends on the injection energy. At low energies, the evolution is faster than in the Kompaneets limit (upper and central panels in Fig.~\ref{fig:x_0.1_10_moment_evol}), while it is slower at higher energies (lower panels in Fig.~\ref{fig:x_0.1_10_moment_evol}). The effect of Doppler boosting and broadening is underestimated in the Kompaneets limit, while Klein-Nishina corrections to the total cross section hinder the evolution at higher energies.

%---------------
\begin{figure}
\centering 
\includegraphics[angle=0,width=0.98\columnwidth]{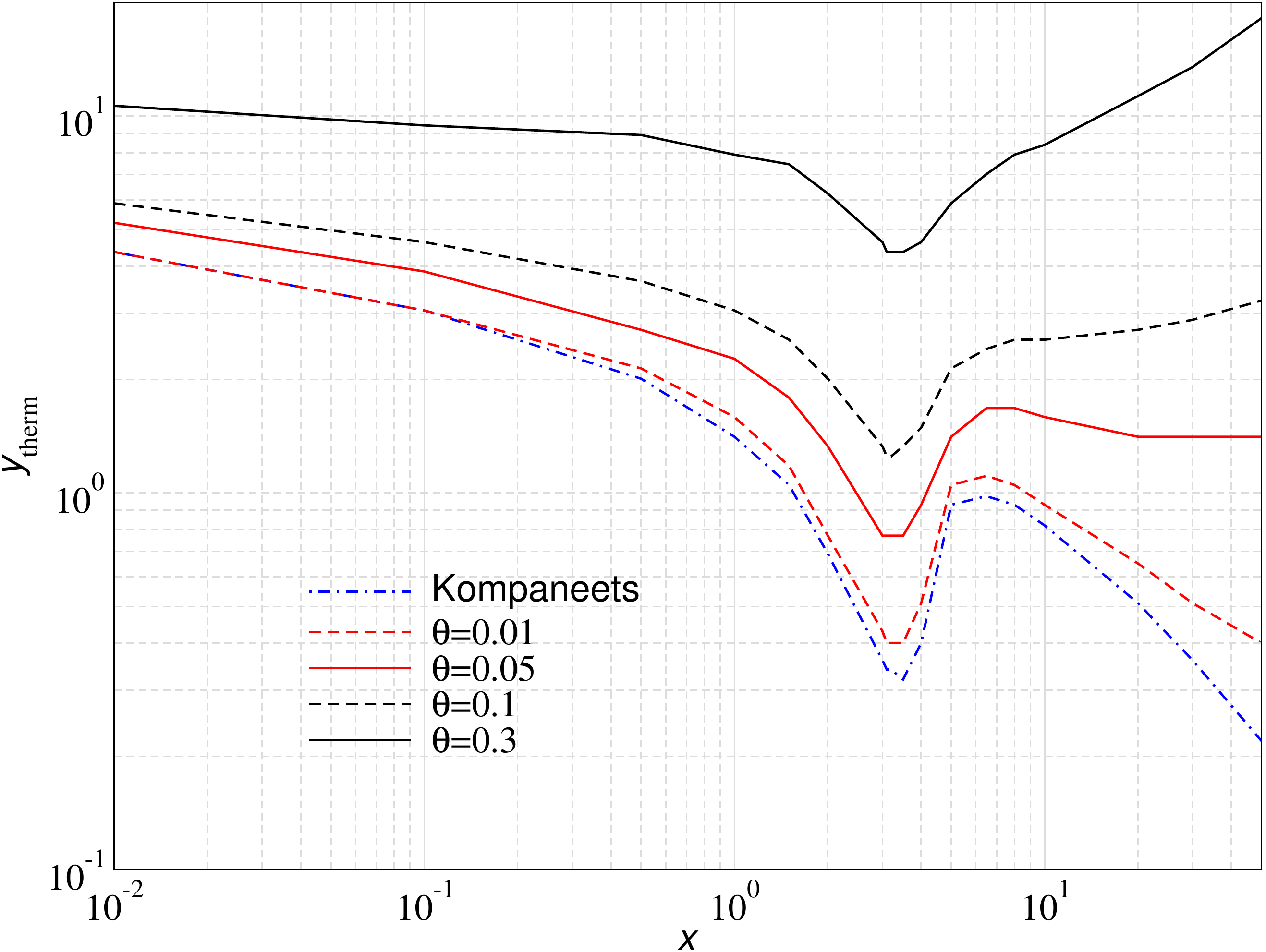}
\\[2mm]
\includegraphics[angle=0,width=0.98\columnwidth]{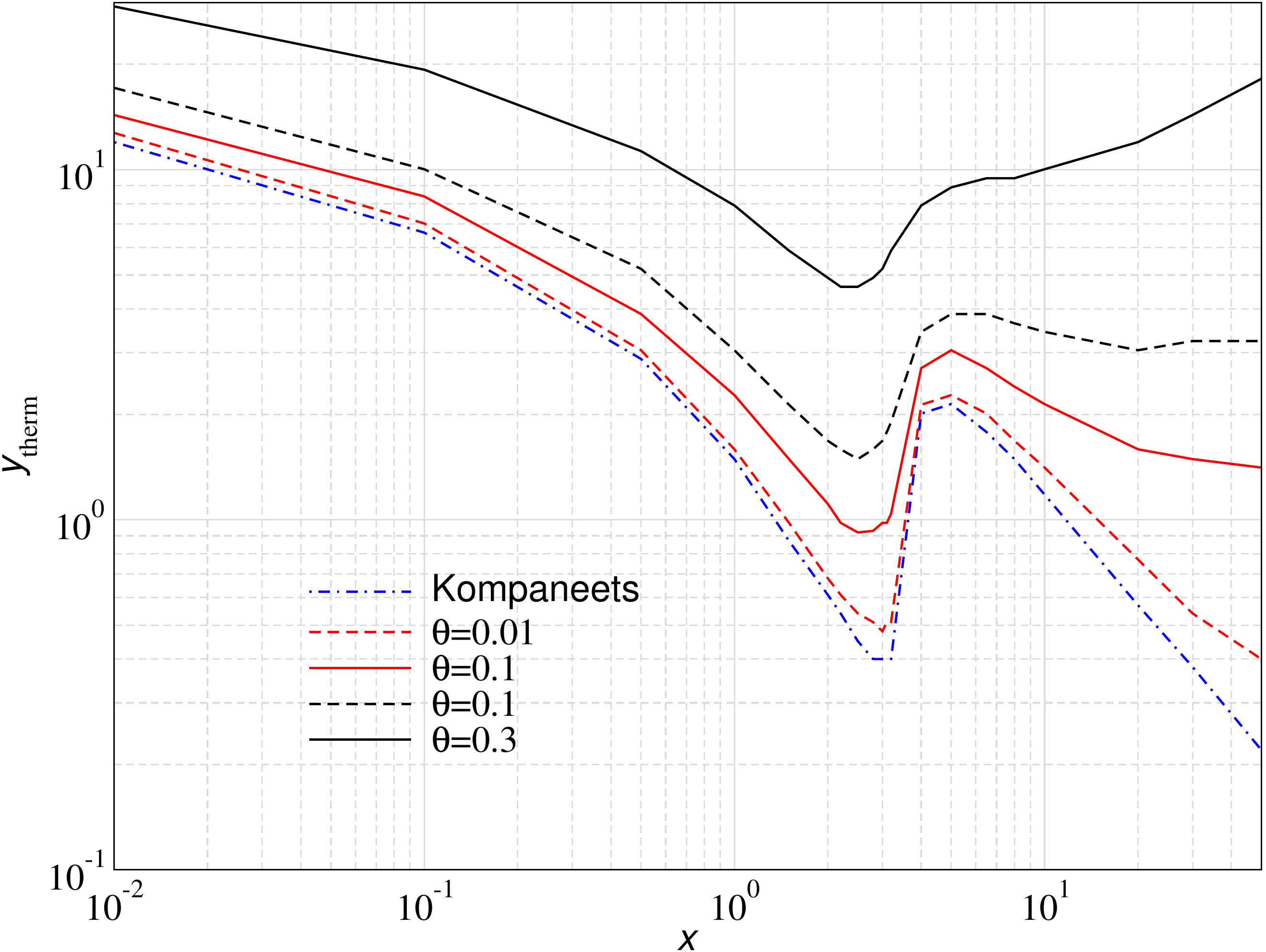}
\\
\caption{Thermalization $y$-parameter, $y_{\rm therm}$ (see text for definition), as a function of the injection frequency for different temperatures. The upper panel omits stimulated scattering while in the lower it is included. The Kompaneets treatment does not capture the temperature dependence of $y_{\rm therm}$. Stimulated scattering slows thermalization for $x_{\rm inj}\lesssim 1$.
}
\label{fig:y_therm}
\end{figure}
%-------------------------------------

%-------------------------------------
\begin{figure}
\centering 
\includegraphics[width=\columnwidth]{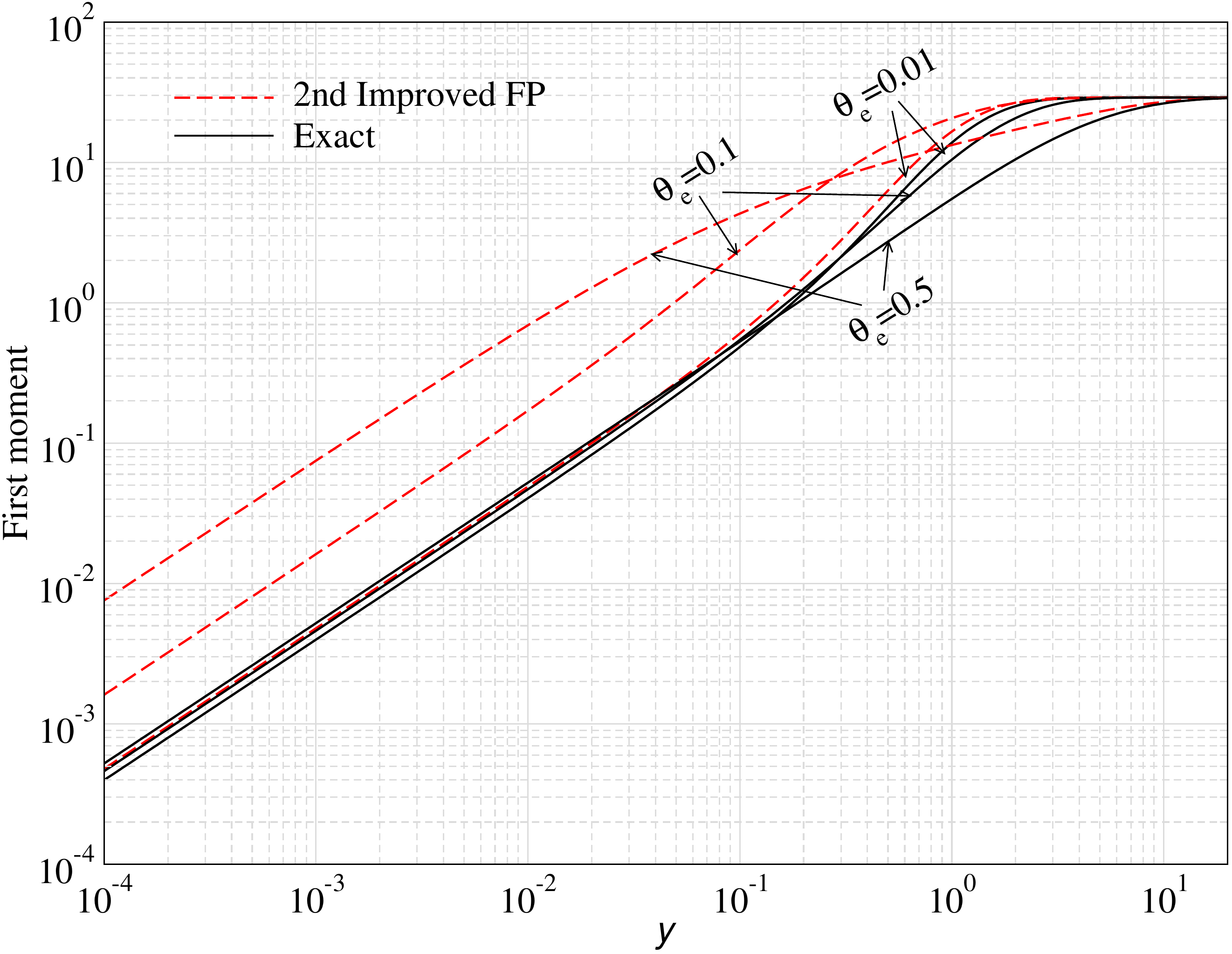}
\\[2mm]
\includegraphics[width=\columnwidth]{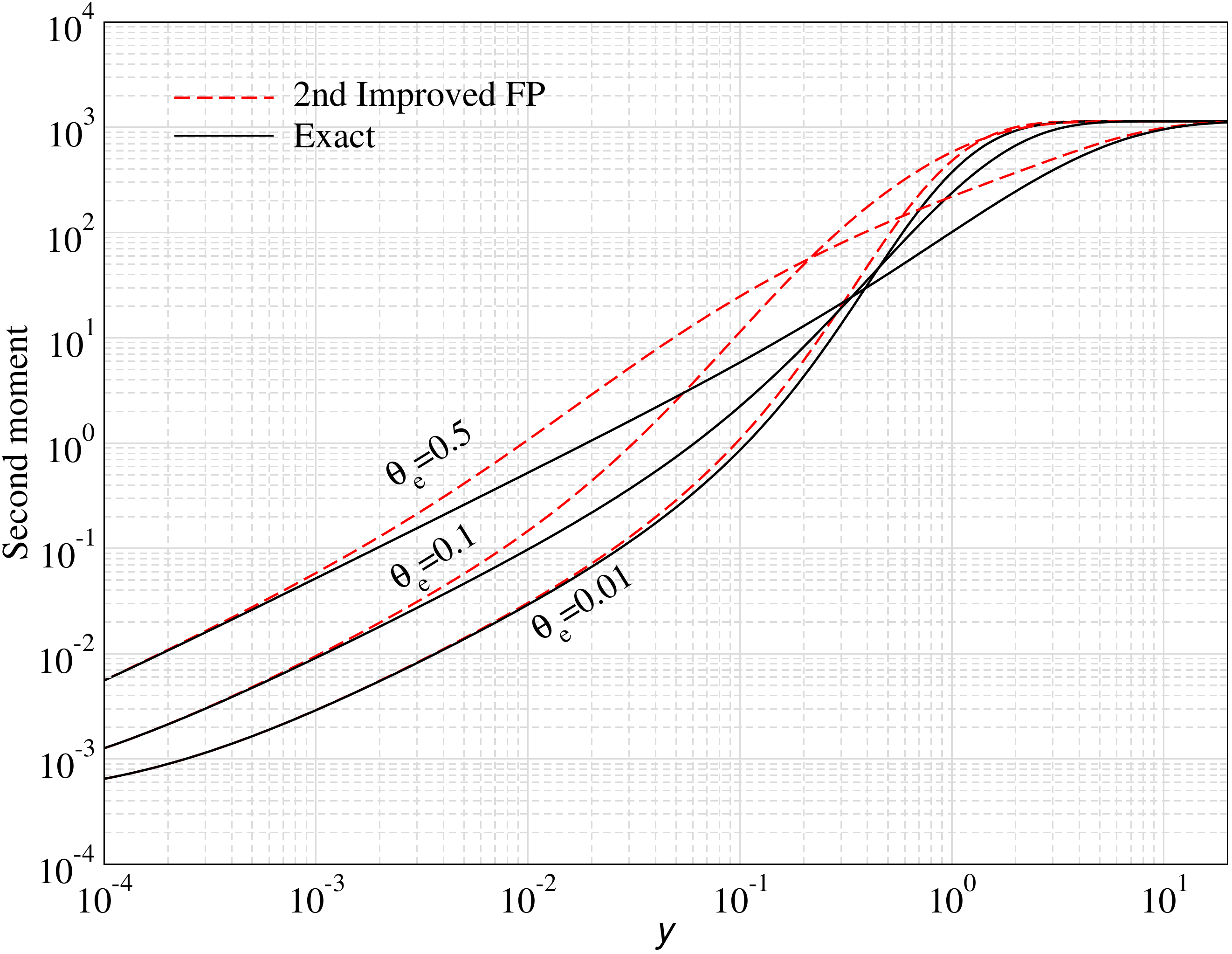}
\\
\caption{Evolution of first and second moment of photon field for injection at $x_{\rm inj}=0.1$ (no stimulated terms). The electrons are kept at a constant temperature $\The$ with values as annotated. We only compare the solutions of the second improved FP approach [see Sect.~\ref{sec:New_attempt_II}] with the exact solution.}
\label{fig:x_0.1_2ndFP_exact_mom}
\end{figure}
%-------------------------------------

Overall, these cases illustrate that the improved FP scheme describes the average evolution of the photon field at $y\lesssim 0.1-1$ fairly well, in some cases even significantly improving over the Kompaneets equation. However, it fails at higher values of the scattering $y$ parameter. The latter is related to the fact that the correct equilibrium solution is not reached, and accurate solutions can only be obtained using the full scattering kernel approach.
It is clear that these statements depend significantly on the initial state of the photon field. For example, if starting close to equilibrium, then the improved FP approach will fail very quickly, even if $y\ll 1$. As we will see below, the solution of none of the FP treatments truly reproduces the detailed evolution of the photon field, and the full treatment is always needed at high temperature and frequency. 

%--------------------------------------------------------------
\subsubsection{Thermalization time-scale}
\label{sec:y_therm}
%--------------------------------------------------------------
Using the evolution of the moments, we can answer the question on which time-scale the photon field relaxes towards the equilibrium solution. As a simple criterion, we define the thermalization $y$-parameter, $y_{\rm therm}$, such that the second moment no longer changes by more than 1 percent\footnote{Monitoring also the first moment gives similar results but is complicated by the fact that for $x_{\rm inj} \simeq 3-4$ it evolves very slowly overall. Alternatively one could directly use the evolution of the photon distribution, but the general picture does not change.}. 
The results of this exercise are summarized in Fig.~\ref{fig:y_therm} for various injection frequencies using the Kompaneets and exact solutions. We typically find $y_{\rm therm}\gtrsim 1$ aside from cases with $x_{\rm inj}\simeq 3$, where it is a little lower. This is related to the fact that redistributing photons from the initial point can be achieved by broadening the distribution (i.e., moving a fraction downwards in energy and a fraction upward) but spending less time moving the mean of the photons distribution.

The Kompaneets treatment does not capture the temperature dependence of $y_{\rm therm}$, nor does it correctly capture the slowing of thermalization at higher energies. The thermalization $y$-parameter increases significantly with temperature owing to the fact that Klein-Nishina corrections become more important. Stimulated scattering further slows thermalization at $x_{\rm inj}\lesssim 1$ because photons up-scatter more slowly (see Sect.~\ref{sec:evol_moments_stim} for more discussion). These finding again highlight how a complete treatment of the problem is needed to obtain accurate results at relativistic energies.

%--------------------------------------------------------------
\begin{figure*}
\centering 
\includegraphics[width=\columnwidth]{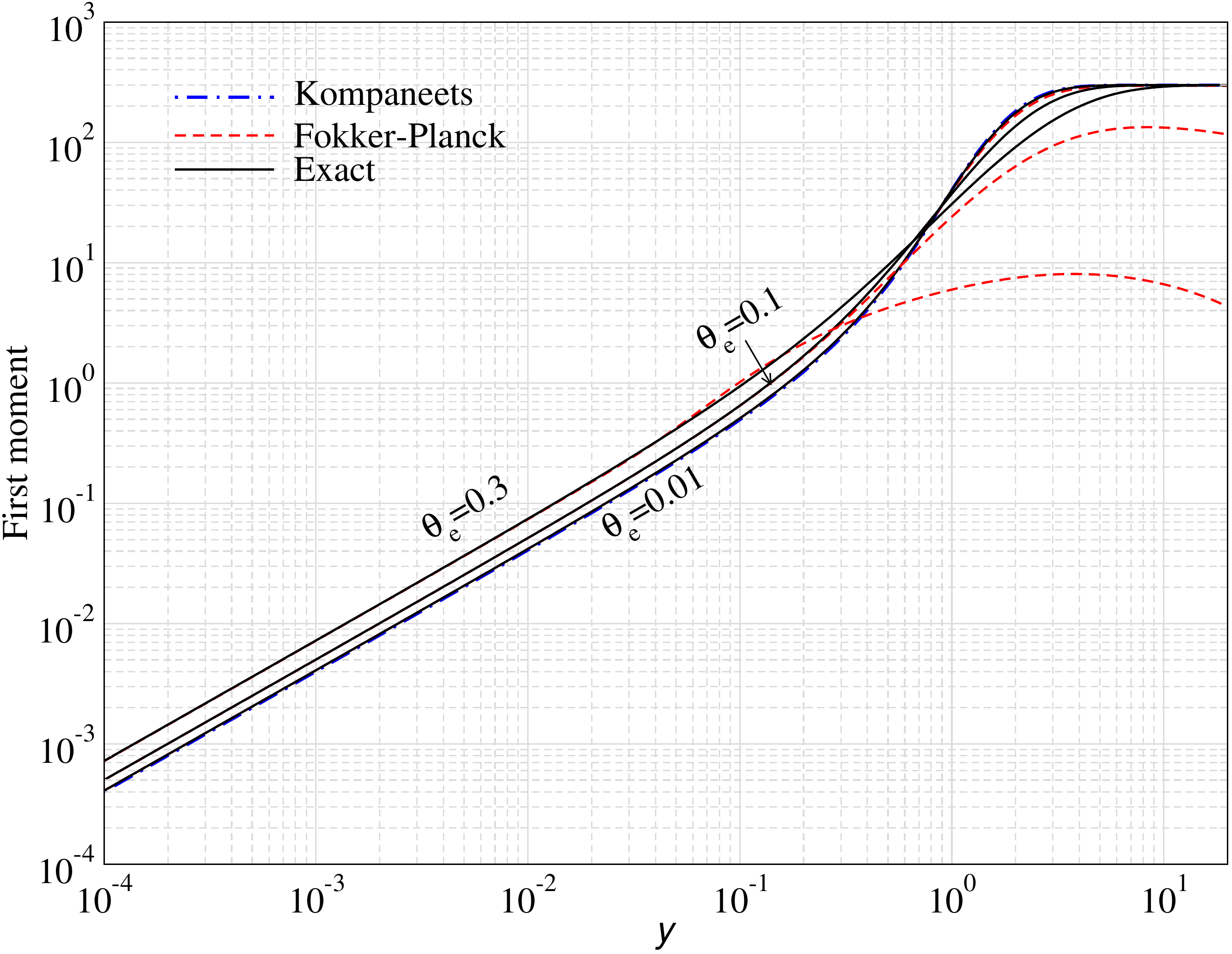}
\hspace{4mm}
\includegraphics[width=\columnwidth]{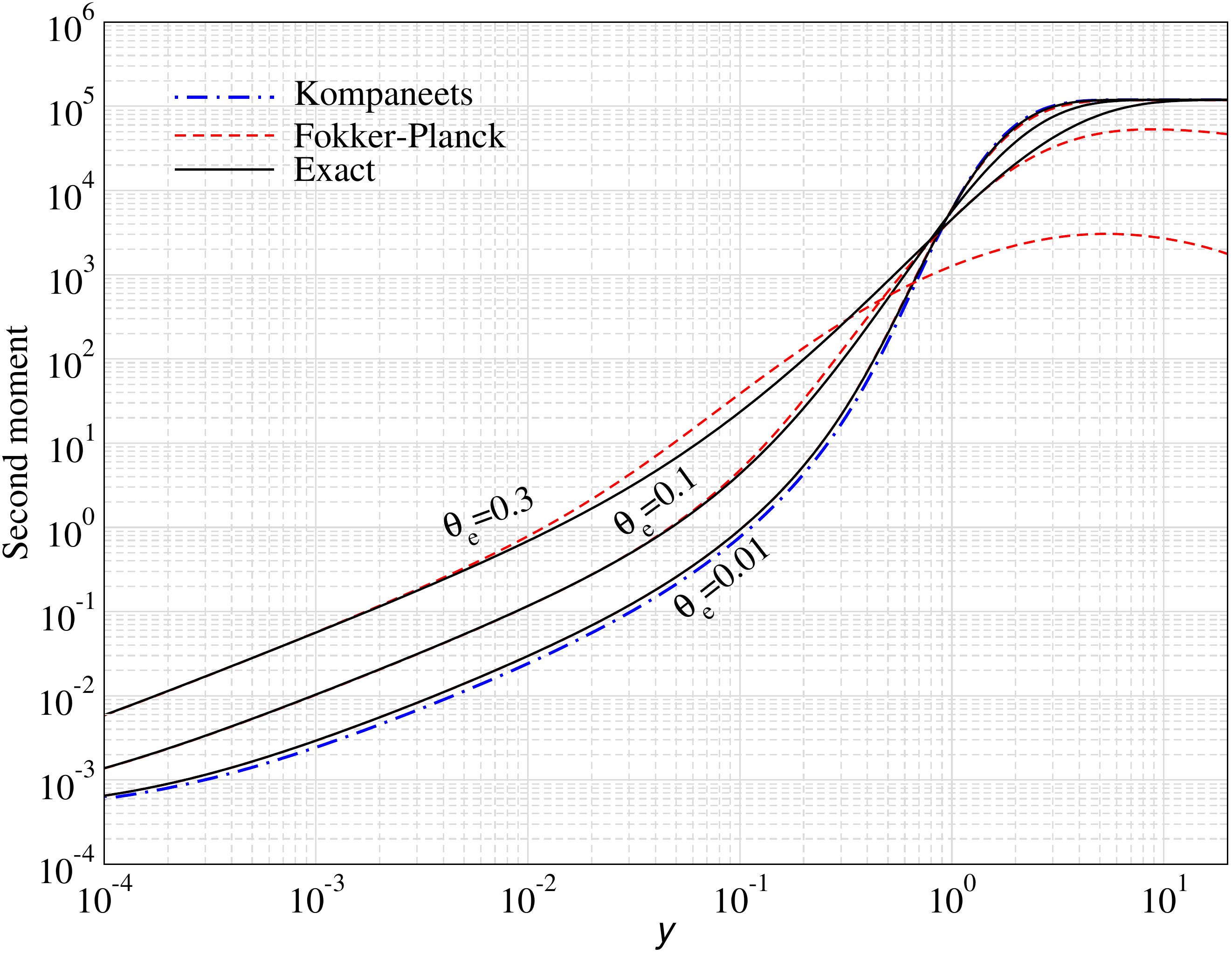}
\\[1mm]
\includegraphics[width=\columnwidth]{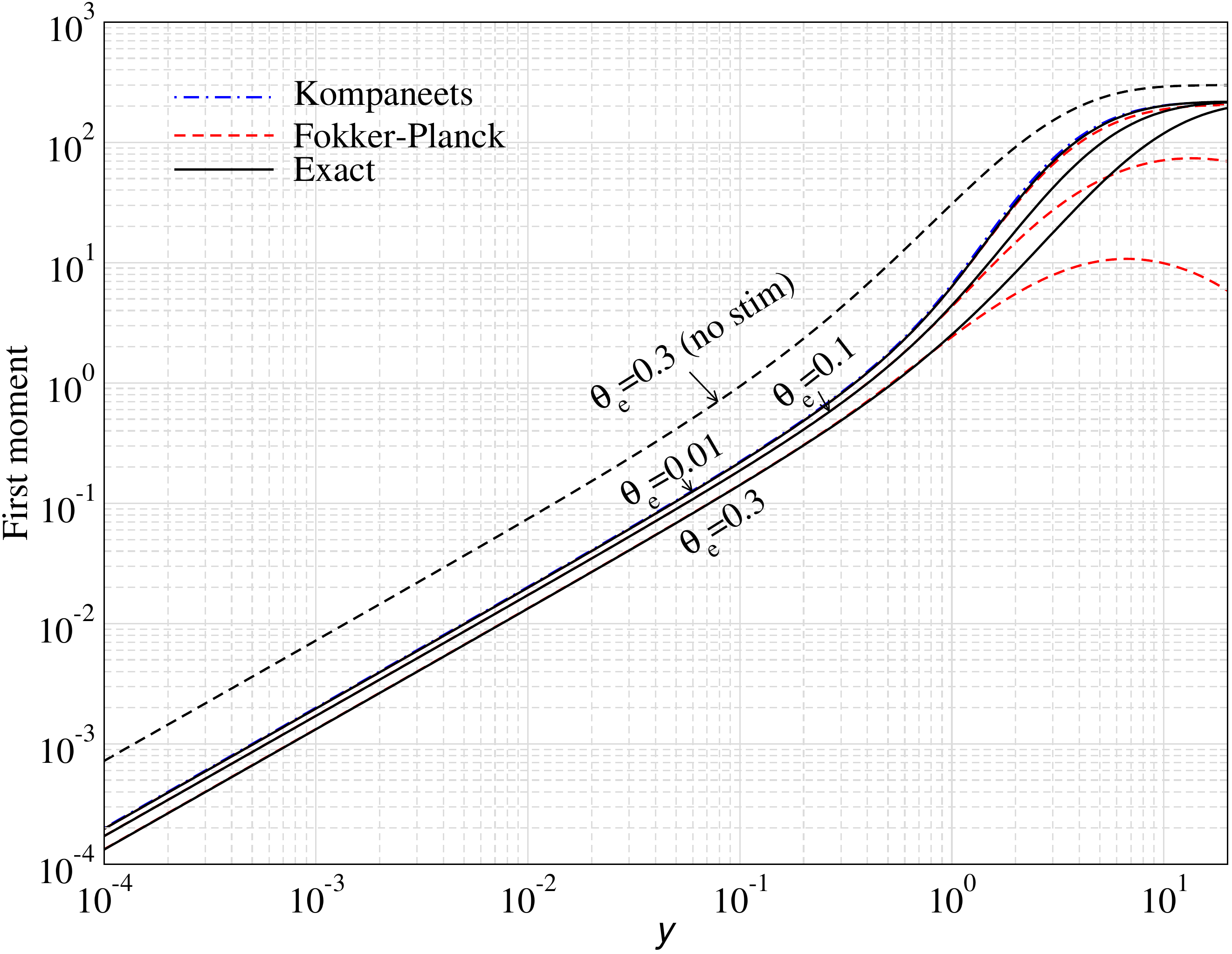}
\hspace{4mm}
\includegraphics[width=\columnwidth]{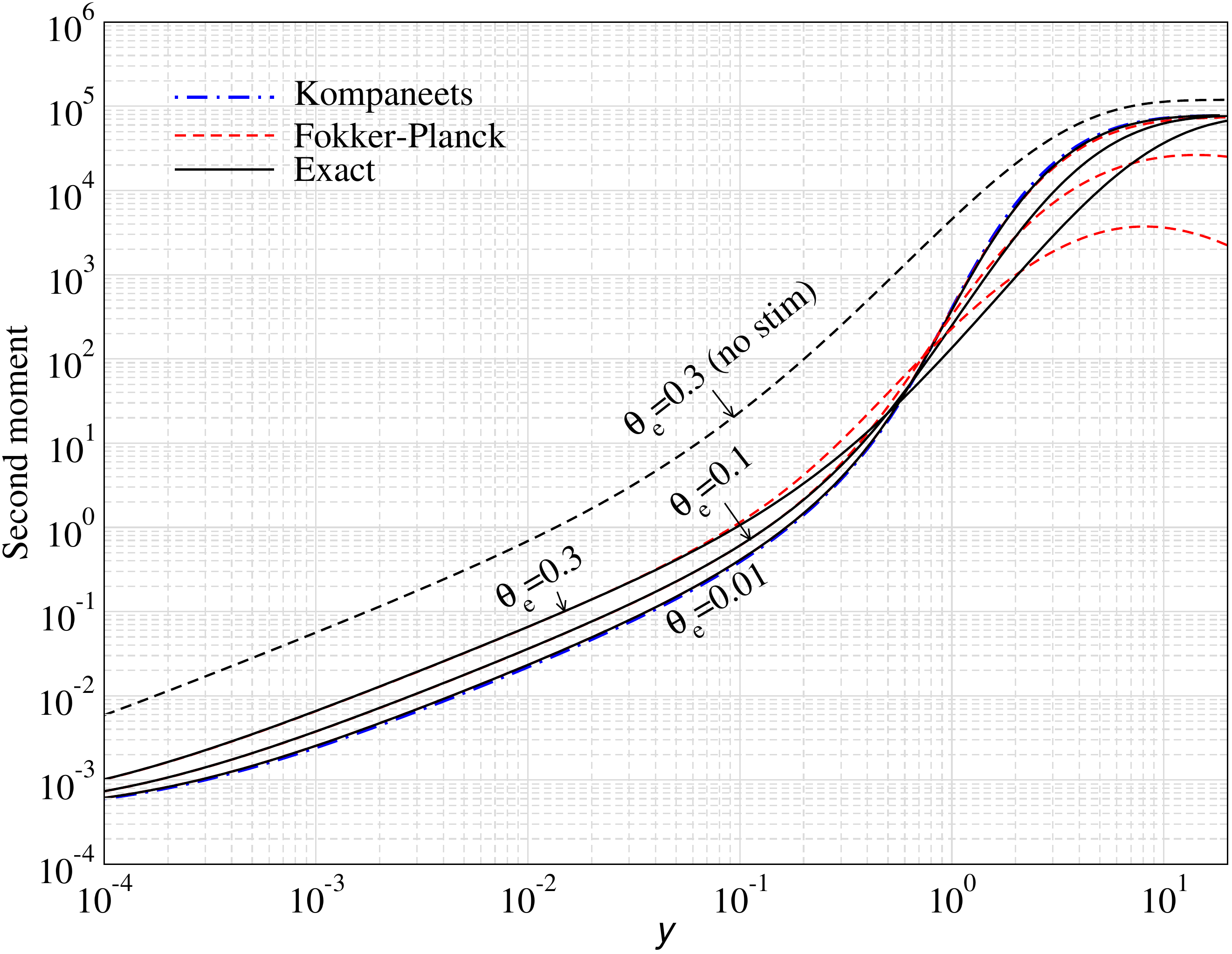}
\\
\caption{Evolution of first (left panels) and second (right panels) moments of photon field for injection at $x_{\rm inj}=0.01$. Stimulated terms were included for the lower panels. For reference, the black dotted line in the lower panel is the result obtained without stimulated scattering and using the exact kernel.
In each panel, the electrons are kept at a constant temperature $\The$ with values as annotated. We compare the solutions of the Kompaneets equation and the improved FP approach [diffusion coefficients as in Eq.~\eqref{eq:imFPI}] with the solution for the exact scattering kernel. For the first moment, stimulated terms change the ordering of the curves with temperature for low $y$ parameter (compare left panels). This effect is not captured by the Kompaneets equation but is reproduced by the improved FP approximation. For the second moment, stimulated scattering terms reduce the temperature-dependence of the solutions.}
\label{fig:x_0.01_moment_nostim_stim}
\end{figure*}
%--------------------------------------------------------------

%--------------------------------------------------------------
\subsubsection{Evolution of moments using the 2nd improved FP approach}
\label{sec:evol_moments_II}
%--------------------------------------------------------------
Before moving on the cases with stimulated scattering, we briefly demonstrate that the 2nd improved FP approach described in Sect.~\ref{sec:New_attempt_II} does not provide a viable alternative for solving the problem. The evolution of the first and second moments of the photon field in this approach is shown in Fig.~\ref{fig:x_0.1_2ndFP_exact_mom} in comparison to the exact kernel result for $x_{\rm inj}=0.1$. While the evolution of the second moment is captured well early on, significant departures are visible for the first moment, even at low values of $y$. This is because the effect of recoil is not correctly included in the diffusion coefficients, hence significantly overestimating the mean shifts per scattering. Although not explicitly illustrated, we find the 2nd improved FP approach to also fail at higher injection energies. Even if in contrast to the first improved FP approach the solution for the 2nd improved FP approach asymptotes to the correct equilibrium, we do not recommend using it. We will therefore not consider it any further in this work. Our analysis also illustrates how naively using detailed balance arguments when constructing a FP representation is not generally possible. However, it works for the Kompaneets limit \citep[e.g.,][]{Rybicki1979} as detailed balance is indeed expected to work order by order in the perturbative parameter, which in this context is the electron temperature.

%%--------------------------------------------------------------
%\begin{figure}
%\centering 
%\includegraphics[width=\columnwidth]{./eps/x_0.01_1st_mom_nostim.pdf}
%\\
%\includegraphics[width=\columnwidth]{./eps/x_0.01_2nd_mom_nostim.pdf}
%\\
%\caption{First and second moment for photon evolution without stimulated scattering for temperature $\theta_e$ as denoted.}
%\label{fig:x_0.01_moment_nostim}
%\end{figure}
%%--------------------------------------------------------------

%%--------------------------------------------------------------
%\begin{figure}
%\centering 
%\includegraphics[width=\columnwidth]{./eps/x_0.01_1st_mom_stim.pdf}
%\\
%\includegraphics[width=\columnwidth]{./eps/x_0.01_2nd_mom_stim.pdf}
%\\
%\caption{First and second moment for photon evolution with stimulated scattering for temperature $\theta_e$ as denoted.}
%\label{fig:x_0.01_moment_stim}
%\end{figure}
%%--------------------------------------------------------------

%-------------------------------------
\begin{figure*}
\centering 
\includegraphics[width=\columnwidth]{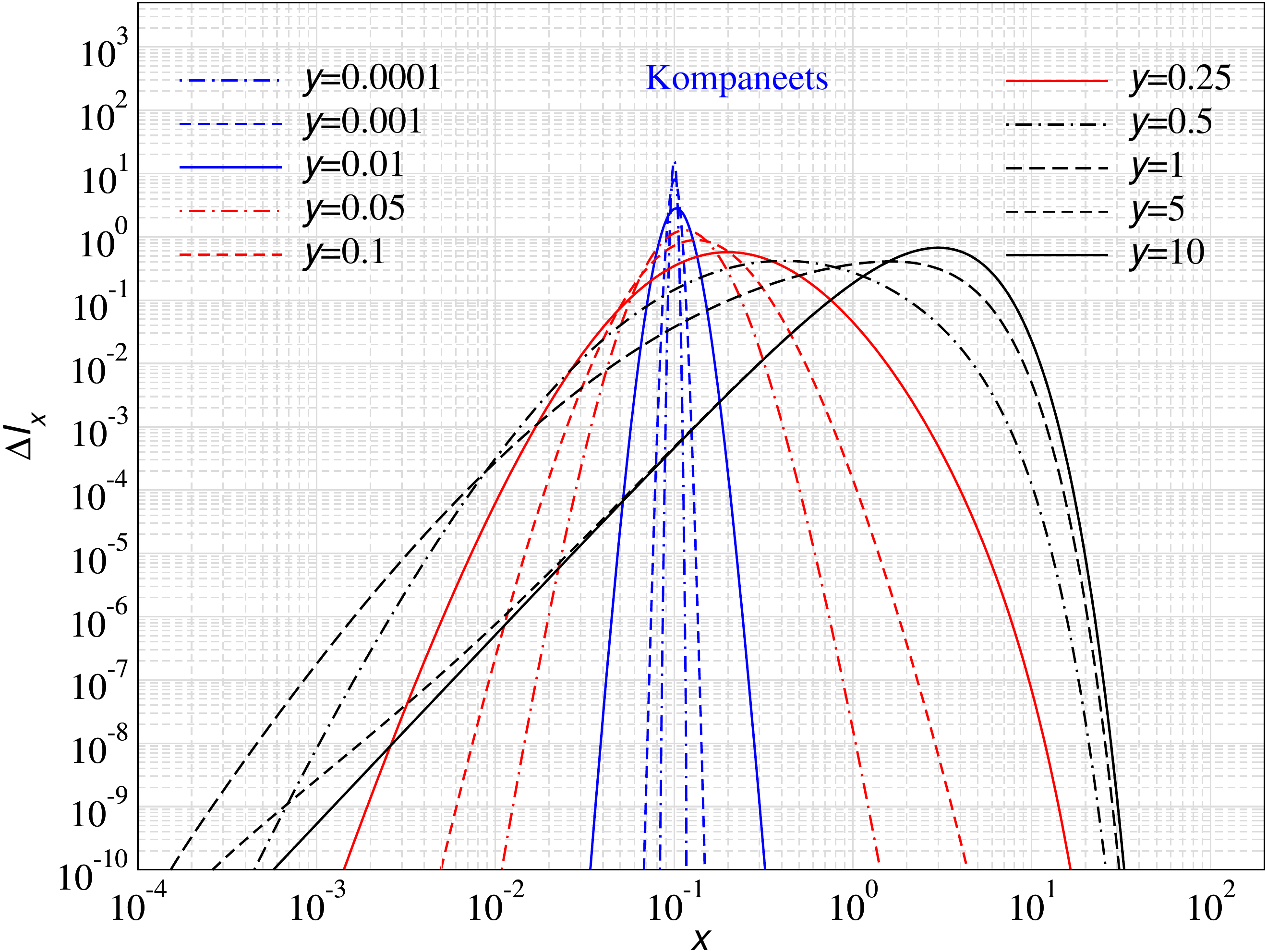}
\hspace{4mm}
\includegraphics[width=\columnwidth]{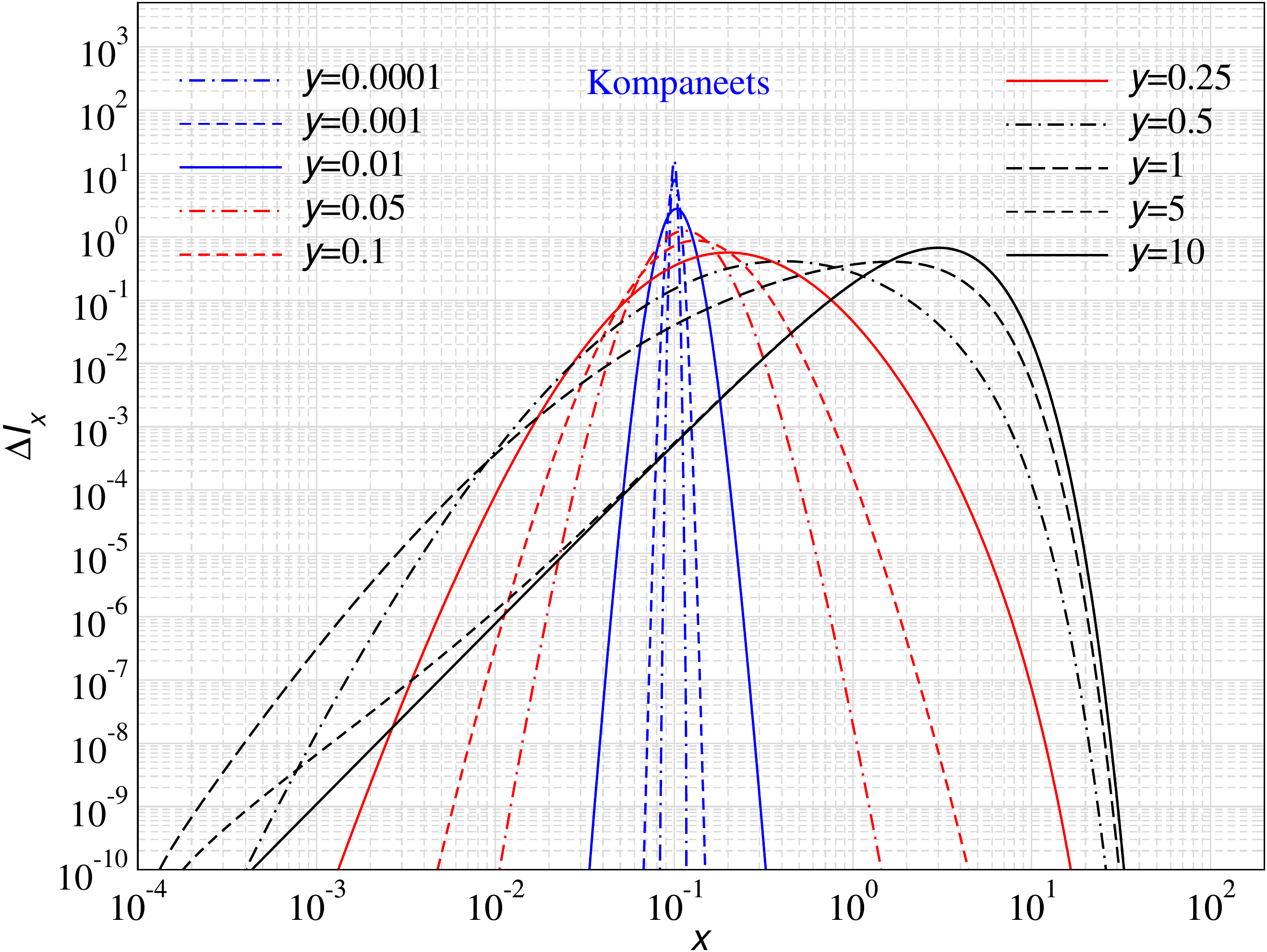}
\\[10mm]
\includegraphics[width=\columnwidth]{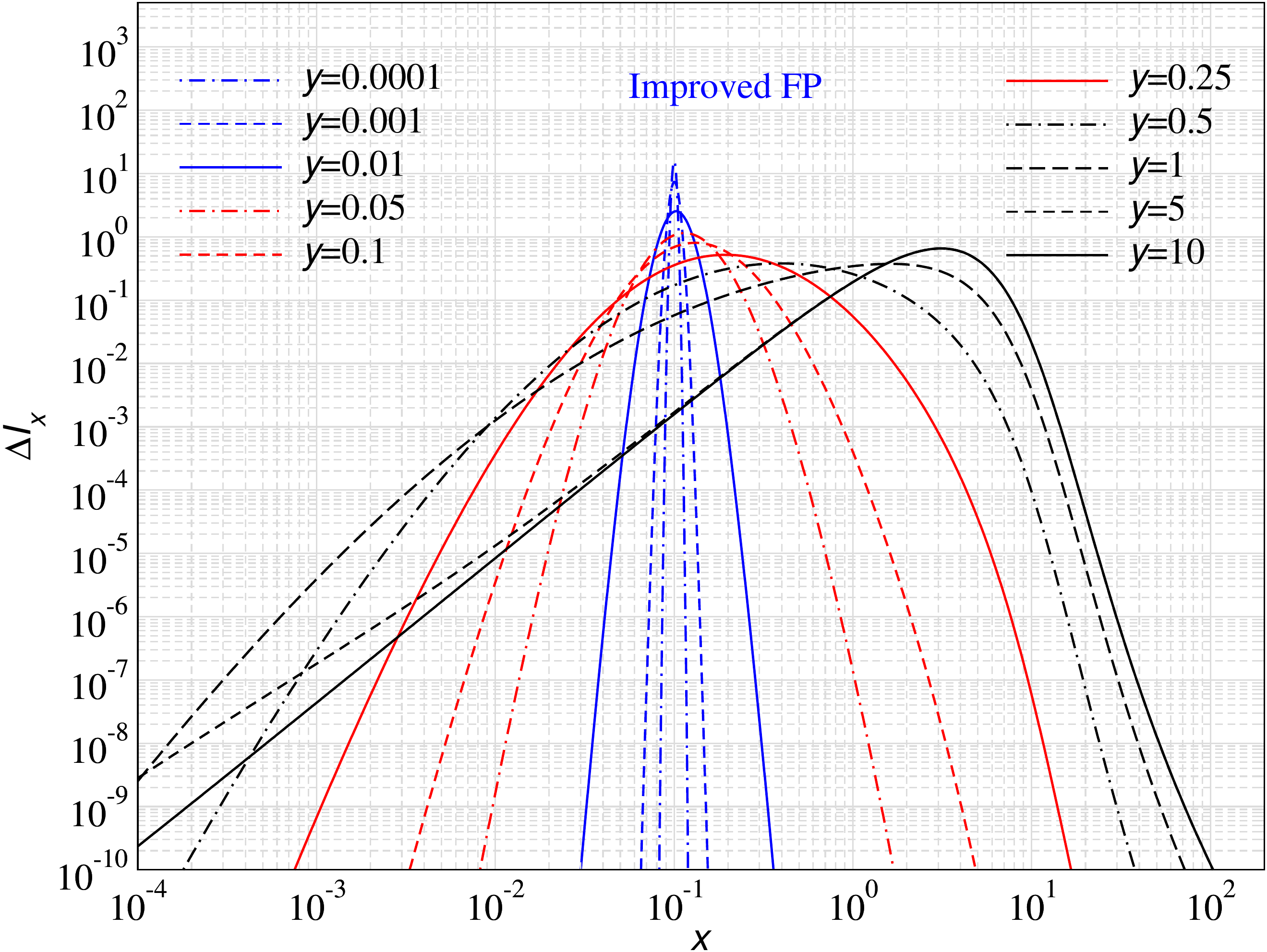}
\hspace{4mm}
\includegraphics[width=\columnwidth]{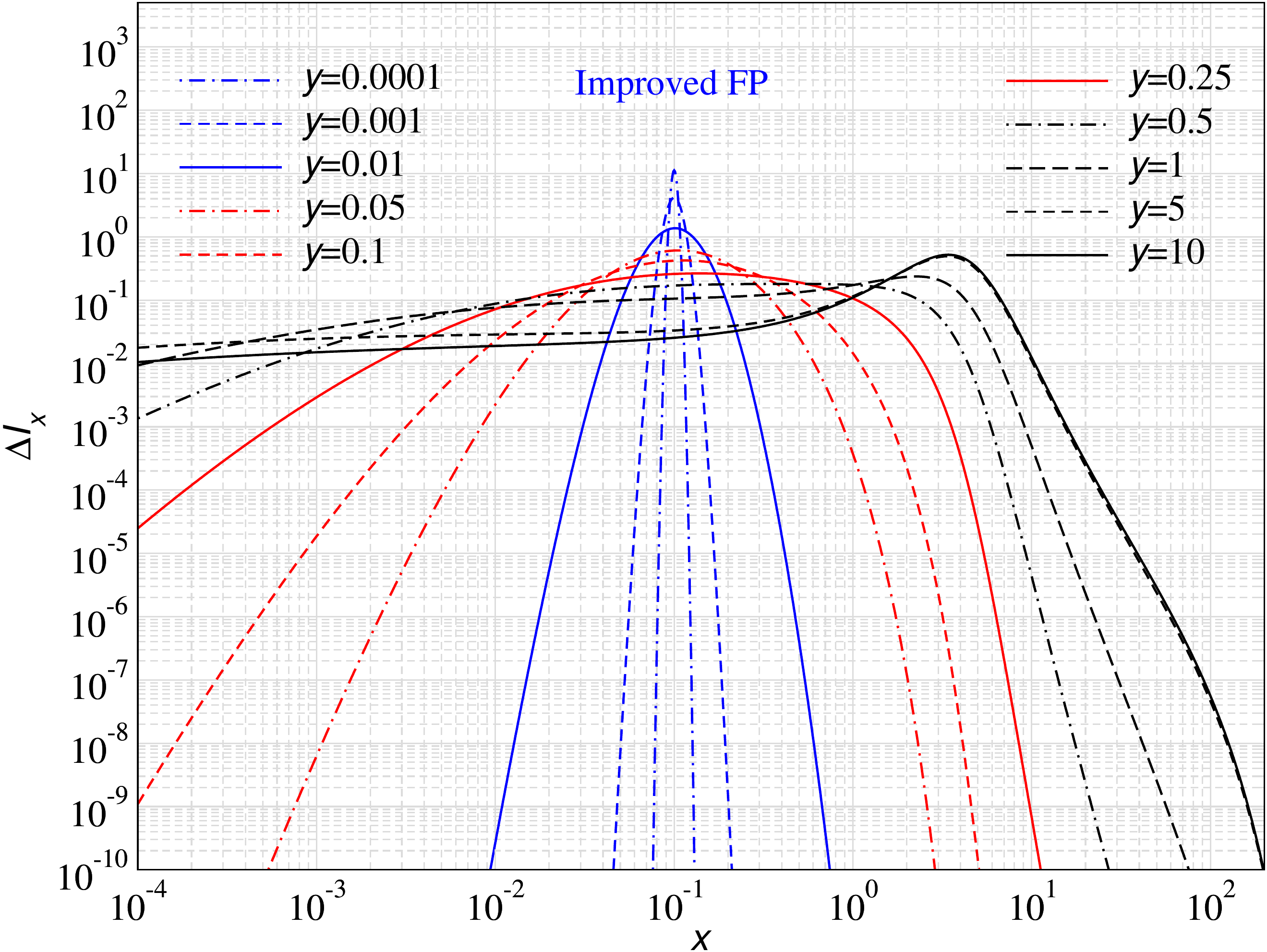}
\\[10mm]
\includegraphics[width=\columnwidth]{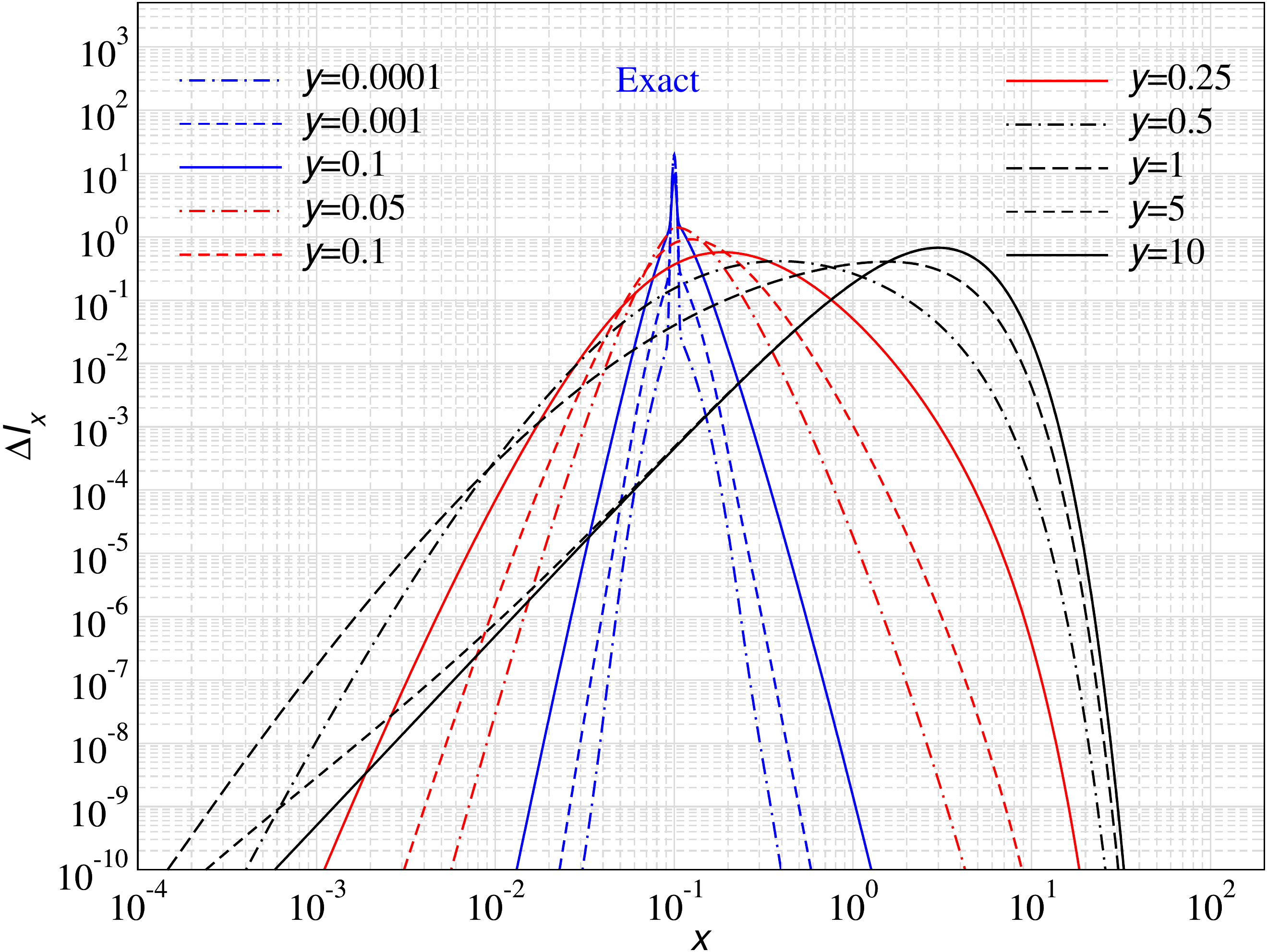}
\hspace{4mm}
\includegraphics[width=\columnwidth]{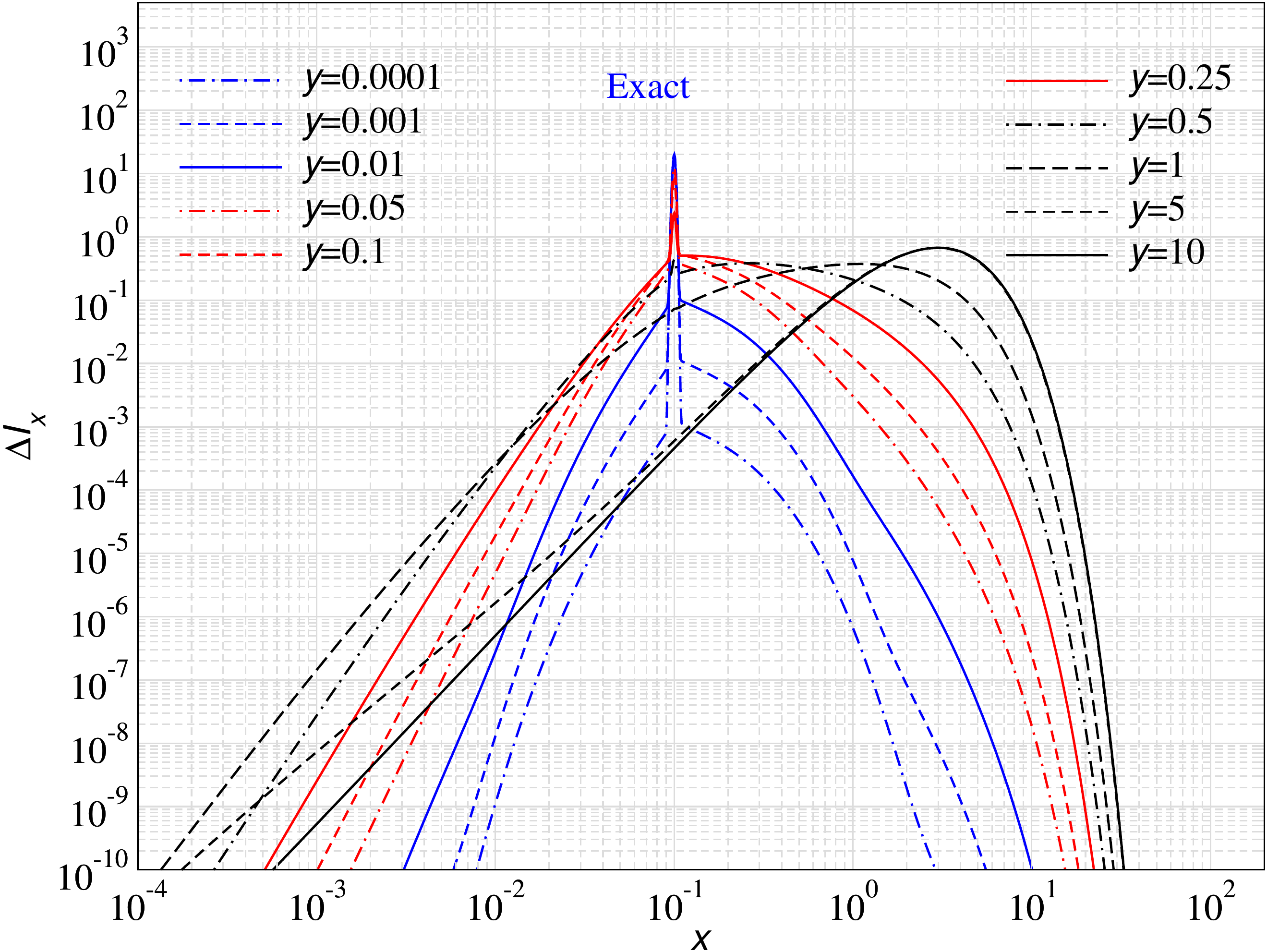}
\\
\caption{Snapshots of solution for photon field as a function of frequency for different $y$ parameters and numerical approaches as annotated. On the y-axis, we plot the dimensionless intensity, $I_x=x^3\,n$. Photon were injected at $x_{\rm inj}=0.1$ at $y=0$.
The left panels are for temperature $\The=0.01$, while the right panels show cases with $\The=0.1$ to illustrate relativistic corrections. Stimulated scatterings were neglected. The improved FP approach describes the average properties of the solution better than the Kompaneets equation until $y$ exceeds $\simeq 0.1$. Afterwards, it generally fails and the Kompaneets approach is preferred. Details of the exact solution are not captured by either of the FP schemes. See text for more discussion.}
\label{fig:snap_x_0.1_theta_0.01_0.1}
\end{figure*}
%-------------------------------------

%-------------------------------------
\begin{figure*}
\centering 
\includegraphics[width=\columnwidth]{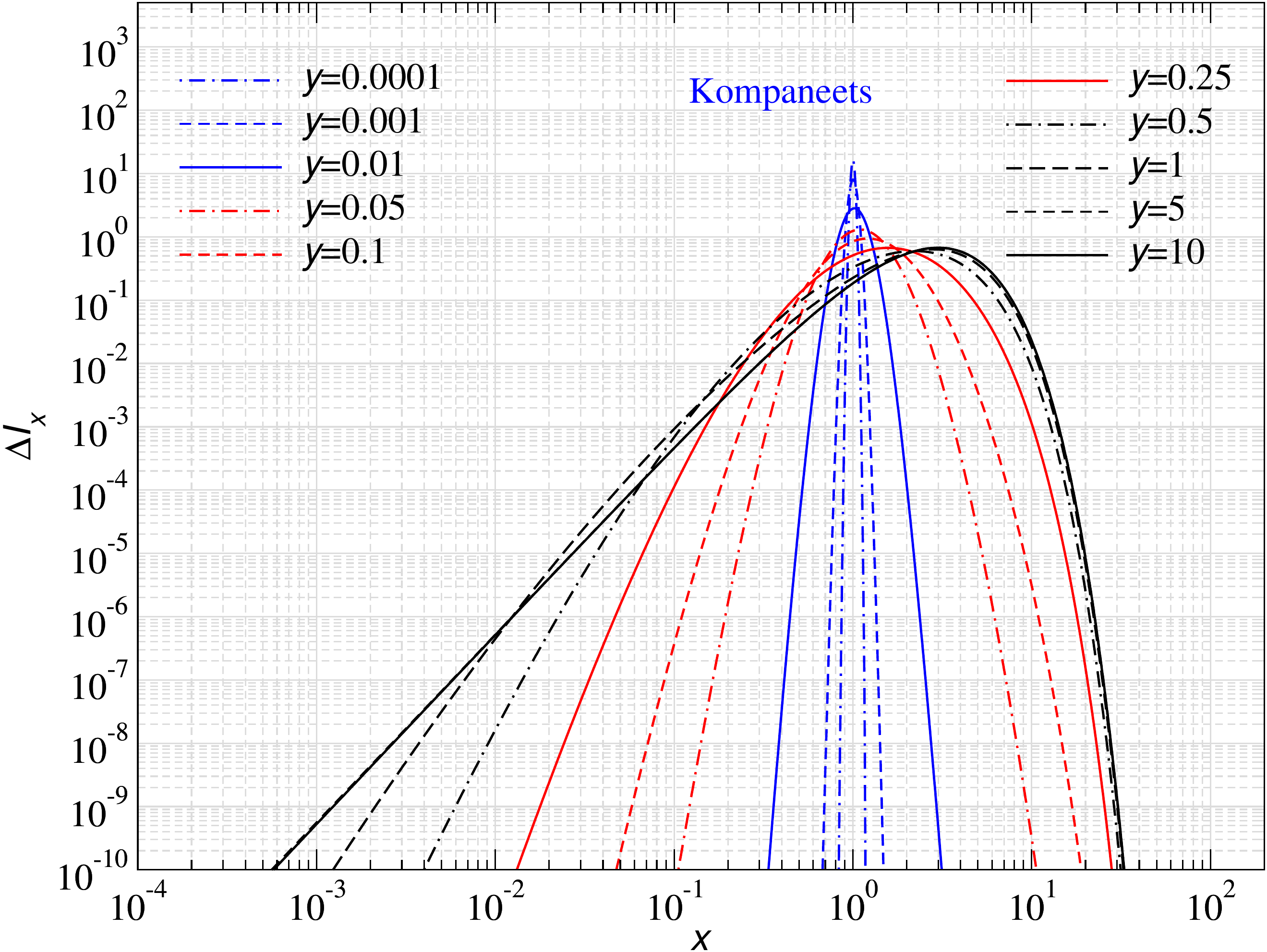}
\hspace{4mm}
\includegraphics[width=\columnwidth]{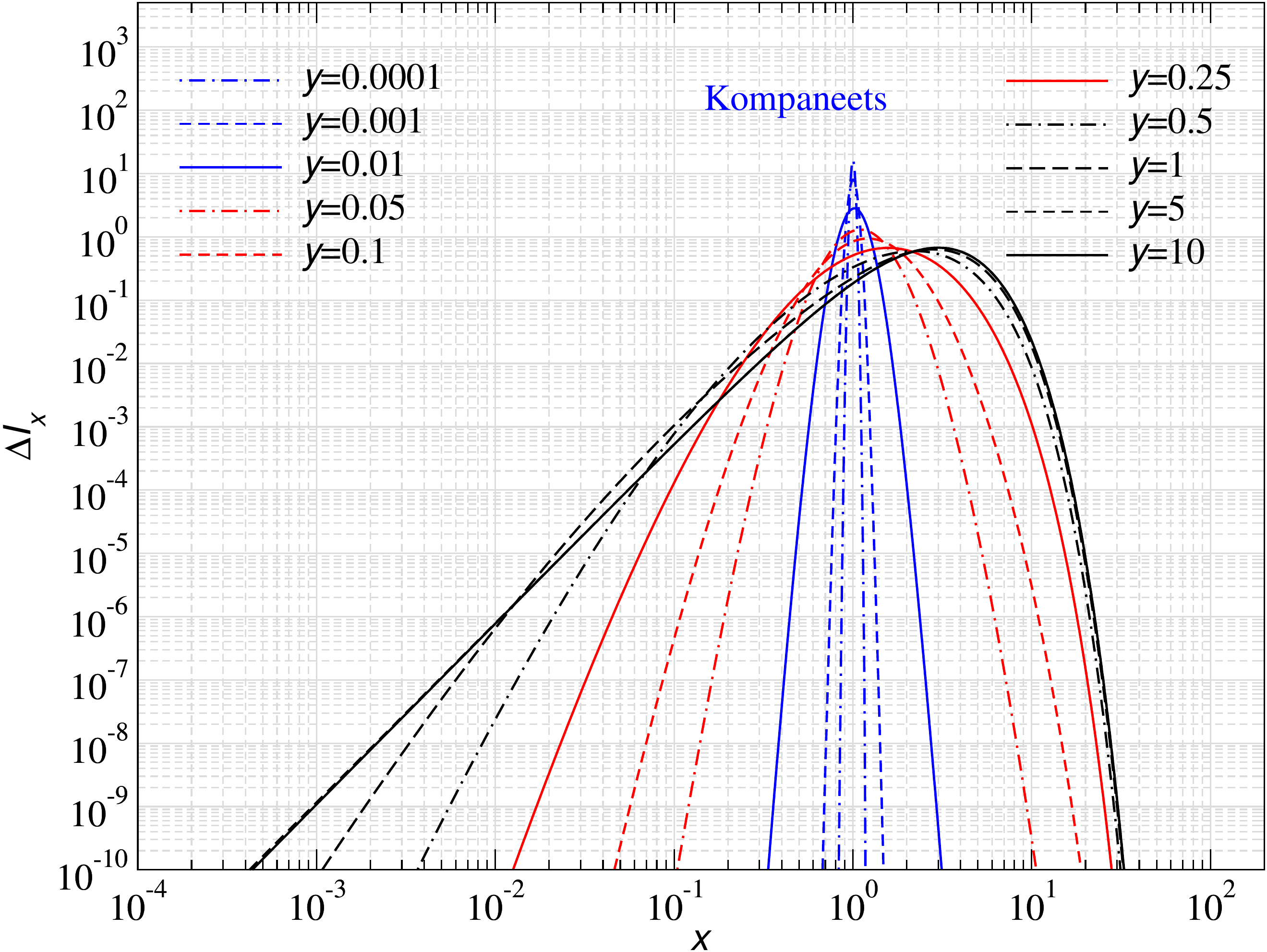}
\\[10mm]
\includegraphics[width=\columnwidth]{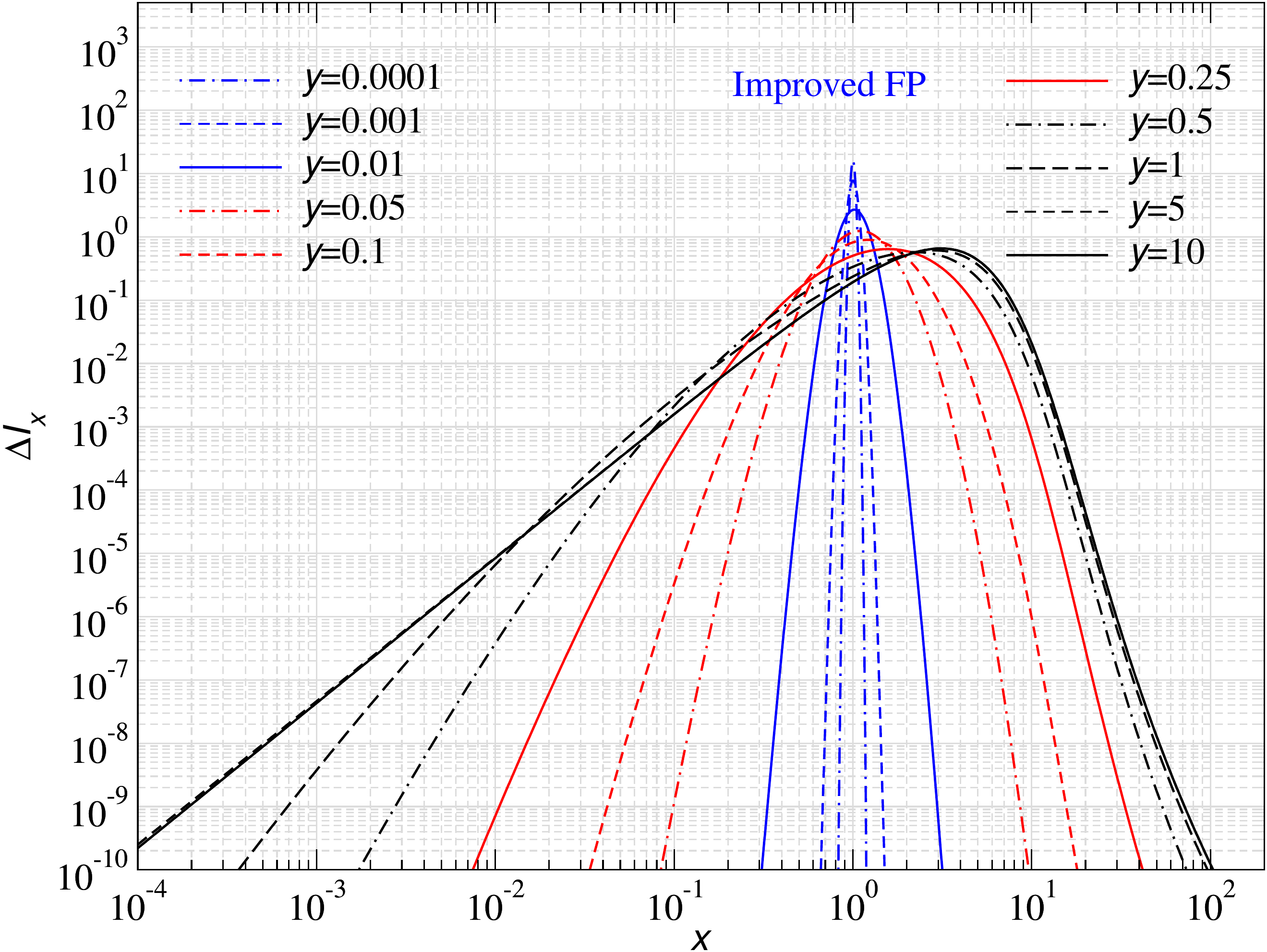}
\hspace{4mm}
\includegraphics[width=\columnwidth]{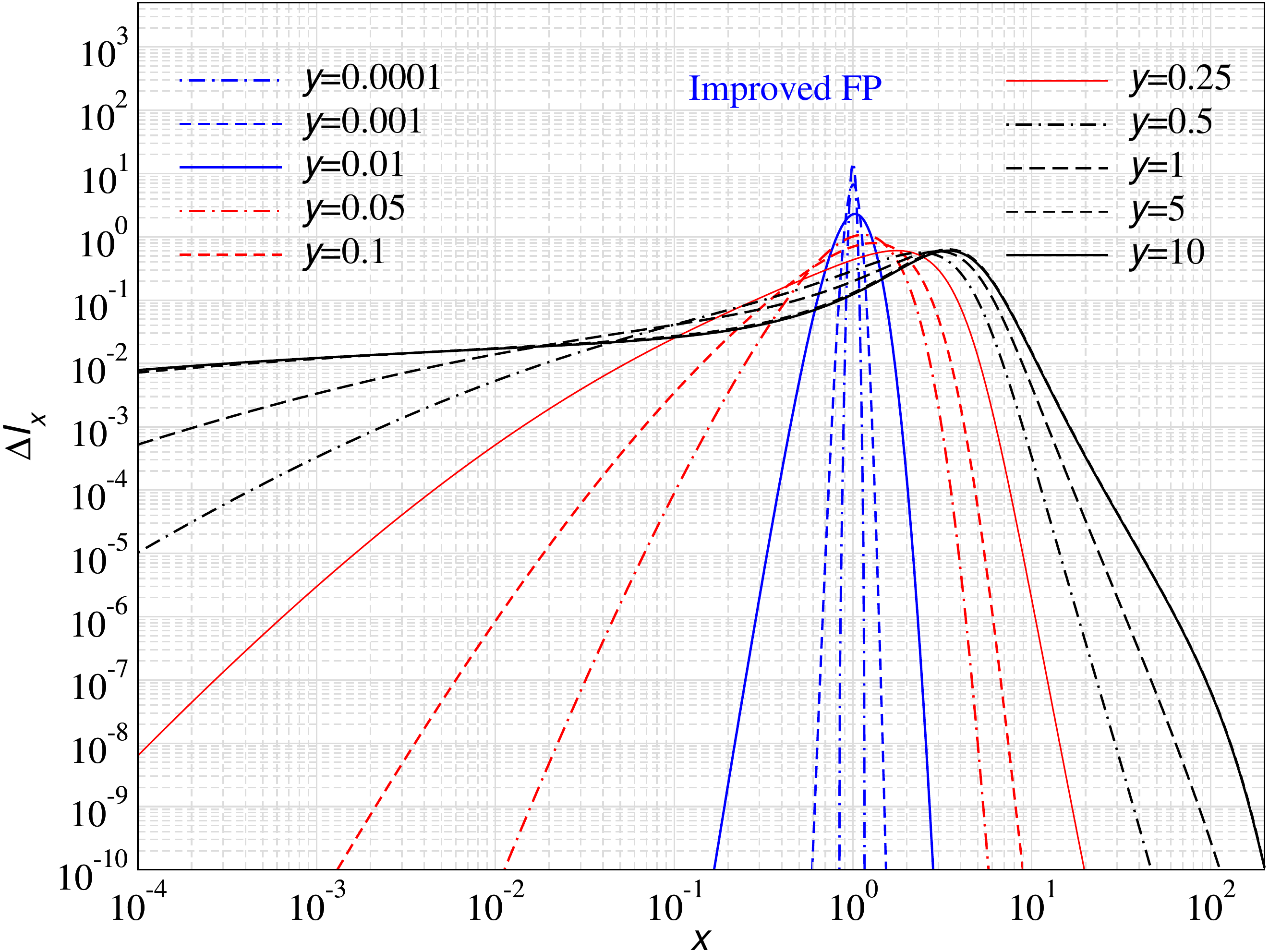}
\\[10mm]
\includegraphics[width=\columnwidth]{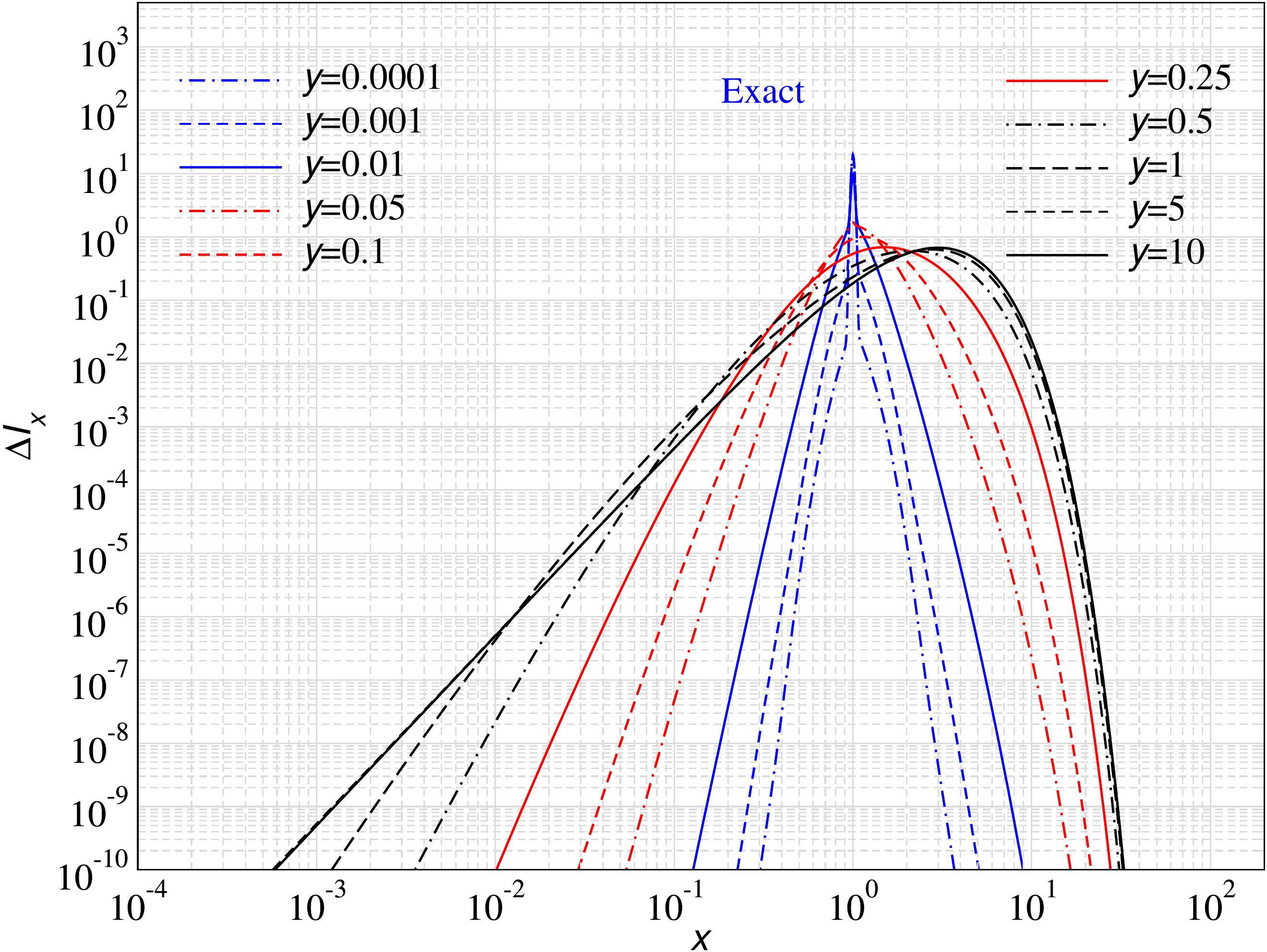}
\hspace{4mm}
\includegraphics[width=\columnwidth]{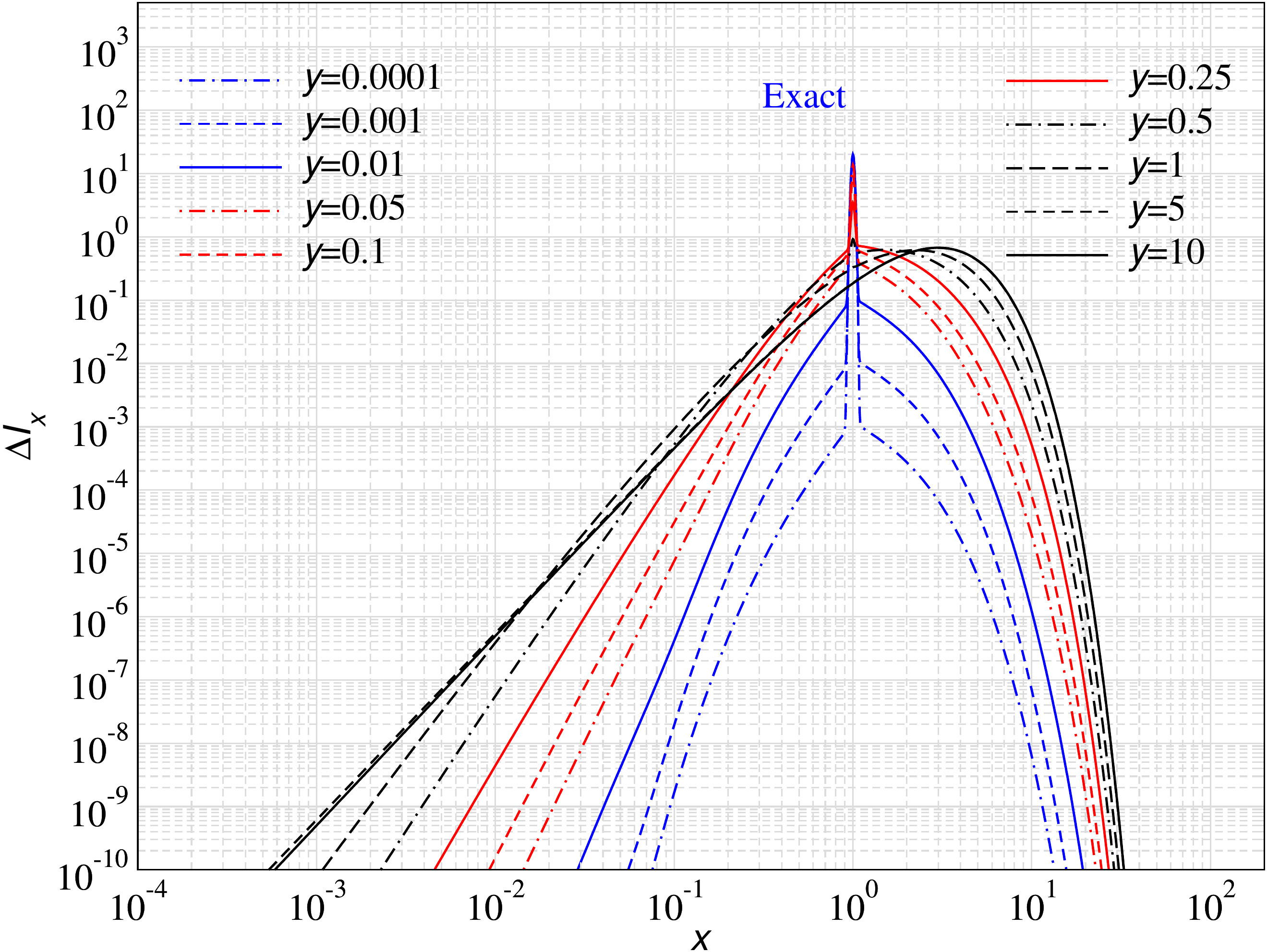}
\\
\caption{Same as Fig.~\ref{fig:snap_x_0.1_theta_0.01_0.1} but for $x_{\rm inj}=1$.
}
\label{fig:snap_x_1.0_theta_0.01_0.1}
\end{figure*}
%-------------------------------------

%-------------------------------------
\begin{figure*}
\centering 
\includegraphics[width=\columnwidth]{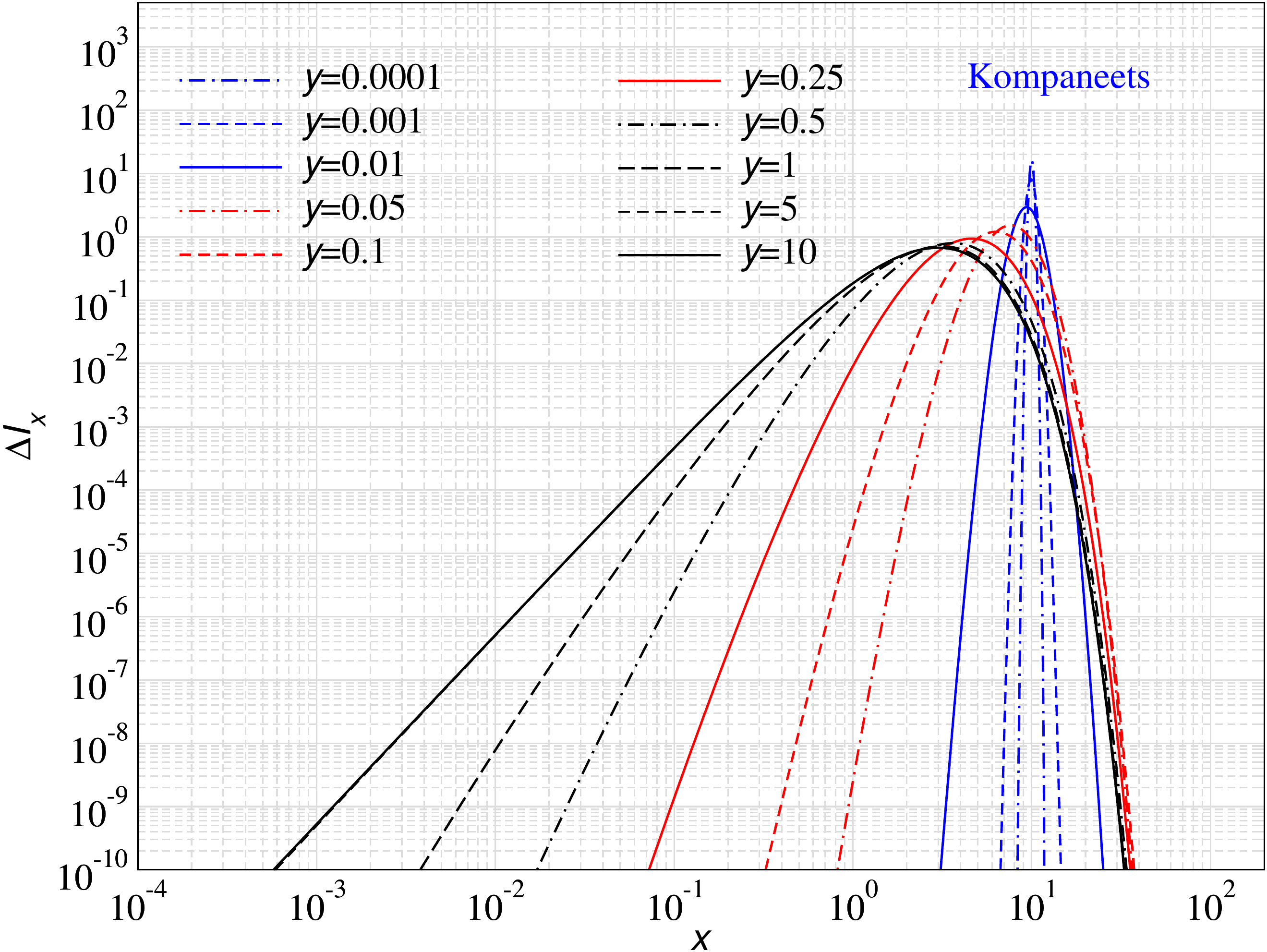}
\hspace{4mm}
\includegraphics[width=\columnwidth]{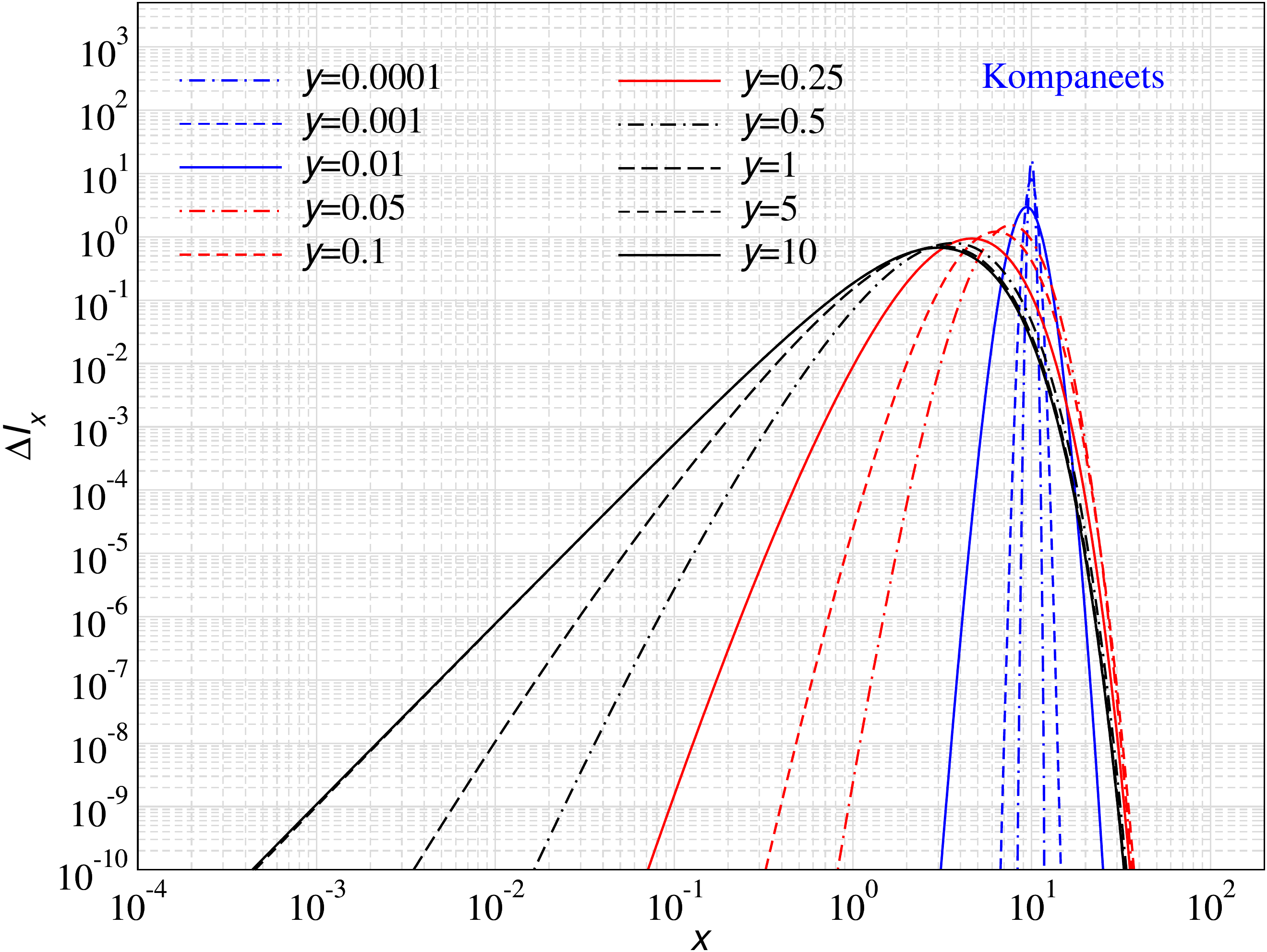}
\\[10mm]
\includegraphics[width=\columnwidth]{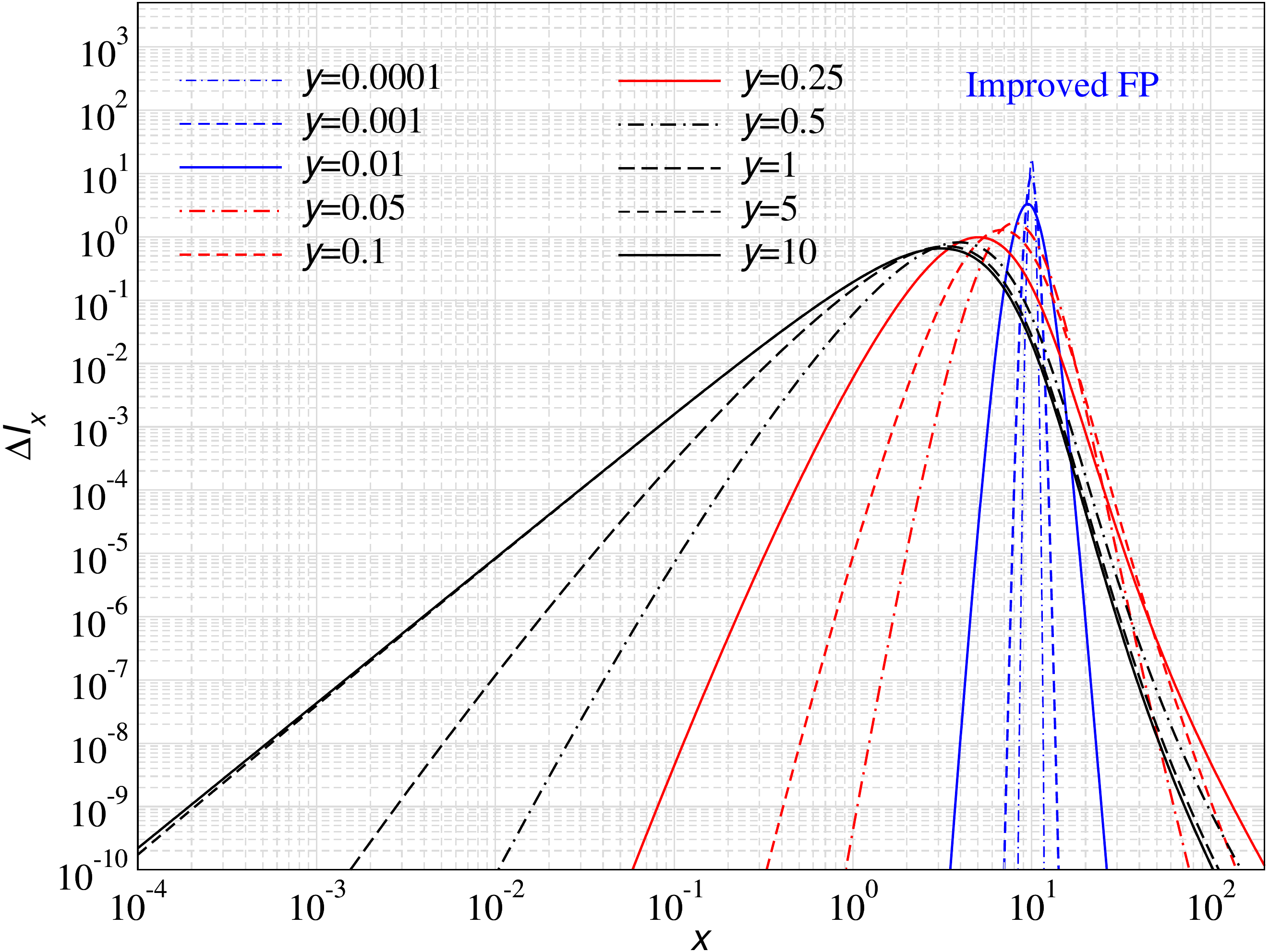}
\hspace{4mm}
\includegraphics[width=\columnwidth]{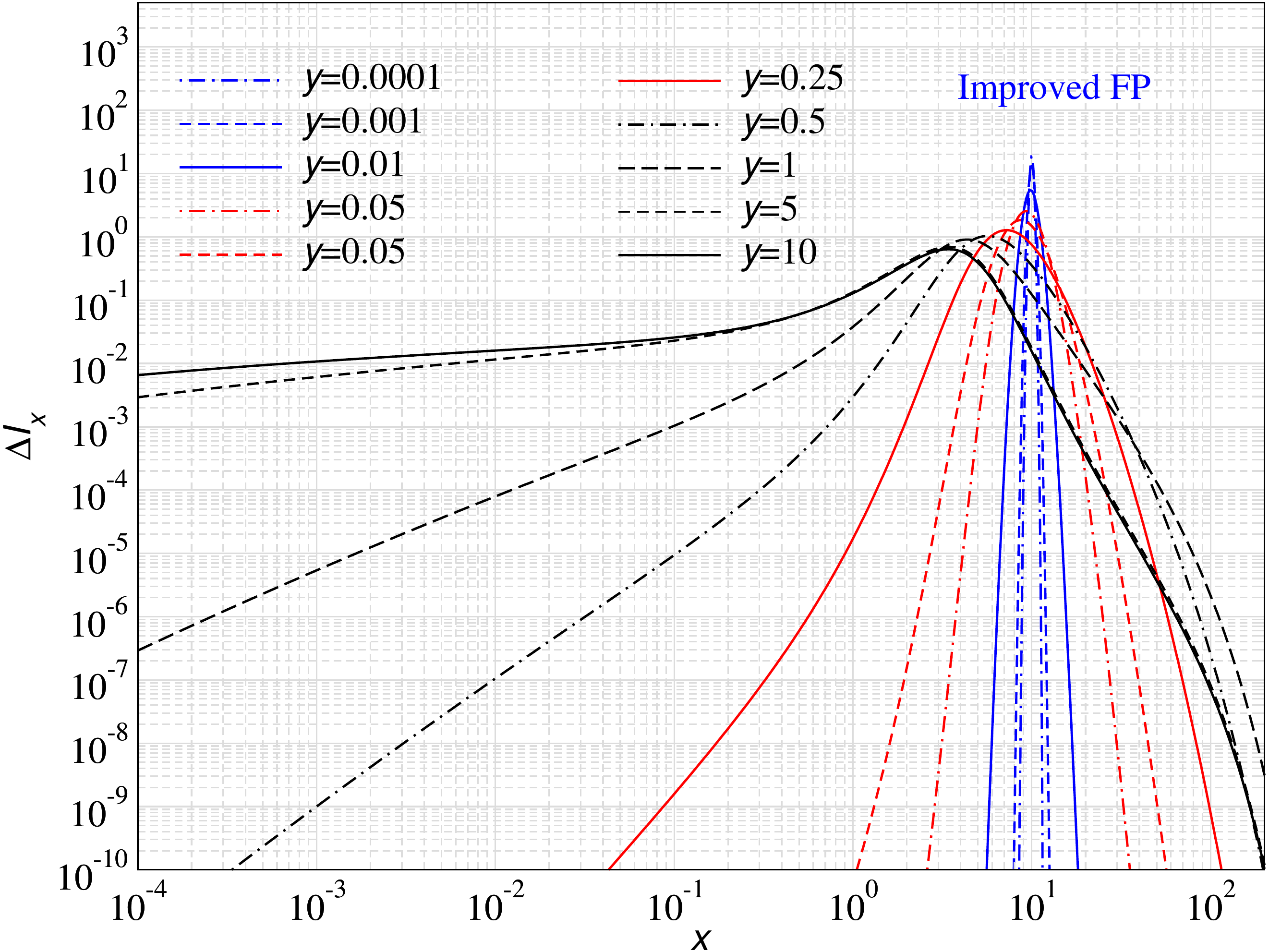}
\\[10mm]
\includegraphics[width=\columnwidth]{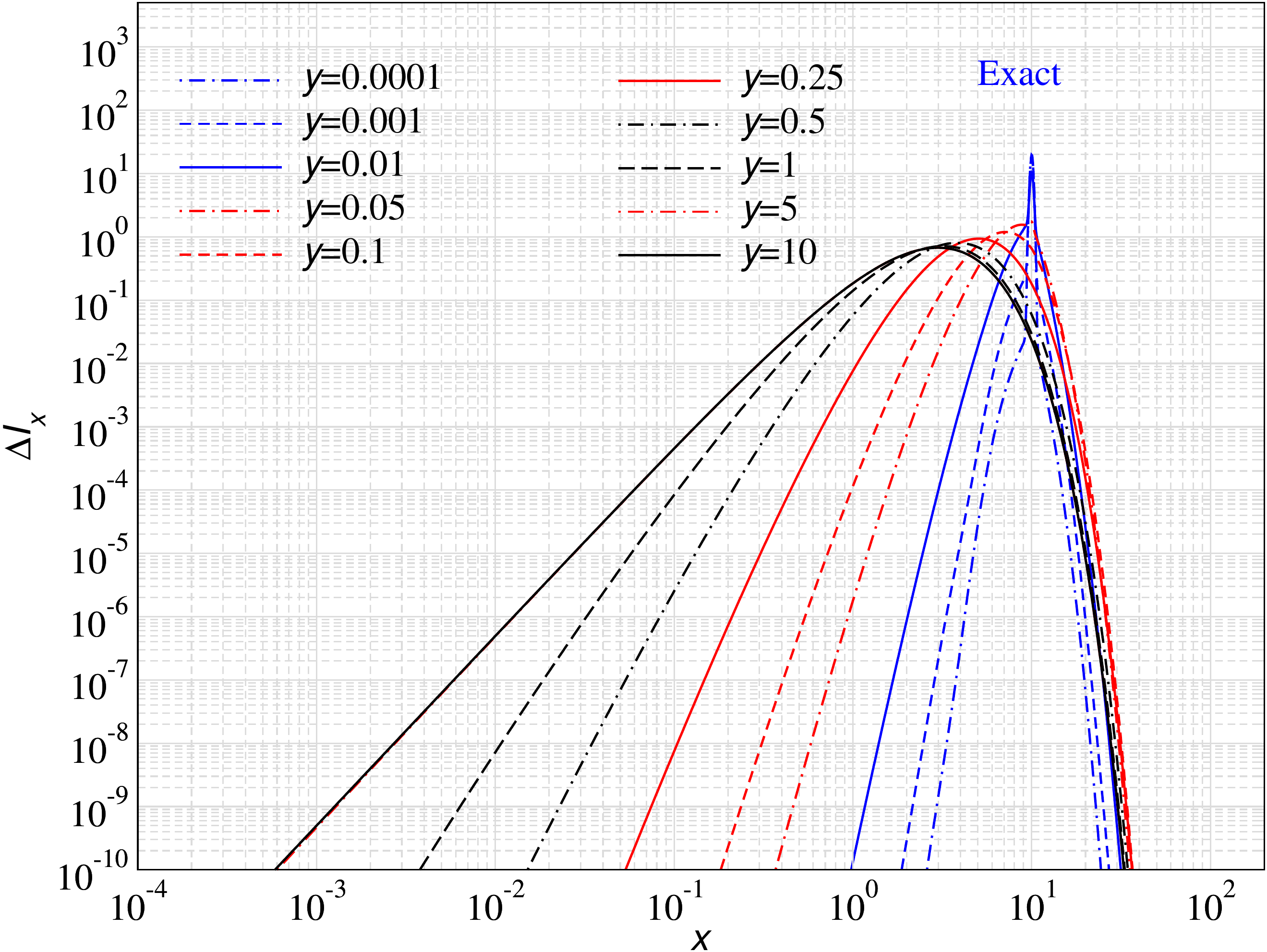}
\hspace{4mm}
\includegraphics[width=\columnwidth]{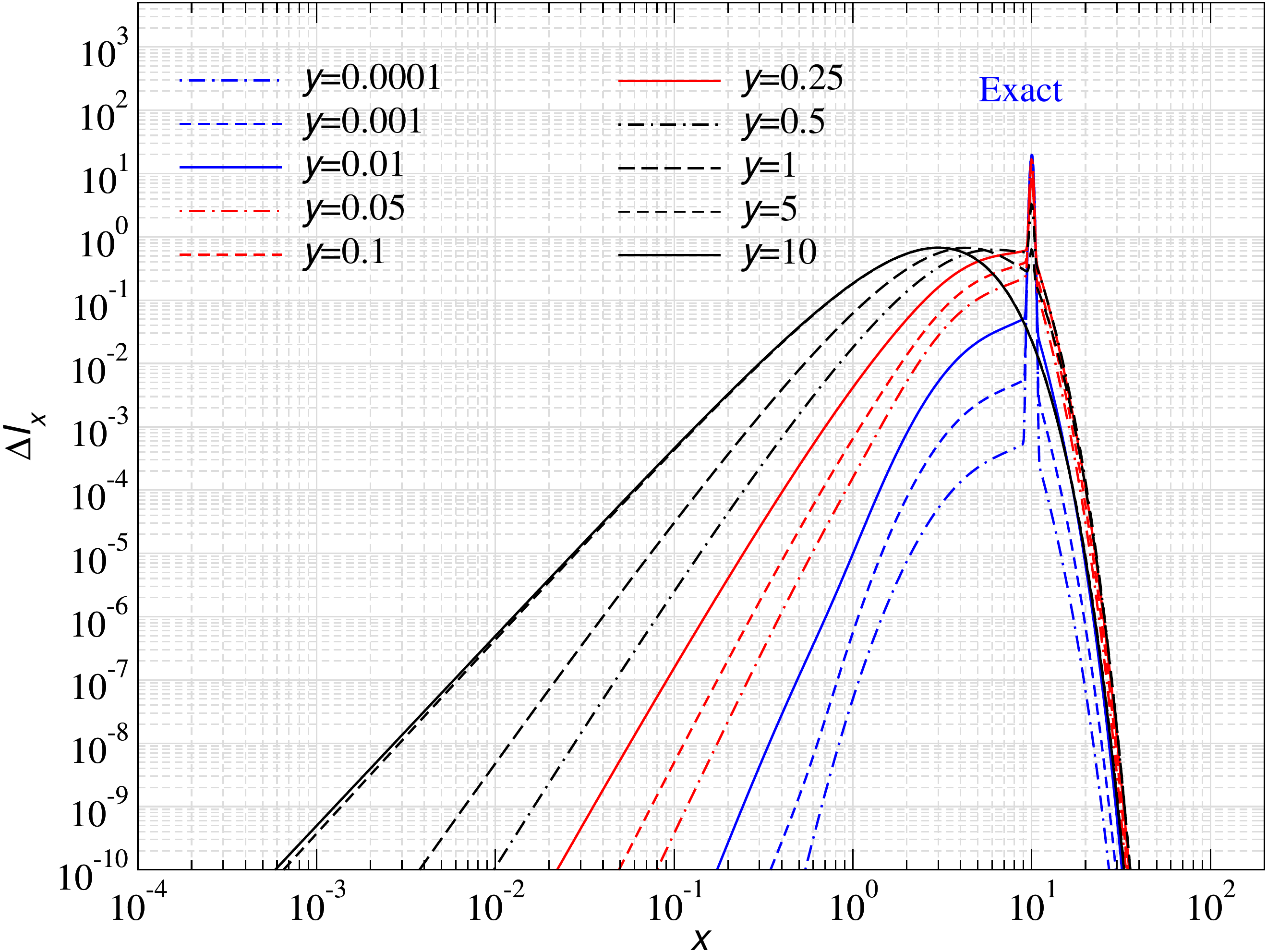}
\\
\caption{Same as Fig.~\ref{fig:snap_x_0.1_theta_0.01_0.1} but for $x_{\rm inj}=10$.
}
\label{fig:snap_x_10.0_theta_0.01_0.1}
\end{figure*}
%-------------------------------------

%-------------------------------------
\begin{figure*}
\centering 
\includegraphics[width=\columnwidth]{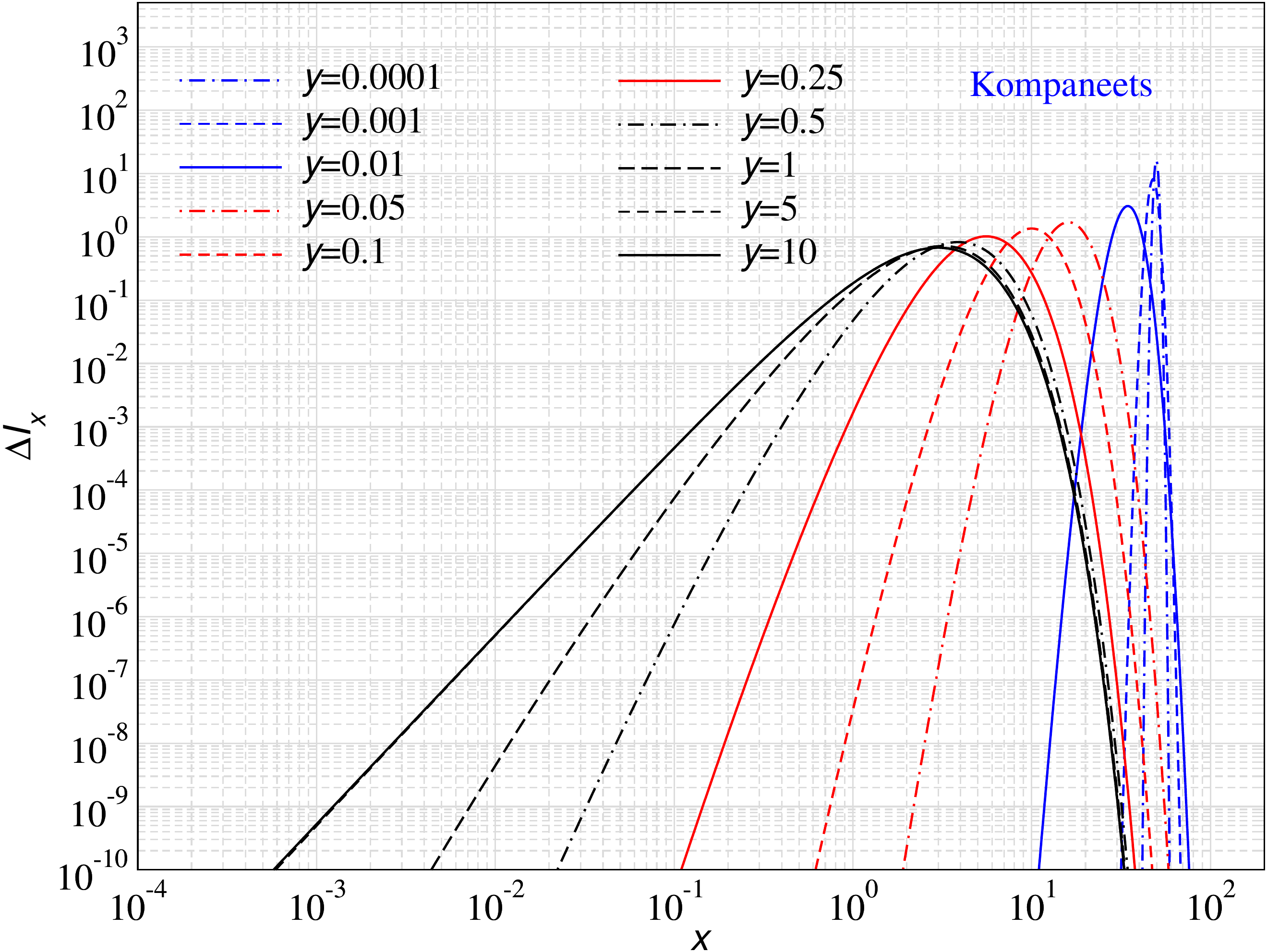}
\hspace{4mm}
\includegraphics[width=\columnwidth]{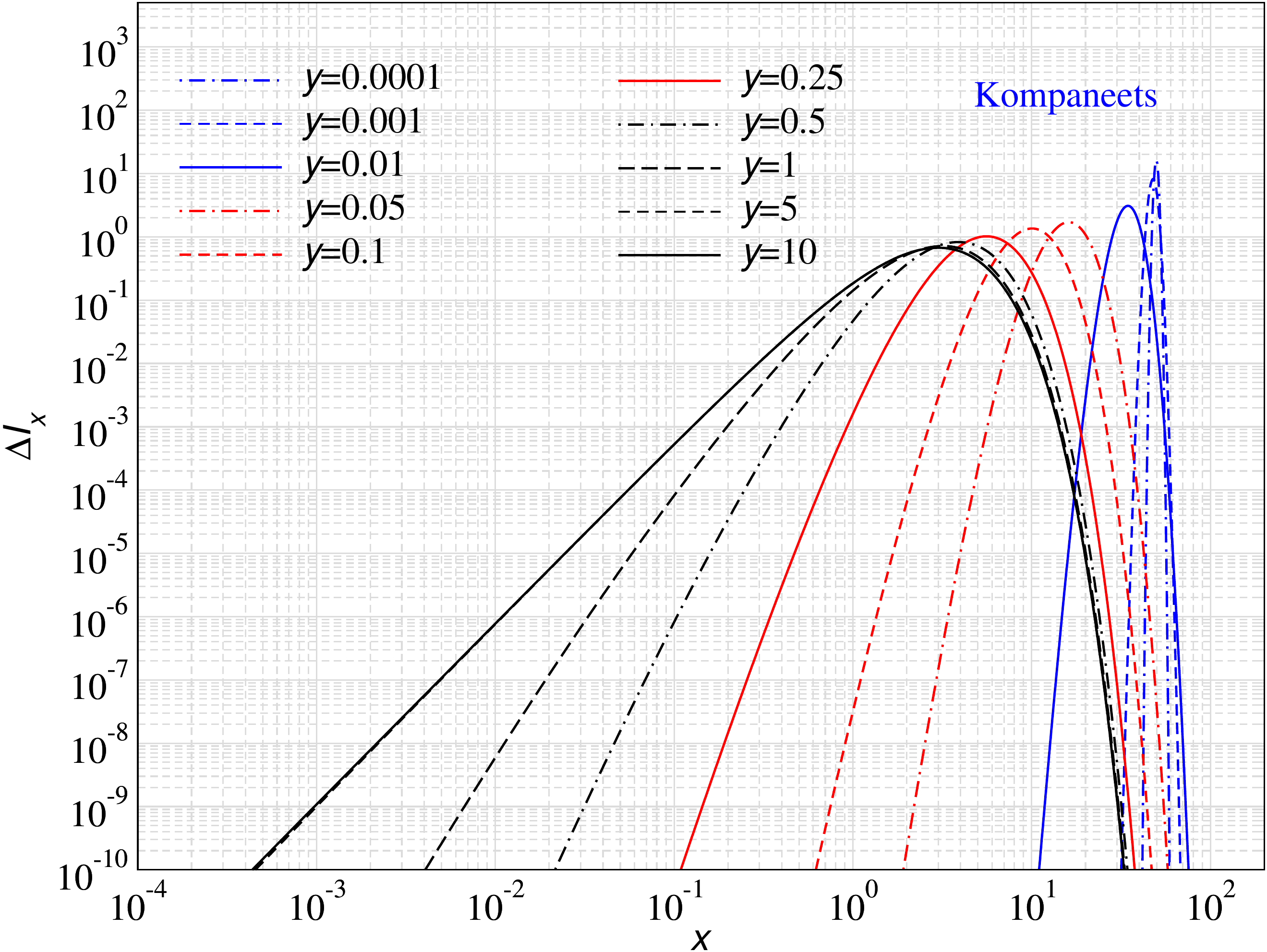}
\\[10mm]
\includegraphics[width=\columnwidth]{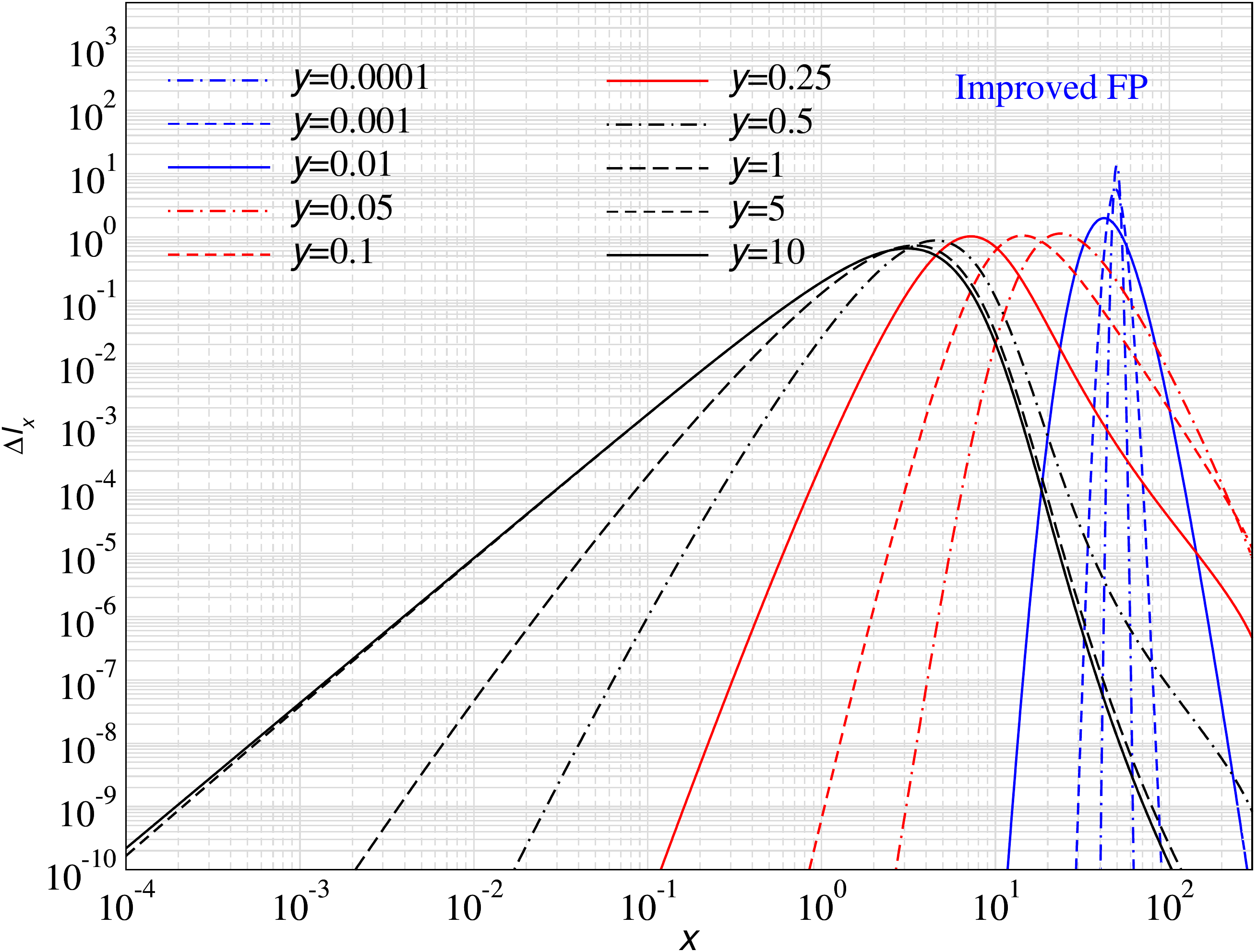}
\hspace{4mm}
\includegraphics[width=\columnwidth]{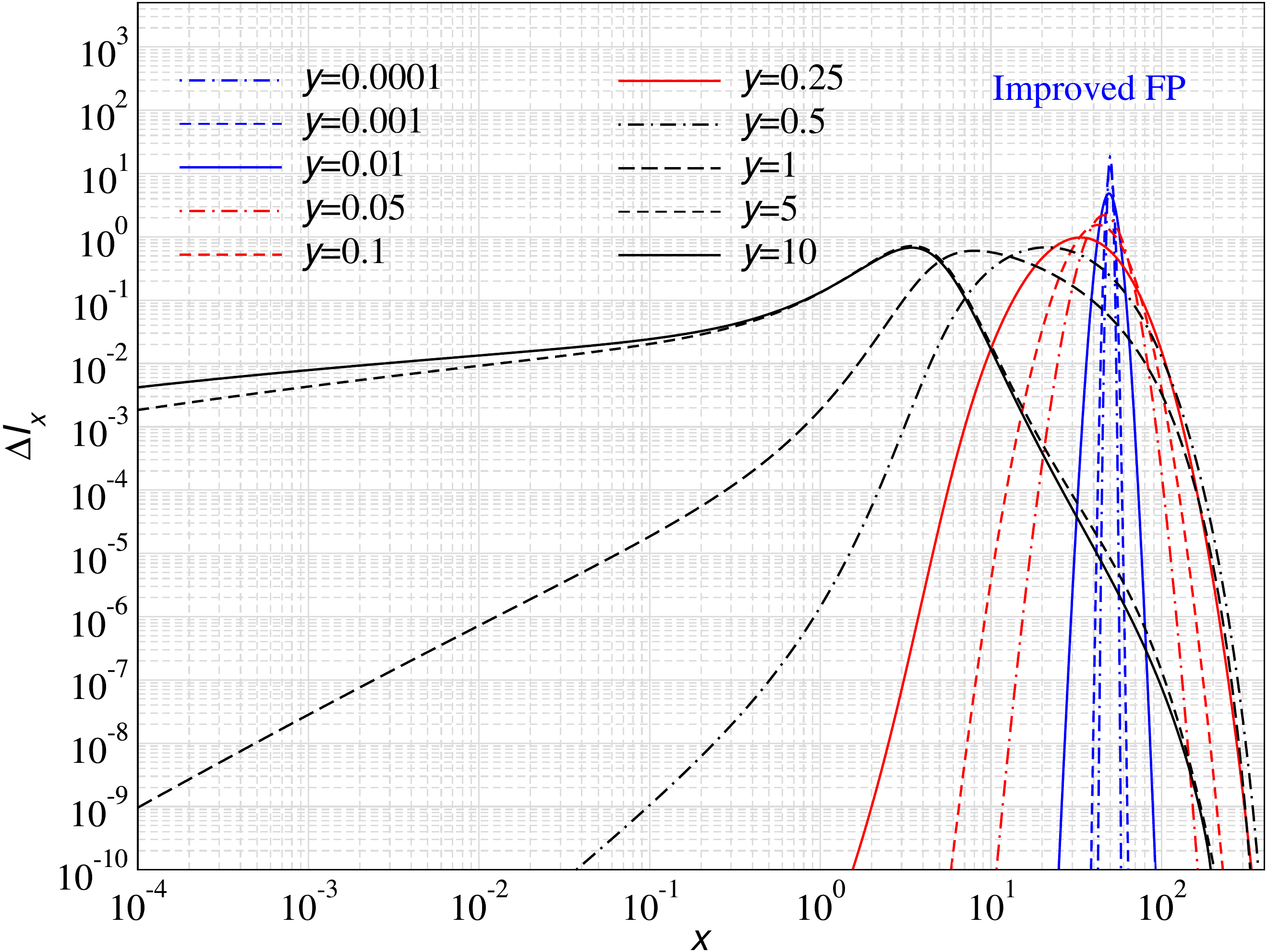}
\\[10mm]
\includegraphics[width=\columnwidth]{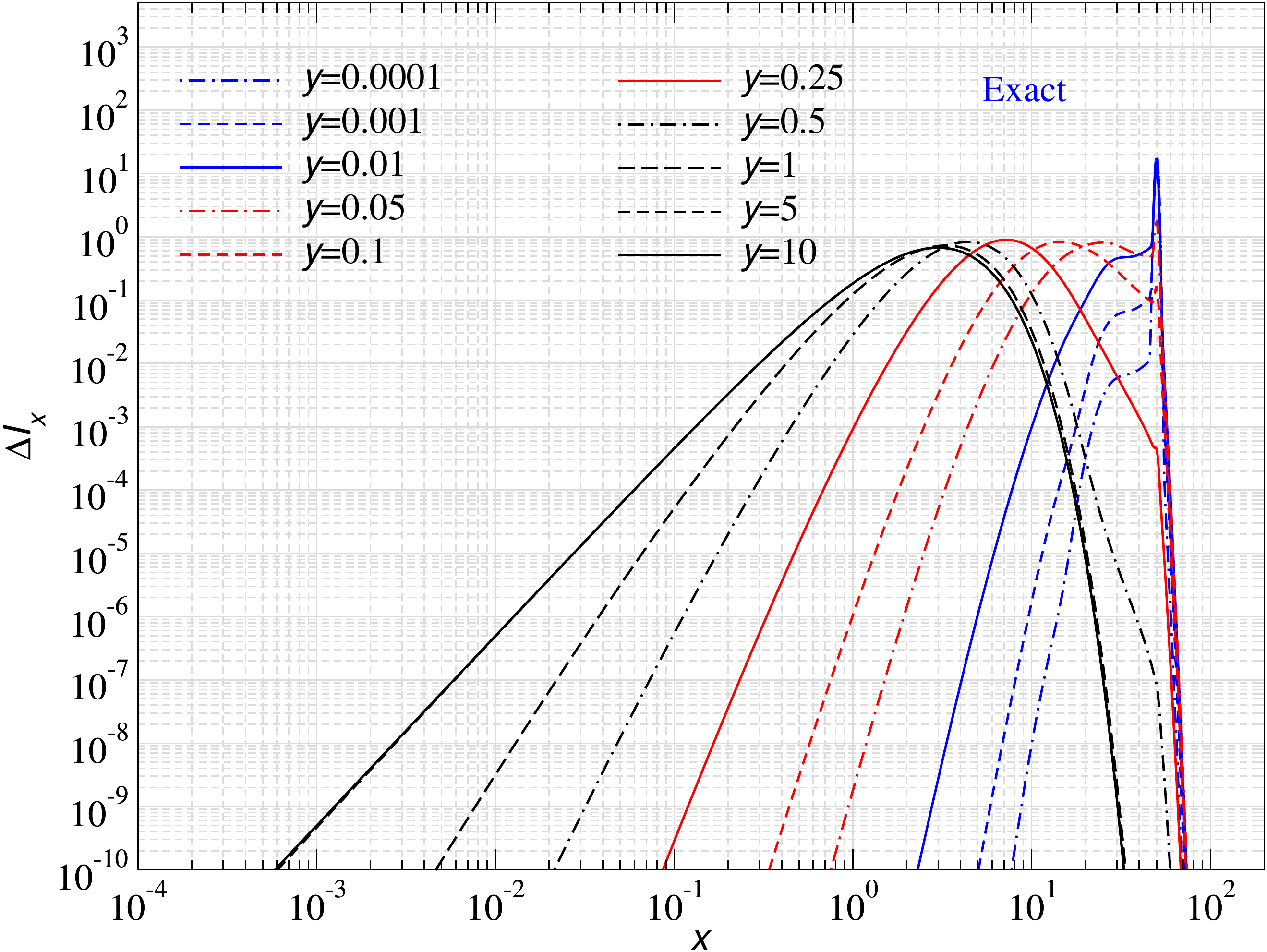}
\hspace{4mm}
\includegraphics[width=\columnwidth]{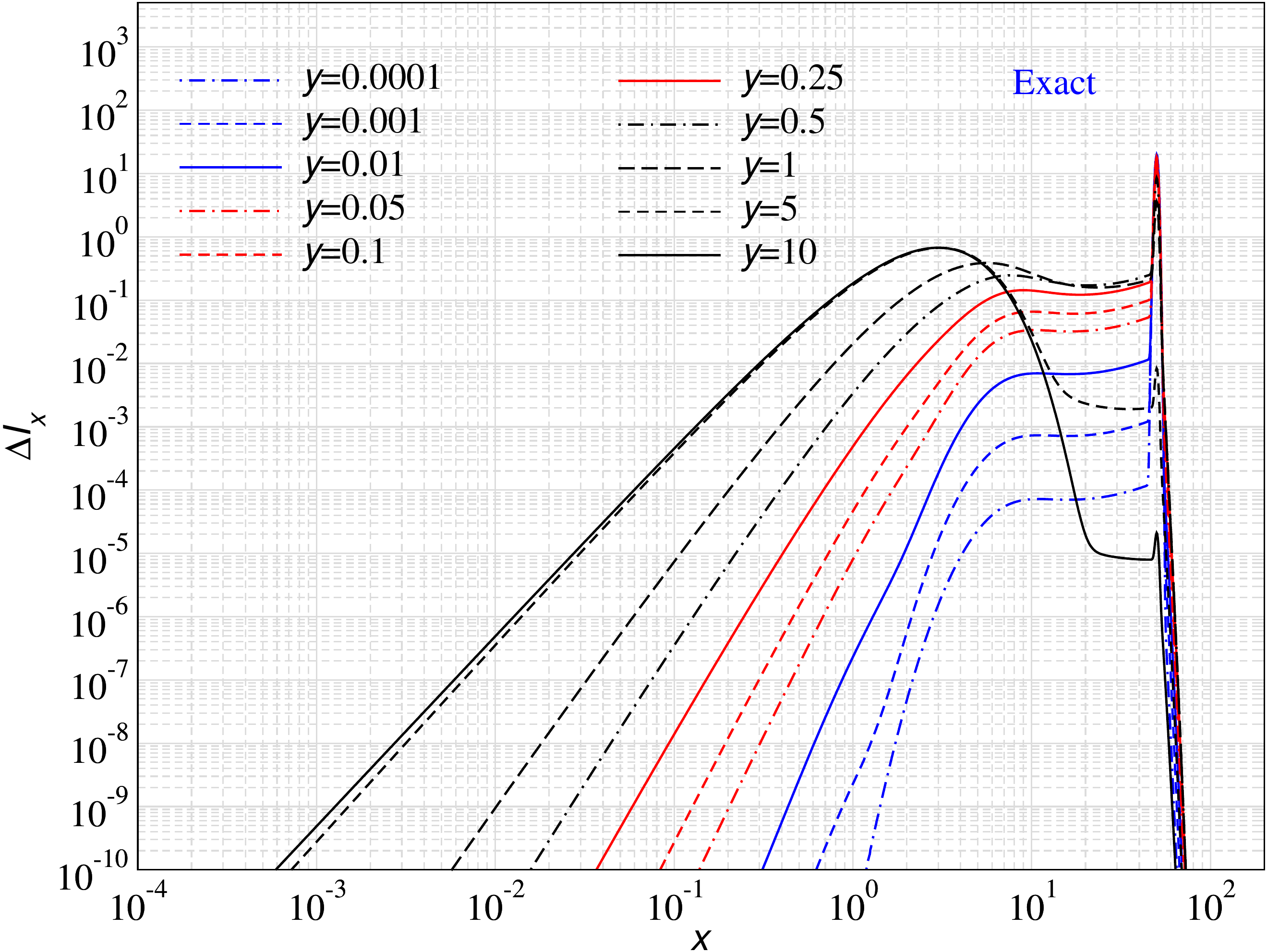}
\\
\caption{Same as Fig.~\ref{fig:snap_x_0.1_theta_0.01_0.1} but for $x_{\rm inj}=50$.
}
\label{fig:snap_x_50.0_theta_0.01_0.1}
\end{figure*}
%-------------------------------------

%\vspace{-3mm}
%--------------------------------------------------------------
\subsubsection{Effect of stimulated scattering}
\label{sec:evol_moments_stim}
%--------------------------------------------------------------
Stimulated effects become important at $x\lesssim 0.1-1$. All cases discussed above are thus not affected very strongly by stimulated scattering. 
In Figure~\ref{fig:x_0.01_moment_nostim_stim}, we illustrate the effect of stimulated scattering on the evolution of first and second moment of the photon field for $x_{\rm inj}=0.01$. The main effect is to resist and slow down Doppler boosting and broadening. This can be seen in the reduction of the first moment (left panels), when stimulated scattering terms are taken into account. In addition, the temperature dependence of the second moment is significantly reduced when stimulated scattering is included.
It also takes longer for photons to thermalize as the photon field reaches equilibrium at $y\lesssim$10 (upper limit being the case of $\The$=0.3) without stimulated scattering which increases to $y\lesssim$20 with stimulated scattering. 

For the Kompaneets limit, the reduction of up-scattering efficiency was previously highlighted \citep{Chluba2008d}. There it was shown that due to stimulated scattering in an ambient blackbody radiation field the mean energy of a photon increases like $E=E_0\expf{y}$ instead of $E=E_0\expf{3y}$ known from the classical solution of \citet{Zeldovich1969}.
However, the reduction in the temperature-dependence of the second moment is not captured by this analysis, given the temperature-independence of second moment in the Kompaneets treatment. This effect is nicely captured by the improved FP approximation at $y\lesssim 0.1$, and essentially causes a sharpening of features at fixed $y$ parameter. As before, the improved FP approach breaks down at larger values of $y$.

%--------------------------------------------------------------
\subsection{Evolution of the photon distribution}
\label{sec:solutions_nx}
%-------------------------------------------------------------
In Figures~\ref{fig:snap_x_0.1_theta_0.01_0.1} - \ref{fig:snap_x_50.0_theta_0.01_0.1}, we plot snapshots of the spectrum of the evolving photon field for several $y$-parameters, injection frequencies $x_{\rm inj}=0.1, 1, 10$ and 50 and two electron temperature $\The=0.01$ and 0.1. Overall it is clear that none of the approximate schemes reproduce the correct solution in detail. However, mean properties and also stationary states are captured with varying level of success.

While the Kompaneets and improved Fokker-Planck approaches have smooth spectra due to the applied diffusion approximation, the exact solution has a spiky feature at the energy of injection while $y\lesssim 1$. For small $y$ parameter, the improved FP approach more closely resembles the evolution of the mean and width of the exact solution, while it fails to converge to the correct equilibrium in the limit of many scattering (i.e., $y\gtrsim 1$), as explained before. 
The Kompaneets and exact treatments both reach very similar spectra in this limit; however, it is important to note that for $\The$=0.1, the exact solution does not fully thermalize even at $y=10$ due to Klein-Nishina corrections, which reduce the probability of scattering of photons with the electrons. This effect is expected to increase with photon energy because Klein-Nishina cross-section become more relevant there. In this regime, the Kompaneets equation overestimates the scattering efficiency significantly, leading to fast evolution via recoil (see case with $x_{\rm inj}=50$ and $\The=0.1$ as most extreme example). This is also expected from our discussion of the thermalization $y$-parameter in Sect.~\ref{sec:y_therm}.

We also carried out similar computations when including stimulated scattering effects. The overall picture is not changed and the important features already discussed in connection with the moments is reproduced. Overall, all our test cases show that while the photon distribution is narrow in comparison with the width of the scattering kernel the approximate schemes work only on average but not in detail. The additional failing of the improved FP approach at large values of $y$ is connected with the inherent construction of the diffusion coefficients and their breaking of detailed balance. While these do improve the description early on in the evolution, one cannot recommend using this approach generally. For robust general computations the Kompaneets equation seems to be the best approximate approach.

%--------------------------------------------------------------
%\vspace{-5mm}
\section{Evolution of CMB spectral distortions after single photon injection}
\label{sec:single}
%--------------------------------------------------------------
After considering problems at fixed electron temperature, we now compute the resulting CMB spectral distortions after single photon injections at various redshifts within the standard $\Lambda$CDM cosmology. For the problem at hand, we need to include the expansion of the Universe and the corresponding dilution of the particle densities. In addition, we have to follow the evolution of the electron temperature, which can be either done by setting it to the Compton equilibrium temperature or by explicitly solving the corresponding differential equation. Here we will simply set the temperature to the Compton equilibrium temperature, which is a valid approximation until rather low redshifts. We shall also assume that the ionization history if the Universe is given by the standard {\tt CosmoRec} solution \citep{Chluba2010b}. Photon emission and absorption processes will be neglected. The extension of the computation from the previous cases is then quite straightforward. We now explain the detailed setup of the problem before presenting solutions.

\vspace{-3mm}
\subsection{Choice of variables and initial conditions}
The effect of the Hubble expansion can be absorbed by using a frequency variable $x=h\nu/k\Tz$, where $\Tz\propto (1+z)$. Since we wish to consider cases that recreate the CMB blackbody at a temperature $\TCMB$ today, we shall simply use $\Tz=\TCMB(1+z)$. We then follow the evolution of the distortion $\Delta n(x)=n(x)-\nbb(x)$ with respect to the equilibrium blackbody $\nbb(x)=1/(\expf{x}-1)$ at the temperature $\Tz$.

We inject some number density $\epsilon_N=\Delta N_\gamma / N_\gamma$ at frequency $x_{\rm inj}$ and redshift $z_{\rm inj}$. This adds $\epsilon_\rho=\Delta \rho_\gamma/\rho_\gamma=\alpha_\rho x_{\rm inj} \epsilon_N$ of energy to the CMB\footnote{The heat capacity of matter is tiny compared to that of the CMB and we will neglect the corresponding corrections to the total energetics of the photon-baryon system.}. Here, $\alpha_\rho\approx 0.3702$ and $\epsilon_N$ and $\epsilon_\rho$ are expressed with respect to the initial blackbody densities at a temperature $T_{\rm in}<\Tz$. By comparing the energy densities we then find \citep[see,][]{Chluba2015GreensII}
%-------------------------------------------------------------------
\begin{align}
\label{eq:Tin}
\epsilon_T=\frac{\Delta T_{\rm in}}{\Tz}&\approx -  \frac{\alpha_\rho}{4} x_{\rm inj} \epsilon_N
\approx - \pot{9.255}{-2}\,x_{\rm inj}\,\epsilon_N
\end{align}
%-------------------------------------------------------------------
assuming that $\epsilon_N \ll 1$. The initial electron temperature is $\Te=T_{\rm in}$. We then have the initial distortion of the occupation number
%-------------------------------------------------------------------
\begin{align}
\label{eq:Dnin}
\Delta n(x)
&=\nbb(x^*)+\mathcal{G}^{\rm pl}_{2}\,\epsilon_N\,\frac{\delta(x^*-x_{\rm inj})}{{x^*}^2}-\nbb(x)
\nonumber \\
&\approx \frac{x \expf{x}}{(\expf{x}-1)^2}\,\epsilon_T+\mathcal{G}^{\rm pl}_{2}\,\epsilon_N\,\frac{\delta(x-x_{\rm inj})}{x^2}
\end{align}
%-------------------------------------------------------------------
with $\mathcal{G}^{\rm pl}_{k}=\int x^k \nbb(x)\id x$ (and thus $\mathcal{G}^{\rm pl}_{2}\approx 2.404$) and $x^*=h\nu/kT_{\rm in}=x/(1+\epsilon_T)$. To confirm the correct conditions, one can compute the integrals $\int x^2 \Delta n(x)\id x$ and  $\int x^3 \Delta n(x)\id x$. This then yields
%-------------------------------------------------------------------
\begin{align}
\label{eq:DN_Drho_f}
\frac{\Delta N_\gamma}{N_\gamma}\Bigg|_{\rm f}
&\approx \left(1-\frac{3}{4}\,\alpha_\rho x_{\rm inj}\right) \epsilon_N,
\qquad \frac{\Delta \rho_\gamma}{\rho_\gamma}\Bigg|_{\rm f}=0
\end{align}
%-------------------------------------------------------------------
with respect to the post-injection equilibrium blackbody. With $\Delta n(x)$ given by Eq.~\eqref{eq:Dnin}, the correct final energy density is already reached. The number density generally needs to be adjusted by double Compton and Bremsstrahlung emission. Only if we inject at $x_{\rm inj}\simeq (4/3)/\alpha_\rho\approx 3.602$, are no extra photons needed to fully restore a blackbody spectrum, and Comptonization alone could allow one to reach this state given enough time \citep[see][]{Chluba2015GreensII}. 

\vspace{-3mm}
\subsection{Compton equilibrium temperature}
To obtain a generalized expression for the Compton equilibrium temperature, we simply have to ask at which electron temperature does the energy exchange with the photons stop. Starting from Eq.~\eqref{eq:target_moments} we may write
%-------------------------------------------------------------------
\begin{align}
%\label{eq:condition_Compton_eq}
\int x^3 \frac{\text{d}n}{\text{d}\tau} \id x
&=\int \Sigma^*_1(\omega, \The)\,x^3 \Delta n_{\rm e} \id x
\nonumber \\
&\approx \int \Sigma^*_1(\omega, \Thz)\,x^3 \Delta n \id x - \frac{\Delta \Te}{\Tz}
\int \Sigma^*_1(\omega, \Thz)\,\frac{x^4 \expf{x}}{(\expf{x}-1)^2} \id x
\nonumber
\end{align}
%-------------------------------------------------------------------
where $\Delta n_{\rm e}=n(x)-\nbb(x \Tz/\Te)$ and $\Sigma^*_1(\omega, \Thz)$ is the first moment of the kernel including stimulated effects in the equilibrium blackbody spectrum given by Eq.~\eqref{eq:moments_stim}.
Equating this expression with zero, we then have
%-------------------------------------------------------------------
\begin{align}
\label{eq:Te_Compton_eq_exact}
\frac{\Delta \Te^{\rm eq}}{\Tz}
&\approx 
\frac{\int \Sigma^*_1(\omega, \Thz)\,x^3 \Delta n \id x}{\int \Sigma^*_1(\omega, \Thz)\,\frac{x^4 \expf{x}}{(\expf{x}-1)^2} \id x}.
\end{align}
%-------------------------------------------------------------------
As shown in Eq.~\eqref{eq:stimulated-moments_approx}, at lowest order in the temperature this yields
%-------------------------------------------------------------------
\begin{align}
\Sigma_1^*
&\approx (4-x)\Thz - 2 x \nbb(x) \Thz = - \Thz \left[ x\frac{\expf{x}+1}{\expf{x}-1} - 4 \right],
\end{align}
%-------------------------------------------------------------------
which defines the spectrum of a $y$-distortion, $Y(x)=x\frac{\expf{x}+1}{\expf{x}-1} - 4$, in terms of the photon occupation number. This then yields
%-------------------------------------------------------------------
\begin{align}
\label{eq:Te_Compton_eq}
\frac{\Delta \Te^{\rm eq}}{\Tz}
&\approx 
\frac{\int Y(x)\,x^3 \Delta n \id x}{\int Y(x)\,\frac{x^4 \expf{x}}{(\expf{x}-1)^2} \id x} =
\frac{\int Y(x)\,x^3 \Delta n \id x}{4\mathcal{G}^{\rm pl}_3}
\end{align}
%-------------------------------------------------------------------
with $\mathcal{G}^{\rm pl}_3\approx 6.494$. This  also follows from the well-known expression \citep{Zeldovich1970TCompton, Chluba2011therm}
%-------------------------------------------------------------------
\begin{align}
\label{eq:Te_Compton_eq}
\Te^{\rm eq}
&\approx 
\frac{\int \nu^4 n (1+n) \id \nu}{4\int \nu^3 n \id \nu} \nonumber \\
&\approx \Tz \left[1+\frac{\int x^4 \Delta n[1+2\nbb(x)]\id x}{4\mathcal{G}^{\rm pl}_3}-\frac{\int x^3 \Delta n \id x}{\mathcal{G}^{\rm pl}_3}\right]
\end{align}
%-------------------------------------------------------------------
which can be derived directly from the Kompaneets equation. 

\vspace{-3mm}
\subsection{Evolution equation for CMB distortion}
The required evolution equation for the CMB distortion is similar to that in Eq.~\eqref{eq:evol_Dn}. However, we now have to include a source term due to the difference in the photon and electron temperatures. Redoing all the steps of the derivation, we then have
%-------------------------------------------------------------------
\begin{align}
\label{eq:evol_Dn_CMB}
\frac{\text{d}\Delta n_i}{\text{d} \tau} 
&\approx \sum_j \mathcal{S}_{ij} \Delta n_j - \mathcal{B}_{i}\,\frac{\Delta \Te^{\rm eq}}{\Tz},
\nonumber
\\[1mm]
\mathcal{S}_{ij}&=w_j\,P(\omega_i \rightarrow \omega_j, \Thz) \,\expf{x_j-x_i} \, f^{\rm eq}_{ji} - \delta_{ij}\,\sigma_i^*,
\nonumber
\\[1mm]
%\mathcal{B}_{i}&=-\sum_j \nbb(x_i)[1+\nbb(x_j)] (x_j-x_i)
\mathcal{B}_{i}
&=\sum_j w_j\,P(\omega_i \rightarrow \omega_j, \Thz)\, f^{\rm eq}_{ij} \,G_i\,\frac{x_j-x_i}{x_i}
\nonumber
\\[1mm]
\sigma_i^*&=\sum_j w_j\,P(\omega_i \rightarrow \omega_j, \Thz)\, f^{\rm eq}_{ij},
\nonumber
\\[1mm]
f^{\rm eq}_{ij}&=\frac{1+\nbb(x_j)}{1+\nbb(x_i)}=\frac{1-\expf{-x_i}}{1-\expf{-x_j}},
\end{align}
%-------------------------------------------------------------------
where $x=h\nu/k\Tz$ in all instances and $G_i=x_i \expf{x_i}/(\expf{x_i}-1)^2$. 

For the computations, it is best to pre-compute all functions that do not depend on the solution, i.e., $G_i$, $\expf{x_j-x_i}$, $f^{\rm eq}_{ij}$ and $\mathcal{B}_{ij}$, which are all fixed once the grid is determined. The scattering kernel, $P(\omega_i \rightarrow \omega_j, \Thz)$, is updated every time $\Tz$ changes\footnote{This can become very expensive, but we implemented an efficient approximation scheme in {\tt CSpack} that pre-tabulates the required matrix elements across a grid of temperatures and then uses interpolation once the difference in temperatures exceeds a certain threshold (e.g., $\Delta T/T\simeq 10^{-4}$).}. Since the Compton equilibrium temperature can be written as 
%-------------------------------------------------------------------
\begin{align}
\label{eq:Te_Compton_eq_disc}
\frac{\Delta \Te^{\rm eq}}{\Tz}
&=\sum_j \mathcal{C}_{j} \Delta n_j
\end{align}
%-------------------------------------------------------------------
it is best to extend the system by this extra algebraic equation, solving for the solution vector $\vek{s}=(\Delta n_i, \Delta \Te^{\rm eq}/\Tz)^{T}$. 
The vector $\mathcal{C}_{i}$ reads
%-------------------------------------------------------------------
\begin{align}
\label{eq:C_approx}
\mathcal{C}_{i}
&= 
\frac{w_i x_i^3 Y_i }{4 \sum_j w_j x_j^3 \nbb_j}
\end{align}
%-------------------------------------------------------------------
when using the low-temperature/low-photon energy limit of the Compton equilibrium temperature, Eq.~\eqref{eq:Te_Compton_eq}, which yields a time-independent coefficient.\footnote{Analytically, the denominator of Eq.~\eqref{eq:C_approx} is $\mathcal{G}^{\rm pl}_3$. However, to ensure better energy conservation, it is best to use the numerical result obtained for the chosen discretization.} To be more precise (e.g., when allowing for larger $\omega$), we can alternatively use 
%-------------------------------------------------------------------
\begin{align}
\label{eq:C_exact}
\mathcal{C}_{i}
&\approx 
\frac{w_i x^3_i \Sigma^*_1(\omega_i, \Thz)}{\sum_j w_j x_j^3 \Sigma^*_1(\omega_j, \Thz)\,G_j},
\end{align}
%-------------------------------------------------------------------
which follows from Eq.~\eqref{eq:Te_Compton_eq_exact}. The stimulated first moment of the kernel, $\Sigma^*_1$, is best computed using {\tt CSpack}. However, both of these choices for $\mathcal{C}_{i}$ do not exactly conserve photon energy due to numerical errors from the $x_i$ discretization. The best choice in the computations therefore is
%-------------------------------------------------------------------
\begin{align}
\label{eq:C_exact_num}
\mathcal{C}_{i}
&\approx 
\frac{\sum_j w_j x^3_j \mathcal{S}_{ji}}{\sum_j w_j x_j^3 \mathcal{B}_j},
\end{align}
%-------------------------------------------------------------------
which directly follows after taking the second moment of the evolution Eq.~\eqref{eq:evol_Dn_CMB} and demanding $\sum_j w_j x_j^3 \id \Delta n_j/\id \tau =0$. The closes the numerical setup of the problem.

To advance the solution from time $t$ to $t+\Delta t$, we can perform a simple implicit Euler step. We then have to solve the system 
%-------------------------------------------------------------------
\bsub
\label{eq:evol_sol}
\begin{align}
\mathcal{J}^{(t+1)}\vek{s}^{(t+1)}&\approx \vek{b}^{(t)}
\\
\mathcal{J}^{(t+1)}&=\left(
\begin{matrix}
\,\delta_{ij}-\Delta \tau S^{(t+1)}_{ij} & \Delta \tau \mathcal{B}_i
\\
-\mathcal{C}^{(t+1)}_{j} & 1
\end{matrix}
\right).
\end{align}
\esub
%-------------------------------------------------------------------
where $\vek{b}^{(t)}=(\Delta n_i^{(t)}, 0)^{T}$. Since the matrix, $\mathcal{J}$, is usually quite sparse, $\vek{s}^{(t+1)}$ can be obtained iteratively using a sparse bi-conjugate gradient method \citep{Chluba2010}. 

We implemented a solver based on this simple Euler step in Eq.~\eqref{eq:evol_sol}. However, for improved performance we used the Gear's method ODE solver implemented to solve the cosmological recombination problem \citep{Chluba2010}. This solver adaptively modifies the time-step and also allows using the solutions from several previous time-steps (up to sixth order). A sparse matrix Gaussian elimination routine turned out to be the most accurate method for obtaining the solution of the problem. The overall computational demand depends on the size of the frequency grid. For typical resolution of a few hundred points per frequency decade in the range $x=10^{-6}-200$ one run takes about 20 seconds on a laptop.

%\vspace{-3mm}
%-------------------------------------
\begin{figure*}
%\centering 
\includegraphics[width=\columnwidth]{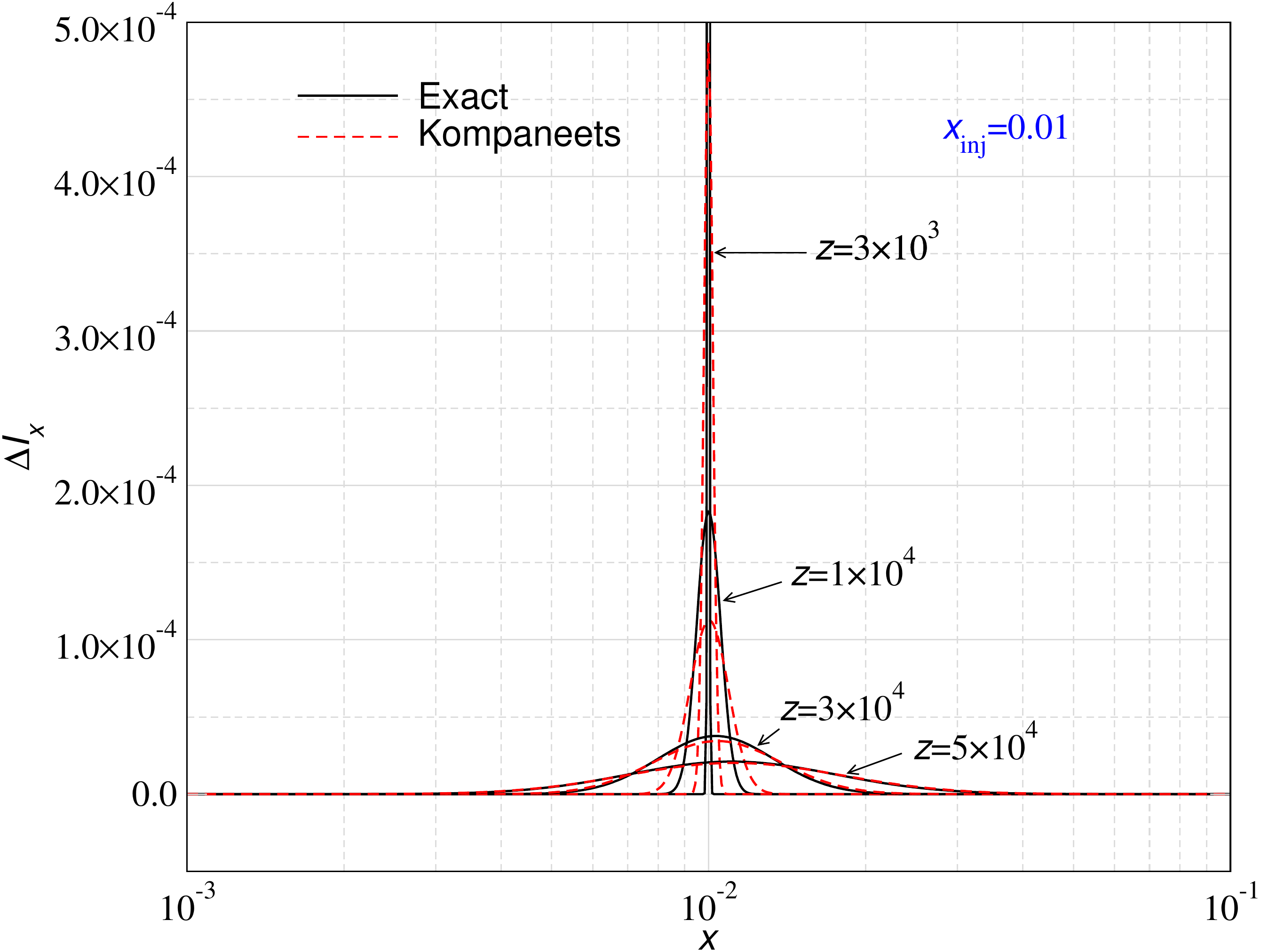}
\hspace{4mm}
\includegraphics[width=\columnwidth]{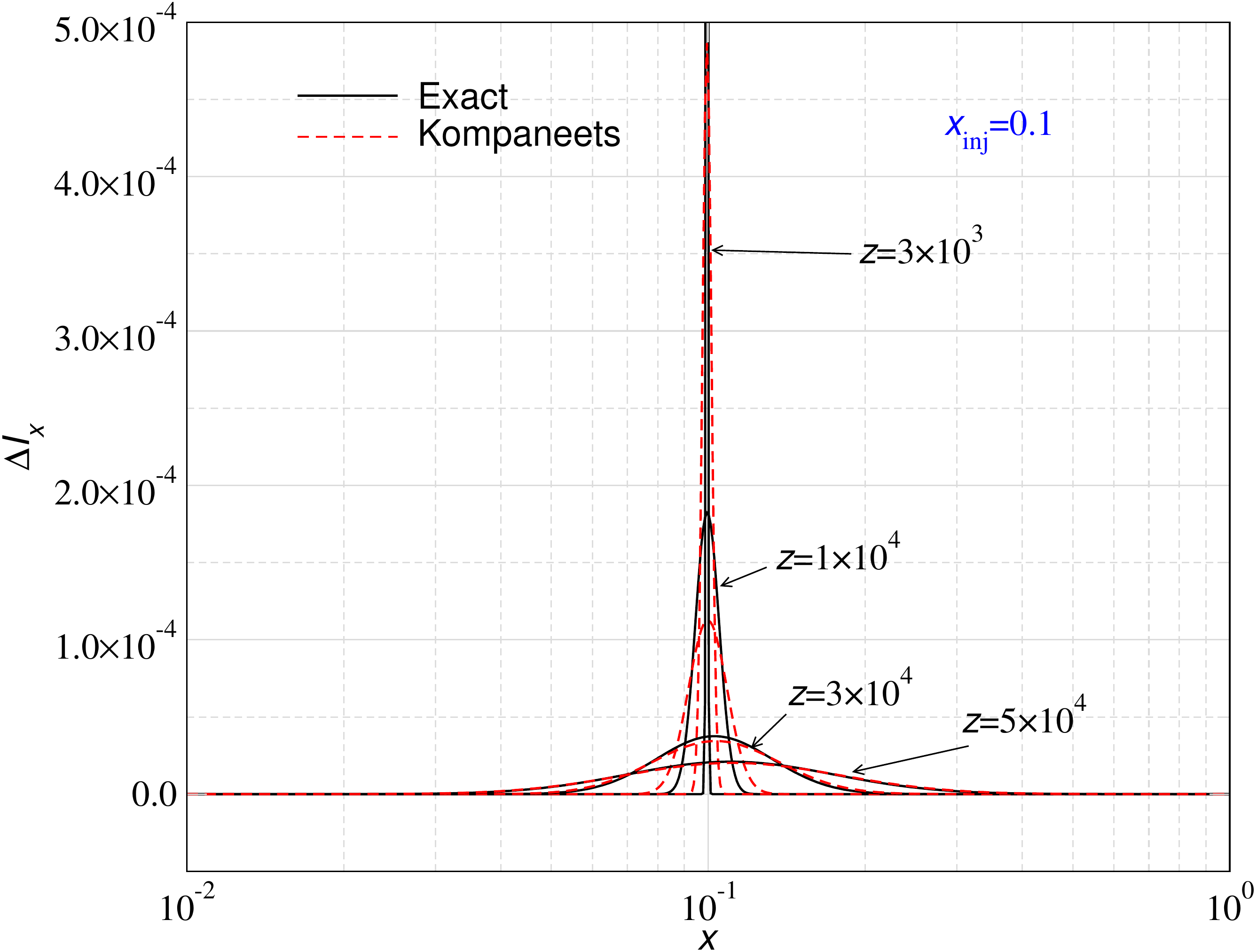}
\\[4mm]
\includegraphics[width=\columnwidth]{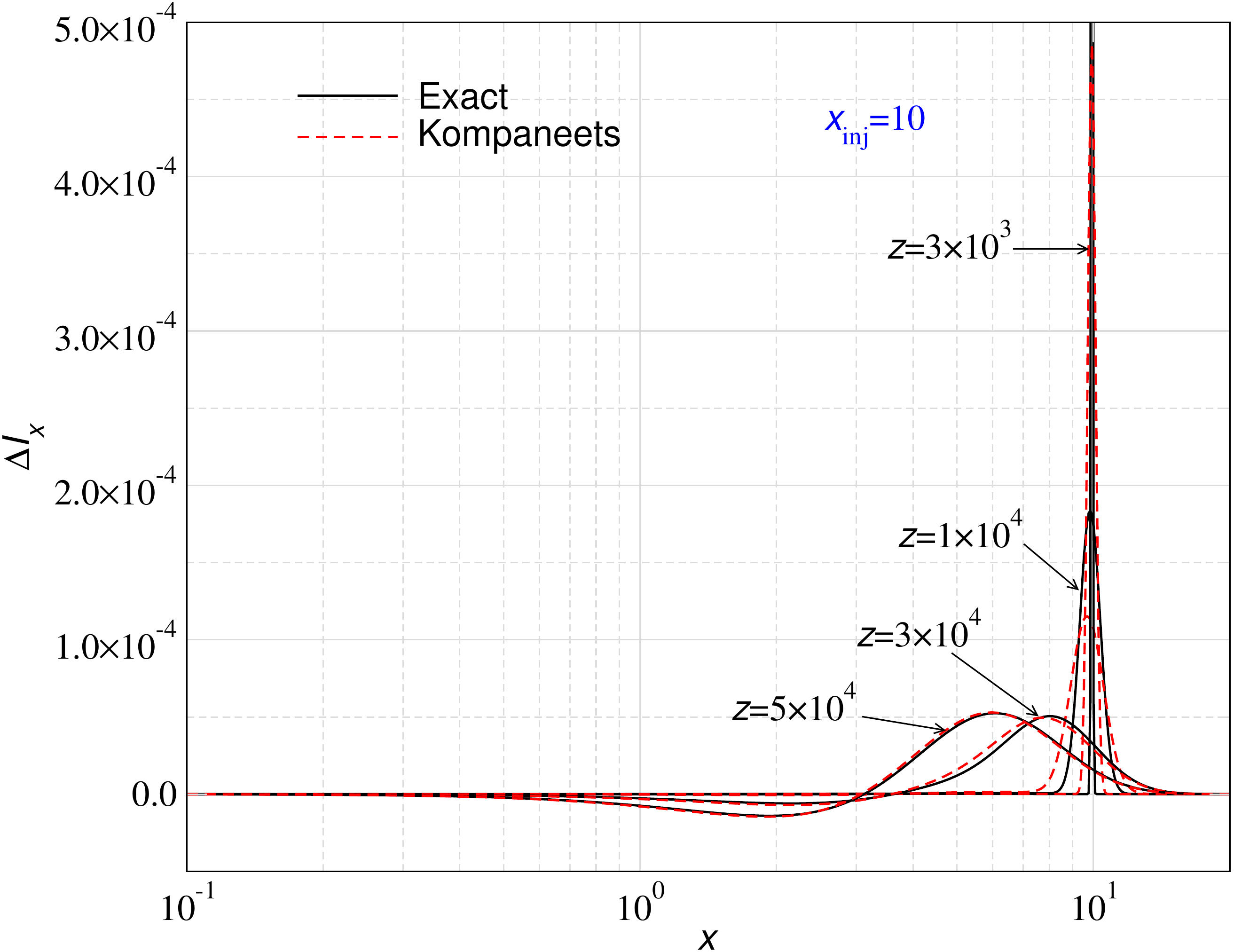}
\hspace{4mm}
\includegraphics[width=\columnwidth]{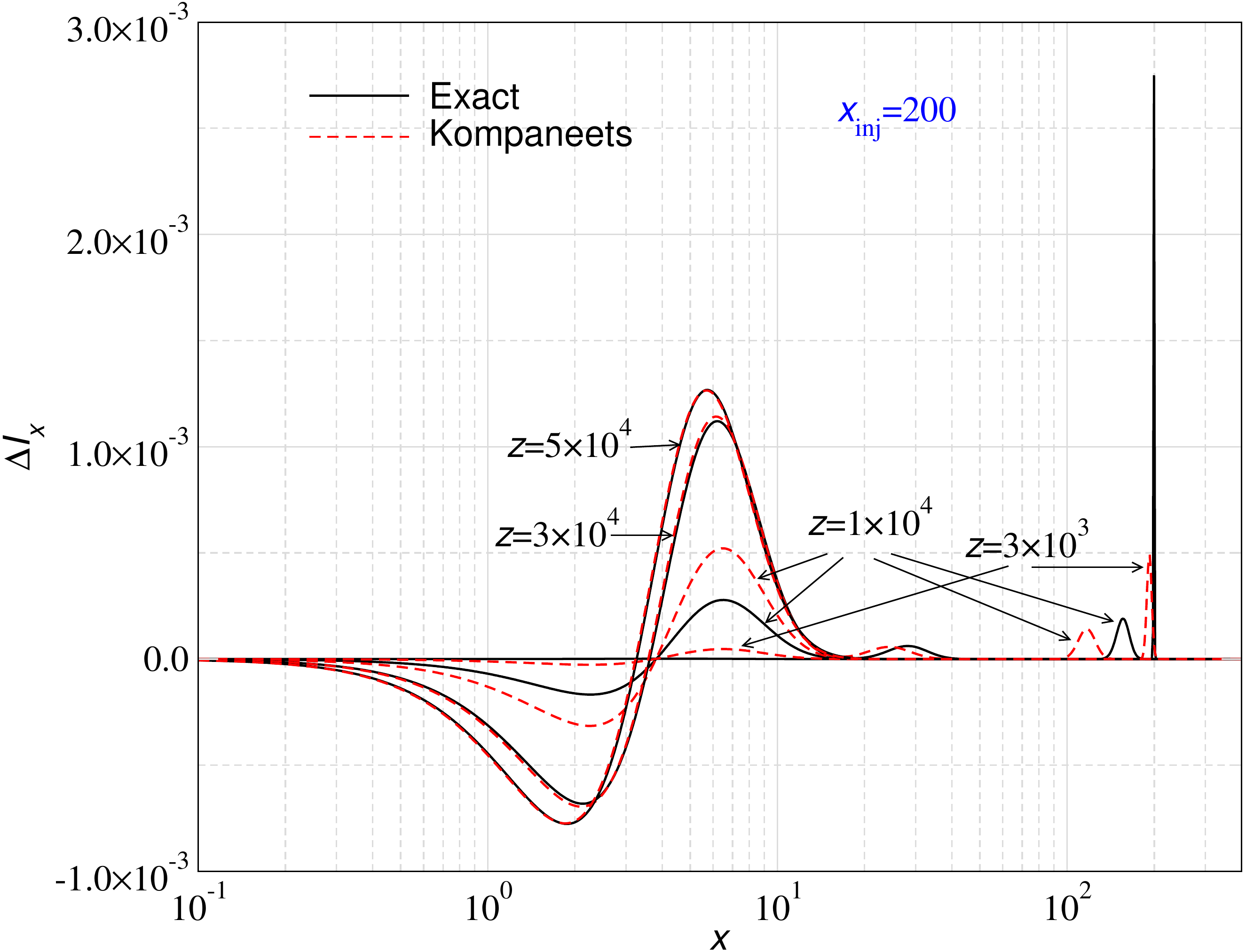}
\\
\caption{Spectral distortion solution for energy injection at different redshifts with frequency of injection as denoted. The Kompaneets solution generally overestimates the broadening and mean shifts in the photon distribution. For injection at $x=10$ and $100$, the exact solution falls back with respect to the Kompaneets treatment, which overestimates the energy exchange with electrons.
}
\label{fig:specdist_soln}
\end{figure*}
%-------------------------------------

%-------------------------------------
\subsection{Solutions for the distortion}
%-------------------------------------
We now have everything together to present a comparison of various solutions for the distortion after single energy injection. The main goal is to highlight some of the differences with the Kompaneets treatment to motivate a more detailed study of the spectral distortion solutions, extending previous investigations \citep{Chluba2015GreensII, Bolliet2020PI} using the exact scattering kernel treatment.

In Fig. \ref{fig:specdist_soln}, we plot the spectral distortion solutions for photon energy injections at different redshifts and injection frequency using the Kompaneets and exact treatments. We focus on $z\lesssim 50,000$ where we can expect partial thermalization, given that the total scattering $y$-parameter is $y\lesssim 0.1$. In addition, we restrict our discussion to $x_{\rm inj}\geq 0.01$, since even at these low redshifts one would expect Bremsstrahlung and DC effects to become important at lower injection frequencies \citep[see Fig.~10 of][for regimes]{Chluba2015GreensII}. 

It is easy to see in Fig.~\ref{fig:specdist_soln} that the Kompaneets treatment generally overestimates the broadening and mean shifts of photon spectrum, when scattering is inefficient. Also, the recoil effect is overestimated by the Kompaneets solution, which is especially visible for $x_{\rm inj}=200$. Recoil leads to heating of electrons which shows up as $y$-distortion around the maximum of the CMB blackbody ($x\simeq 1-10$). This explains the significantly higher amplitude for the Kompaneets solution at $1<x<10$ for this case. Even though, for the exact solution we have photons at higher energy compared to the Kompaneets case at $z=10^4$, it eventually catches up at $z=3\times10^4$. As the photons move leftward (or to lower energy), eventually Kompaneets approximation for energy loss by recoil ($\Delta E/E\simeq -h\nu/m_{\rm e}\rm c^2$) takes over and the photons at higher energy lose energy more efficiently and move leftward faster. This also leads to efficient thermalization and in all considered cases, the two solutions are practically same for injections at $z\gtrsim 3\times 10^4$. 

Our calculations illustrate that at low redshifts, where the temperature of the electrons is well inside the non-relativistic regime, the largest corrections are found for (i) small scattering $y$-parameter, (ii) injection at high frequency and (iii) in the detailed time-dependence of the energy exchange with the electrons.
The first and second effects directly manifest themselves in changes to the distortion signal, and hence affects the interpretation of the signal in terms of injection frequency and time.
For $y\gtrsim y_{\rm therm}\simeq 1$, the third effect can become important when including photon production processes (at earlier times) and ionization of atoms (at late times). This motivates a more detailed study to refine earlier calculations \citep{Chluba2015GreensII, Bolliet2020PI}, however, this is beyond the scope of this paper.

We also highlight that injection of high-energy photons can cause the production of non-thermal electron populations \citep[e.g.,][]{Slatyer2015, Acharya2019a}, which then cause scattering corrections that are not modeled here. The threshold energy for non-thermal electron production by photons is a strong function of redshift \citep[see Fig.~5 of][]{Bolliet2020PI}, usually requiring $h\nu_{\rm inj} \gtrsim {\rm few}\times {10\,{\rm keV}}$. At $z=10^5$, this means $x_{\rm inj} \gtrsim 500$, such that none of the cases presented here fall into this domain. It will be extremely interesting to explicitly treat the scattering problem with thermal and non-thermal electron populations, following their detailed evolution. Accounting for the exact Compton energy exchange using {\tt CSpack} will play a big role in these calculations.

\section{Conclusion}
\label{sec:conclusion}

%-------------------------------------------------------
In this paper, we carried out a detailed study of evolution of photon spectrum in contact with a thermal distribution of electrons and exchanging energy via repeated Compton scatterings. 
The main goal of our study is to illustrate the accuracy of various numerical treatments of the problem in comparison with the exact scattering kernel solution. We especially focus on problems that start with narrow photon distributions (i.e., line injection), as these are expected to most strongly violate the assumptions going into commonly used approximation schemes.

Initially, we consider scenarios in which the temperature of electrons remains unchanged even after energy exchange with photons while photon number is conserved identically. We investigate the behaviour of evolved photon spectrum derived under various Fokker-Planck treatments (i.e. Kompaneets, first and second improved) and compare with the exact solution (e.g., see Fig.~\ref{fig:snap_x_0.1_theta_0.01_0.1}-\ref{fig:snap_x_50.0_theta_0.01_0.1}). 
We find that in general Fokker-Planck approximations do not capture the evolution of the photon spectrum well.
The Kompaneets equation works well for $\The\lesssim 0.01$ and $x\lesssim 10$, but Klein-Nishina, recoil and boosting corrections become important at higher temperature and for $x\gtrsim 10$. These corrections are also important for the photon thermalization timescale which can be significantly underestimated by the Kompaneets solution (see Fig.~\ref{fig:y_therm}). We also explicitly demonstrate that the inclusion of stimulated scattering slows down the timescale of thermalization and also changes the temperature dependence of moment evolution of photon field. 

The first improved Fokker-Planck approximation, which is tuned to capture the correct first and second moment of the photon spectrum (see Sect.~\ref{sec:New_attempt} for details) and was also considered by \citet{Belmont2008}, fails to give the correct equilibrium solution due to second and higher order temperature corrections to the exact scattering kernel (see Fig.~\ref{fig:equilibrium_soln}). Nevertheless, it precisely reproduces the evolution of the first and second moments at $y\lesssim 0.1$ (e.g., Fig.~\ref{fig:x_0.1_10_moment_evol}). We extend this approach to include stimulated scattering term (see Sect.~\ref{sec:New_attempt_II}).
While during the early phases of the evolution, the effect of stimulated terms are correctly reproduced by the first improved FP approach (e.g., Fig.~\ref{fig:x_0.01_moment_nostim_stim}), this does not cure the problems in the limit of many scatterings.

We then consider a second improved Fokker-Planck approximation (see Sect.~\ref{sec:New_attempt_II} for details), which indeed ensure the correct equilibrium solution but fails to capture the correct evolution of the first and second moments of the photon spectrum (see Fig.~\ref{fig:x_0.1_2ndFP_exact_mom}), and hence is not recommended. 

Overall, this first part of our study shows that the Kompaneets equation provides the most robust general approximation to the scattering problem. However, in detail it fails and is unable to account for several important aspects that are relevant at higher electron temperature and photon energy. In this regime, the exact scattering kernel approach developed here has to be applied.

We comment that in principle one could further improve the precision of the FP treatments by including higher order frequency derivatives of the photon distribution. To achieve exact consistency at second order in the electron temperature, this would mean adding the derivatives $\partial^3_\nu \Delta n$ and $\partial^4_\nu \Delta n$. However, numerically this would be harder to solve and two additional boundary conditions would have to be given. It is also likely that this will only lead to minor improvements to the solution, as it is well-known that corresponding approaches for computing relativistic temperature corrections to the Sunyaev-Zeldovich effect only converge asymptotically \citep{Sazonov1998, Itoh98, Chluba2012SZpack}.
The numerical scheme presented here using {\tt CSpack} avoids these difficulties.

In the second part of paper, we study the evolution of CMB spectral distortion signals from photon energy injections at $z\lesssim 10^5$. For this, we accurately take into account the energetics of the coupling between CMB photons and the background electrons while photon number is identically conserved as before (see Sect.~\ref{sec:single} for details). We compare the spectral distortion signal obtained from the Kompaneets equation and the exact evolution equation. Since the temperature of the Universe at $z\lesssim 10^5$ is $\lesssim 20\,{\rm eV}$, the Fokker-Planck approximation reduces to Kompaneets limit in non-relativistic regime. We find that Kompaneets solution generally overestimates the broadening of the photon spectrum and the recoil effect for photon with energy $x\gtrsim10$ (see Fig.~\ref{fig:specdist_soln}). There are visible differences between the two solutions at $z\lesssim 3\times 10^4$, while the difference become less pronounced at higher redshifts due to efficient thermalization. However, the timescales on which the photon distribution evolves are misestimated in the Kompaneets treatment, implying that the mapping between the final distortion and the underlying source of the distortion becomes inaccurate.

Our work motivates the more detailed study for the evolution of the CMB spectral distortions after injection of photons and energy at $z\gtrsim 2\times 10^6$ when photon non-conserving processes become very efficient. The rate of photon production/absorption by these processes depends on the shape of the spectral distortion itself for which Kompaneets solution is inadequate as we have shown here. This will affect the thermalization process and can change the $\mu$-visibility function which captures the survival probability of spectral distortion signature until today. Furthermore, deriving accurate spectral distortion constraints including non-thermal particle cascades is deferred to a more detailed work in future. The exact treatment presented here provides an important first step towards a detailed modeling of Compton scattering in these scenarios.

\vspace{-3mm}
{\small
\section*{Acknowledgments}
The authors would like to thank Geoff Vasil for useful discussions about diffusion problems.
This work was supported by the ERC Consolidator Grant {\it CMBSPEC} (No.~725456).
JC was furthermore supported by the Royal Society as a Royal Society University Research Fellow at the University of Manchester, UK.
}

{\small
\vspace{-3mm}
\bibliographystyle{mn2e}
\bibliography{Lit}
}

\begin{appendix}

\section{Intermediate steps in derivations}
\label{app:intermediate_step}
%-------------------------------------------------------------------

\subsection{Intermediate steps for Eq.~\eqref{eq:col_CS_Dn_lin}}
\label{app:intermediate_Eq5}
%-------------------------------------------------------------------
Introducing the kernel average of a quantity $X(\omega)$ as
%-----------------------------------------------------------
\begin{align}
\left<X(\omega)\right>&=\int X(\omega)\,P^{\rm th}(\omega_0 \rightarrow \omega) \id\omega.
\end{align}
%-----------------------------------------------------------
and neglecting terms $\mathcal{O}(\Delta n^2)$, we can rewrite the Boltzmann equation~\eqref{eq:kin_eq2_starting_point} in the compact form
%-------------------------------------------------------------------
\bsub
\begin{align}
\frac{\text{d}n_0}{\text{d}\tau} 
&=\bigg<n(1+n_0)\,\expf{x-x_0}\bigg>-\bigg<n_0(1+n)\bigg>
\nonumber\\
&\approx \bigg<\left[\Delta n(1+n^{\rm eq}_0)+n^{\rm eq}\,\Delta n_0\right]\expf{x-x_0}\bigg>
-\bigg<\Delta n_0(1+n^{\rm eq})+n^{\rm eq}_0\,\Delta n\bigg>
\nonumber\\
&= \bigg<\Delta n\left[(1+n^{\rm eq}_0)\,\expf{x-x_0}-n^{\rm eq}_0\right]\bigg>
-\bigg<\Delta n_0(1+n^{\rm eq}-n^{\rm eq}\,\expf{x-x_0})\bigg>
\nonumber\\
&= \bigg<\Delta n\,\expf{x-x_0} f_{\omega,\omega_0}\bigg>-\bigg<\Delta n_0\,f_{\omega_0,\omega}\bigg>
\nonumber\\ \nonumber
&\equiv \int P(\omega_0 \rightarrow \omega, \The) \,\Big[\expf{x-x_0} \,\Delta n \, f_{\omega,\omega_0} - \Delta n_0\, f_{\omega_0,\omega}\Big]\,\id\omega
\\[1mm]
f_{\omega_0,\omega}&=\frac{1+n^{\rm eq}}{1+n^{\rm eq}_0}=\frac{1-\expf{-x_0-\muc}}{1-\expf{-x-\muc}}.
\nonumber
\end{align}
\esub
%-------------------------------------------------------------------
Here, we used the identity $\expf{\frac{\omega-\omega_0}{\The}}\equiv[(1+n^{\rm eq})/n^{\rm eq}]\,[n^{\rm eq}_0/(1+n^{\rm eq}_0)]$ to obtain the factors of $f_{\omega_0,\omega}$.

\subsection{Intermediate steps for Eq.~\eqref{eq:target_moments_kin}}
\label{app:intermediate_Eq8}
%-------------------------------------------------------------------
The moments of the Boltzmann equation~\eqref{eq:kin_eq2_starting_point} can be simplified in the following manner: 
%-------------------------------------------------------------------
\begin{align}
\int \omega_0^{k+2}\frac{\text{d}n_0}{\text{d}\tau} \id\omega_0
&= 
\int \omega_0^{k+2} P^{\rm th}(\omega_0 \rightarrow \omega) \,\expf{\frac{\omega-\omega_0}{\The}} \,n(1+n_0) \id\omega\id\omega_0
\nonumber\\\nonumber
&\qquad
- \int \omega_0^{k+2} P^{\rm th}(\omega_0 \rightarrow \omega) \,n_0(1+n)\id\omega\id\omega_0
\nonumber\\\nonumber
&= 
\int \omega_0^{k+2} \frac{\omega^2}{\omega_0^2}\,P^{\rm th}(\omega \rightarrow \omega_0) \,n(1+n_0) \id\omega\id\omega_0
\nonumber\\\nonumber
&\qquad
- \int \omega_0^{k+2} \left[\int P^{\rm th}(\omega_0 \rightarrow \omega) (1+n)\id\omega\right]\,n_0\id\omega_0
\nonumber\\\nonumber
&= 
\int \omega^{k+2} \left[\frac{\omega^k_0}{\omega^k}\,P^{\rm th}(\omega \rightarrow \omega_0) (1+n_0) \id\omega_0\right]
\,n\id\omega
\nonumber\\\nonumber
&\qquad
- \int \omega_0^{k+2} \left<(1+n)\right>\,n_0\id\omega_0
\nonumber\\\nonumber
&\equiv
\int \omega_0^{k+2} \left< \left[\frac{\omega^k}{\omega_0^k}-1\right](1+n)\right>\,n_0\id\omega_0.
\nonumber\\\nonumber
\end{align}
%-------------------------------------------------------------------
In the last step, we switched the roles of $\omega_0$ and $\omega$ to group terms together. This then yields Eq.~\eqref{eq:target_moments_kin}.

\subsection{Intermediate steps for Eq.~\eqref{eq:target_moments_kin_lin_sim}}
\label{app:intermediate_Eq11}
%-------------------------------------------------------------------
%-------------------------------------------------------------------
\begin{align}
\int \omega_0^{k+2} \frac{\text{d}n_0}{\text{d}\tau} \id \omega_0
&\approx
\int \omega_0^{k+2} \,\left<\left[\frac{\omega^k}{\omega_0^k}  - 1\right] (1+n^{\rm eq}) \right>\,\Delta n_0 \id \omega_0
\nonumber
\\
&\qquad +
\int \omega_0^{k+2} \,\left<\left[\frac{\omega^k}{\omega_0^k}  - 1\right] \Delta n \right>\,n^{\rm eq}_0 \id \omega_0
\nonumber\\
&\equiv 
\int \omega_0^{k+2} 
\left<\left[\frac{\omega^{k}}{\omega_0^{k}}-1\right] (1+n^{\rm eq}) \right> 
\,\Delta n_0 \,\text{d}\omega_0
\nonumber\\
&\qquad-
\int \omega_0^{k+2}\left<\left[\frac{\omega^{k}}{\omega_0^{k}}-1\right] n^{\rm eq} \expf{\frac{\omega-\omega_0}{\The}}\right> 
\,\Delta n_0 \,\text{d}\omega_0
\nonumber\\[1mm]
&= 
\int \omega_0^{k+2}\left<\left[\frac{\omega^{k}}{\omega_0^{k}}-1\right] f_{\omega_0,\omega}\right> 
\,\Delta n_0 \,\text{d}\omega_0.
\nonumber
\end{align}
%-------------------------------------------------------------------
In the intermediate step we used the detailed balance relation and the identity $\expf{\frac{\omega-\omega_0}{\The}}\equiv[(1+n^{\rm eq})/n^{\rm eq}]\,[n^{\rm eq}_0/(1+n^{\rm eq}_0)]$ to switch the roles of $\omega_0$ and $\omega$.

\section{Diffusion solver coefficients}
\label{app:diff_solver_coefficients}
%-------------------------------------------------------------------
For the Kompaneets limit, the required coefficients for our diffusion solver are \citep[see also][]{Chluba2011therm}
%-------------------------------------------------------------------
\bsub
\begin{align}
A&= x_0^2,\;
B= x_0^2\left(1+\frac{4}{x_0}\right),
\; C=4 x_0
\\
A^*&= x_0^2,\;
B^*= x_0^2\left(f+\frac{4}{x_0}\right),
\;
C^*=x^2_0f\left(\frac{4}{x_0}-\frac{2\expf{-x}}{1-\expf{-2x}}\right)
\end{align}
\esub
%-------------------------------------------------------------------
with $f=1+2n_0^{\rm eq}$.
The required coefficients for our improved FP schemes are given by
%-------------------------------------------------------------------
\bsub
\label{eq:Coefficients}
\begin{align}
A(\omega, \The)&=x_0^2 \,\frac{\Sigma_2}{2\The}
\\
B(\omega, \The)
&=x_0 \left\{\frac{4\Sigma_2-\Sigma_1}{\The}+\frac{x_0\Sigma'_2}{\The}\right\}
\\
C(\omega, \The)&=\frac{3(2\Sigma_2-\Sigma_1)}{\The}+\frac{x_0(4\Sigma'_2-\Sigma'_1)}{\The}+x_0^2\frac{\Sigma''_2}{2\The},
\end{align}
\esub
%-------------------------------------------------------------------
where the primes denote derivatives with respect to $x_0$. These expressions agree with the formulation of \citet{Belmont2008} when stimulated terms are neglected. The derivatives of the kernel moments are computed numerically values of $\Sigma_1$ and $\Sigma_2$ on the grid of frequency points. Stimulated terms are easily included by replacing $\Sigma_k \rightarrow \Sigma_k^*$, as explained in Sect.~\ref{sec:New_attempt_stim}.

\end{appendix}

\end{document}